\documentclass[prfluids,onecolumn,groupedaddress]{revtex4-2}
\usepackage{graphicx}
\usepackage{dcolumn}
\usepackage{bm}
\usepackage{siunitx}
\usepackage{amsmath}
\usepackage{amssymb}
\usepackage{xcolor}

\begin{document}


\title{Pathways from nucleation to raindrops.}


\author{F. Poydenot, B. Andreotti}
\affiliation{Laboratoire de Physique de l’École normale supérieure, ENS, Université PSL, CNRS, Sorbonne Université, Université Paris Cité, F-75005 Paris, France}

\date{\today}

\begin{abstract}
Cloud droplets grow via vapor condensation and collisional aggregation. Upon reaching approximately $\approx 100~{\rm \mu m}$, their inertia allows them to capture smaller droplets during descent, initiating rain. Here, we show that raindrop formation is not primarily governed by gravity or thermal diffusion, but by a critical range of drop sizes ($3-30~{\rm \mu m}$) where collisions are largely ineffective and controlled by van der Waals and electrostatic interactions. We identify four pathways to rain. The \emph{coalescence pathway}, which is slow, involves the broadening of the drop size distribution across the $3-30~{\rm \mu m}$ low-efficiency gap through collisions, until enough large individual droplets achieving efficient collisions have formed. The \emph{mixing pathway}, which is faster, requires mixing at the cloud top with drop-free, cold, humid air to create locally supersaturated conditions that grow droplets above the low-efficiency gap. The \emph{electrostatic pathway} bypasses the gap through a static vertical field creating attractive interactions between droplets. The \emph{turbulence pathway} relies on air turbulence to bring the droplets together at an increased rate, but we show that this pathway is unlikely. For all dynamical mechanisms, we demonstrate that the initiation time for rainfall occurs at the crossover between the broadening of the drop size distribution and the emergence of individual droplets large enough to trigger the onset of the rainfall cascade.
\end{abstract}

\maketitle

\section{Introduction}
\subsection{Clouds and global warming}
Anthropogenic climate change is a pressing issue that requires urgent attention and action \cite{lee_ipcc_2023}. The complexity of the problem arises from the intricate interplay between human activities and natural phenomena across a broad range of spatial and temporal scales. Among the various components of the Earth's climate system, clouds play a pivotal role in regulating precipitation patterns, water resources, and agriculture. Moreover, clouds exhibit radiative effects that can vary depending on their geometry, clustering, and altitude \cite{tobin_observational_2012}. Despite the overwhelming evidence for global warming, predicting its consequences on regional to global scales and the system's responses to potential mitigation or adaptation efforts remains challenging.

\begin{figure}[h]
\centering
\includegraphics{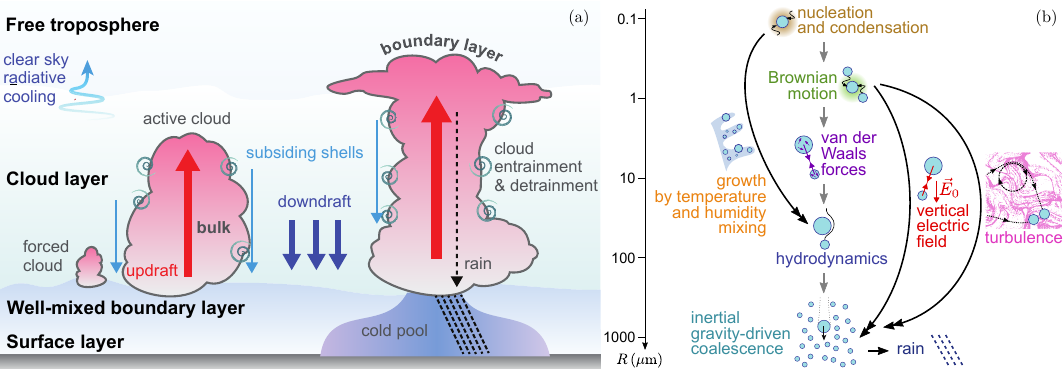}
\caption{(a) Schematic of the structure of the atmosphere in the subtropics, illustrating the key physical processes that lead to self-organization at the mesoscopic scale (adapted from \cite{muller_spontaneous_2022}). (b) Pathways to rain investigated in this paper. Along the slow pathway, shown vertically in the center, droplets first grow by condensation and then from micron size to rain through successive collisional regimes involving Brownian motion, van der Waals forces, hydrodynamic interactions, and finally a highly efficient coalescence cascade to rain. This pathway requires a relatively large initial liquid water content to trigger rain rapidly enough. It can be bypassed if temperature and humidity mixing produce very large drops (orange, left), if there is a vertical electric field that increases the collision rate in the regime between Brownian and gravity-driven inertial collisions (red, center right), where the collision efficiency is very low, or if turbulence can increase the collision rate (pink, right) (adapted from \cite{bec_multifractal_2005} and \cite{falkovich_rain_2006}).}
\label{fig:schematic}
\end{figure}

One of the significant sources of uncertainty in climate projections is the response of clouds to warming \cite{bony_clouds_2015,boucher_presentation_2020,zhang_sensitivity_1995,schmidt_ceresmip_2023}. Warm trade wind cumulus and stratocumulus clouds, which are prevalent in the subtropics, have a substantial impact on the global planetary albedo. These clouds are highly reflective and can cool the Earth's surface by reflecting a significant amount of incoming solar radiation \cite{stevens_sugar_2020}. On the other hand, deep convective clouds, which are common in the tropics, can affect the atmospheric energy balance, circulation, and precipitation patterns. The behavior of these clouds under global warming is critical for understanding and predicting the Earth's energy budget, hydrological cycle, and climate sensitivity \cite{schneider_climate_2017,stephens_radiation_1978,bony_observed_2020,muller_spontaneous_2022}. The atmosphere can be conceptualized as having a four-layer structure in the lowest portion of the troposphere (Fig.~\ref{fig:schematic}). The surface layer, where thermals, i.e. hot, humid, buoyant parcels form, is characterized by strong temperature gradients and turbulent mixing. Above the surface layer, a well-mixed convective layer is present, which is in the ultimate convective turbulent regime \cite{ahlers_heat_2009}. Forced, passive clouds form at the top of thermals, above the condensation level. This layer interacts with the weakly stratified cloud layer, where the convective ascent of active clouds is partly driven by latent heat exchange during condensation and evaporation processes. This layer is characteristic of cumulus-type clouds. In its absence, clouds rise within a background of turbulent convection, rather than within a stably stratified background, which is characteristic of stratocumulus clouds. Finally, the cloud layer is capped by a temperature inversion, inhibiting the ascent of clouds that would rise to this altitude and exhibiting significant radiative cooling. The strongly stratified free troposphere sits above this capping layer. This vertical structure determines the fate of a moist, buoyant parcel rising from the surface layer that could form a cloud and initiate rain. First, the parcel encounters colder air above it through the mixing layer, enhancing its updraft. In the cloud layer, latent heat release through condensation warms the parcel so that its top stays hotter than the surrounding air and keeps rising. Once it reaches the inversion, the parcel becomes capped by warmer air and stops ascending.

\subsection{Warm clouds}
In warm clouds, where ice is absent, the formation of droplets occurs on solid particles known as cloud condensation nuclei (CCN), as moist air ascends via convection and reaches the condensation level. The concentration of CCN can significantly impact the number and size of cloud droplets, which in turn can affect the cloud's radiative properties and precipitation potential. The growth of these droplets via condensation ceases when the surrounding air becomes saturated \citep{twomey_nuclei_1959,ghan_droplet_2011}. To better understand this process, let us consider droplets within a closed, isothermal system, assumed to be well-mixed. This system initially contains a certain number of soluble nuclei per unit volume, denoted as $\psi$, and a supersaturated water vapor density, denoted as $\rho_v^0$. The mass growth rate of a droplet of radius $R$ condensing in an atmosphere with vapor density $\rho_v$ is given by \citep[chap. 13]{pruppacher_microphysics_2010}:
\begin{equation}
\frac{dm}{dt}=\frac{4\pi}{3} \rho_\ell \frac{\mathrm d R^3}{\mathrm d t} = 4\pi D (\rho_v - \rho_\mathrm{sat}) R,
\label{eq:radiusevolution}
\end{equation}
Here, $\rho_\ell$ represents the density of liquid water, and $D$ is an effective diffusion coefficient of water molecules in air. Assuming that all $\psi$ droplets per unit volume have the same radius, the water vapor density can be expressed as $\rho_v = \rho_v^0 - 4\pi \psi \rho_\ell R^3/3$, where $\rho_\ell$ is the water liquid density. The evolution equation governing the drop size $R$ can then be rewritten as:
\begin{equation}
\frac{\mathrm d R^2}{\mathrm d t} = \frac{R_\infty^2}{\tau_D}\left(1 - \frac{R^3}{R_\infty^3}\right)
\end{equation}
In this equation, $R_\infty$ is the equilibrium radius determined by mass conservation, given by
\begin{equation}
\frac 43 \pi\psi \rho_\ell R_\infty^3 = \rho_v^0 - \rho_\mathrm{sat}
\label{eq:revolutionpsi}
\end{equation}
As a consequence, the volume fraction of liquid water in clouds, typically less than $10^{-6}$, is thermodynamically controlled by the difference between $\rho_v^0$ and $\rho_\mathrm{sat}$.
$\tau_D$ is a characteristic growth time defined as:
\begin{equation}
\tau_D=\frac{\rho_\ell}{\rho_v^0 - \rho_\mathrm{sat}}\;\frac{R_\infty^2}{D}.
\end{equation}
This results in a trade-off between the typical drop size and the number of drops per unit volume. A higher concentration of condensation nuclei favors a large number of small drops over a small number of large drops \citep{krueger_technical_2020}. The number of drops per unit volume in warm clouds is typically around $\psi \sim 10^8\;{\rm m^{-3}}$, leading to micron-sized droplets as a result of condensation growth \citep{hess_optical_1998}. In clouds, $\rho_v^0$ is on the order of $1\;{\rm g.m^{-3}}$, and $R_\infty$ is in the range of $1$ to $10\;{\rm \mu m}$, resulting in $\tau_D$ ranging from seconds to minutes.

Cloud models, as well as weather and climate models, typically include a set of equations derived from thermofluidic equations, such as mass, momentum, and energy conservation. In this context, the droplet size distribution is a function of droplet size, space, and time. While some models directly solve the evolution equation of the distribution, allowing its shape to be obtained without imposing it a priori \cite{khain_notes_2000,seifert_comparison_2006,khain_representation_2015,kreidenweis_100_2019} these methods can be numerically costly for simulated domains exceeding a few kilometers \cite{flossmann_review_2010}. To solve larger domains, including some climate models, bulk parameterization schemes divide the size distribution into a few classes \cite{cotton_storm_2011,hansen_global_2023,schmidt_ceresmip_2023}. Based on the work of \citet{kessler_distribution_1969} and \citet{berry_analysis_1974}, two drop populations are typically considered: cloud drops and raindrops, with a separation around $30\,{\rm \mu m}$. Each class is represented by a small number of its moments, typically concentration and total mass. Transition rates between classes are then introduced phenomenologically, including autoconversion (cloud droplets becoming rain), accretion (rain droplets collecting cloud droplets), and autocollection (rain droplets collecting rain droplets)
 \cite{kessler_distribution_1969,berry_analysis_1974-3,hall_detailed_1980,seifert_double-moment_2001,morrison_confronting_2020}. Some recent models simulate a small number of representative "super-droplets" in a Lagrangian description to avoid the cost of an Eulerian description of the droplet field \cite{shima_super-droplet_2009,grabowski_modeling_2019,shima_predicting_2020}. However, in all cases, the aggregation coefficients still need to be calculated a priori with good mechanistic knowledge of the processes involved to accurately describe the microphysics at the population level. For example, the collision rates between two cloud drops reported in the literature are sometimes inconsistent and calculated over narrow size ranges, leading to poorly resolved crossovers between different mechanisms \cite{khain_notes_2000}.

\subsection{Rain formation}
Rain formation in warm clouds is a complex dynamic process that remains an active research topic in atmospheric sciences. The initiation of rain in warm clouds has been studied from two different perspectives in the literature. The first approach, proposed by \citet{bowen_fomation_1950}, focuses on the growth of a single drop coalescing with a cloud of fixed droplets. The second approach involves numerically integrating the droplet distribution as a whole, which allows for the consideration of collective effects that can reduce the growth time \citep{telford_new_1955,berry_analysis_1974,falkovich_rain_2006}. Various explanatory mechanisms have been proposed to account for the observed dynamics of rain formation in warm clouds. These include stochastic effects of "lucky drops" that are able to grow faster than their neighbors \cite{kostinski_fluctuations_2005,wilkinson_large_2016}, turbulence-induced rare but intense collisions \cite{ghosh_how_2005,wilkinson_caustic_2006,wang_role_2009,devenish_droplet_2012,grabowski_growth_2013,pumir_collisional_2016}, the presence of giant condensation nuclei \cite{szumowski_microphysical_1999,mechem_bulk_2008,posselt_influence_2008,barahona_comprehensively_2010}, the break up of centimer-scale drops into smaller droplets \cite{blanchard_behavior_1950,villermaux_fragmentation_2020} and even radiative effects \cite{zeng_modeling_2018}.

Despite these proposed mechanisms, the rapid appearance of rain within just $15$ to $30$ minutes in warm clouds remains largely unexplained and poses a significant challenge to understanding the underlying mechanisms \cite{beard_warm-rain_1993}. One of the main complexities of rain formation in warm clouds is the vast separation between the concentration of raindrops and that of micron-sized droplets. The concentration of raindrops in clouds is typically $10^{-5}$ smaller than that of the smaller droplets \citep{mcfarquhar_rainfall_2022}. Due to this substantial size difference, rain in warm clouds must be formed through the collision and coalescence of cloud droplets. This process involves the merging of numerous droplets: one million droplets of radius $10\;{\rm \mu m}$ are required to form a single millimeter-sized raindrop. Therefore, understanding the mechanisms behind the rapid appearance of rain in warm clouds remains an important research question in atmospheric physics.
\begin{figure}[h]
\centering
\includegraphics{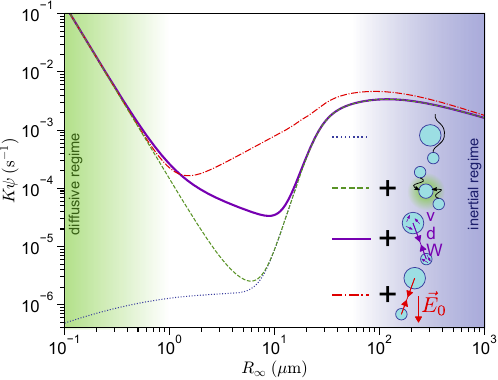}
\vspace{-3 mm}
\caption{Collision rate of a drop of radius $1.1 R_\infty$ with a homogeneous spray of radius $R_\infty$ and liquid water content $\rho_d = 4\pi \rho_\ell \psi R_\infty^3/3 = 1~{\rm g.m^{-3}}$, which sets the density $\psi$. Dotted blue line: hydrodynamics alone; dashed green line: adding thermal diffusion; solid purple line: adding furthermore van der Waals interactions; dashed dotted red line: adding finally a static electric field. The asymptotic thermal diffusion regime is shown in shaded green, and the asymptotic inertial regime is shown in shaded blue.}
\vspace{-3 mm}
\label{fig:collisionrate}
\end{figure}

Here, we propose a new method for mechanistically understanding the different contributions to growth towards rain. In Sec.~\ref{sec:model}, we describe the mean-field coagulation equation modelling the evolution of the drop size distribution, the properties of the collision kernel describing collisions due to thermal diffusion and gravity, taking into account inertial effects in the gas flow, non-continuum lubrication, flow inside the drops and electrostatic interactions, as well as the numerical integration scheme. We investigate in Sec.~\ref{sec:pathway} the interplay between the various microphysical mechanisms and their effects on the drop population growth dynamics. Finally in Sec.~\ref{sec:raininitiation}, we discuss under which conditions rain can be initiated rapidly enough and which mechanisms are significant in this process. Given the structure of the collision kernel, we isolate four possible scenarios for rain formation in warm clouds.

\section{Model}
\label{sec:model}

\subsection{Smulochowksi equation}
The modeling of the coagulation of droplets in warm clouds is based on the Smoluchowski equation, which describes the evolution of an ensemble of drops uniformly distributed in space and growing through successive pair coalescence \cite{von_smoluchowski_three_1916}. This non-linear, integro-differential partial differential equation is mathematically complex and requires the expression of coefficients that describe the transition from one size to another. Denoting by $n(m,t)$ the density of drops of mass $m$, this evolution equation reads \cite{von_smoluchowski_three_1916}:
\begin{equation}
\frac{\partial n}{\partial t} = \frac{1}{2} \int_0^m \mathrm d x\, n(m-x) n(x) K(m-x, x) - n(m) \int_0^\infty \mathrm d x\, n(x) K(m, x)
\label{eq:smolu}
\end{equation}
where $K$ is the collisional kernel \cite{gillespie_stochastic_1972,gillespie_three_1975,pruppacher_microphysics_2010,khain_physical_2018}. $K(x, m)n(x)\mathrm{d}x$ is the collision rate of a single drop of mass $m$ with the population of drops of size $x$. However, there is considerable uncertainty as to the details of the processes involved, particularly for droplets of commensurable sizes between $3$ and $30\,{\rm \mu m}$. The droplet size distribution in clouds broadens over time as droplets grow \citep{brenguier_droplet_2001}, which initially involves the interaction of micron-sized droplets of similar sizes. Fogs exhibit similar behaviors, evolving over time through condensation and collision. Identifying the mechanism of droplet growth from observations remains an active research question \cite{mazoyer_experimental_2022}. The complexity of dynamic measurements of drop size distributions in situ and the difficulty of separating out microphysical processes make numerically integrated theoretical models an important means of understanding cloud physics.

\subsection{Collisional kernel}
The Smoluchowski equation (\ref{eq:smolu}) admits simple solutions when the kernel $K$ follows self-similarity relations, which are defined by two conditions. The first condition is the homogeneity property $K(c x, c y) = c^\lambda K(x, y)$ for all $c$ \cite{van_dongen_solutions_1987,aldous_deterministic_1999,fournier_existence_2005,herrmann_instabilities_2016}. The second condition is the asymptotic behavior $K \propto x^\mu y^\nu,$ which applies for $x \ll y$ and symmetrically when $x \gg y$ by interchanging $\mu$ and $\nu$. These conditions are met by certain physically relevant kernels, as discussed in Appendix~\ref{sec:selfsimilarkernels}. In this generic case, the distribution evolves regularly, with the largest drops growing according to power laws of time. However, this evolution lacks the distinctive characteristic of rain: the abrupt emergence of a few very large drops.

To elucidate the formation of rain, it is essential to account for the collective mechanisms governing the coalescence of drops of varying sizes, which renders the kernel $K$ not a homogeneous function. In the cloud physics literature, the nonhomogeneity effects in the kernel are typically represented by the collision efficiency $E_d$, defined as the ratio between the full kernel and its asymptotic limit of ballistic gravity-driven collisions \cite{langmuir_production_1948,pruppacher_microphysics_2010}. In Ref.~\cite{poydenot_gap_2024}, we have extended this definition of the efficiency to include the small-size behavior of Brownian collisions in the asymptotic kernel. The complete kernel for two drops of radii $R_i$, falling at their terminal velocities $U^t_i$ and diffusing with the diffusion coefficient $D_i$, reads in this case
\begin{equation}
\label{eq:ed}
K = E_d \pi (R_1 + R_2)^2 \left|U^t_1-U^t_2\right| q_\mathrm{diff}.
\end{equation}
$q_\mathrm{diff}$ is a function of the Péclet number $\mathrm{Pe} = \left|U^t_1-U^t_2\right| (R_1+R_2)/(D_1+D_2)$ only, which quantifies the relative importance of diffusion and gravitational settling. When settling is much stronger than diffusion ($\mathrm{Pe} \gg 1$), $q_\mathrm{diff} \sim 1$ and the asymptotic kernel is purely ballistic. When diffusion is much stronger than settling ($\mathrm{Pe} \ll 1$), $q_\mathrm{diff} \sim 4/\mathrm{Pe}$, which corresponds to purely Brownian-driven collisions. In general, $E_d < 1$: the mechanisms creating nonhomogeneity in the kernel tend to inhibit collisions compared to idealized behaviors. However, as we shall see, it is not so much the value of the kernel that determines the growth dynamics, but rather its functional dependence with size.

The specifics of the kernel employed in this study are detailed in Ref.~\cite{poydenot_gap_2024}. This kernel encompasses Brownian motion, droplet inertia, inertial effects in the gas flow, non-continuum lubrication, flow inside the drops, van der Waals interactions, and induced dipole forces in the presence of a static electric field. It also incorporates reference models for the effects of turbulence \cite{pumir_collisional_2016}, which have a minor impact on the kernel under typical cloud conditions; their influence is discussed in Sec.~\ref{sec:raininitiation}. Figure~\ref{fig:collisionrate} illustrates the collision rate $K \psi$ of a single drop in a monodisperse spray of droplets with radius $R_\infty$ and number concentration $\psi$, including a drop slightly larger than the spray droplets. The liquid water content $\rho_d = \frac 43 \rho_\ell \psi R_\infty^3$ is kept constant, causing $\psi$ to vary with $R_\infty$ as $\psi \propto R_\infty^{-3}$. The collision kernel $K$ exhibits two homogeneous asymptotic regimes. For small sizes below $0.4 \,{\rm \mu m}$, Brownian diffusion dominates the collision rate. Collisions between droplets create a local concentration gradient, driving further collisions. Although larger droplets diffuse slower due to the Stokes-Einstein diffusion coefficient, their increased surface area allows for more collisions. These size dependencies balance each other, resulting in a homogeneity degree of $\lambda = 0$ in this regime so that $K \psi$ decreases as $R_\infty^{-3}$. In the regime above $200 \,{\rm \mu m}$, droplet inertia is large enough for them to fall in straight lines, overcoming the lubrication layer. The drag force on the droplets is also inertial, causing the terminal velocity $U^t$ to increase with their radius $R$ as $U^t \propto \sqrt{R}$. This results in a homogeneity degree of $\lambda = 5/6$ in this asymptotic regime. Contrary to conventional understanding, where the collision frequency transitions between a Brownian regime at small sizes and an inertial regime at large sizes \cite{pruppacher_microphysics_2010}, Fig.~\ref{fig:collisionrate} reveals a range of drop sizes between $3\,{\rm \mu m}$ and $30\,{\rm \mu m}$ where both thermal diffusion and gravity are inefficient at inducing coagulation. The collision rate displays a minimum in this transitional regime. There is a small subregime between $2 \,{\rm \mu m}$ and $10 \,{\rm \mu m}$ where van der Waals interactions dominate the kernel. The attractive effect of these forces counteracts the repulsion of the lubrication layer, leading to an effective scaling law $\lambda \approx 0.77$ in this size range (for a more comprehensive discussion on this effect, see~\citep{poydenot_gap_2024}). Around $20 \,{\rm \mu m}$, the collision rate significantly increases with size as droplet inertia overcomes the lubrication layer. In this narrow subregime, the kernel loses all homogeneity. We therefore hypothesize that the sudden emergence of a few large drops and the transition to rain in warm clouds originate from the complex structure of the collision kernel.
\begin{figure}[h]
\centering
\includegraphics{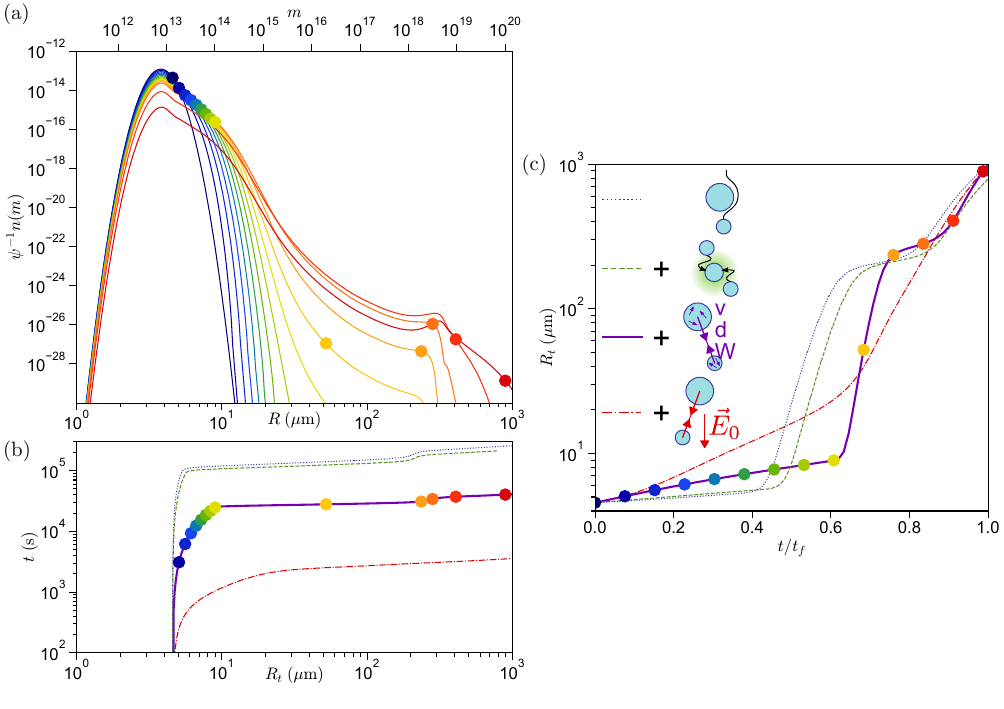}
\caption{(a) Mass distribution $n(m)$ at different times, obtained by integration of the Smoluchowski equation with the reference kernel $K$. Masses are expressed in units of the mass of a water molecule. (b) Characteristic drop size $R_t$ in the distribution tail as a function of time, taking into account hydrodynamics alone but neither thermal noise nor electrostatics (dotted blue line), including thermal diffusion (dashed green line), including furthermore van der Waals interactions (solid purple) and adding finally a static electric field parallel to gravity (dashed dotted red line). $\rho_d = 1~{\rm g.m^{-3}}$. (c) Characteristic drop size $R_t$ in the distribution tail as a function of the relative time $t/t_f$, where $t_f$ is the time needed to reach $R_t = 1~{\rm mm}$. The different curves take into account the same mechanisms as before. For all collisional mechanisms, growth until $10~{\rm \mu m}$ is slow, and once enough droplets have crossed $30~{\rm \mu m}$, droplets growth by an order of magnitude in radius very rapidly. Growth then slows down relatively due to concentration effects, as the mass of these large drops around $100~{\rm \mu m}$ is much larger than the mass of the droplets that make up most of the distribution, and large drops of comparable size are much fewer.
}
\label{fig:distributionmoments}
\end{figure}

\subsection{Numerical integration}
We revisit the problem of collisional aggregation of water droplets, considering the system at times larger than the condensation timescale $\tau_D$. Drops are initially distributed at a size around $R_\infty$, and the air between drops is saturated in water. A size-binning algorithm \cite{debry_solving_2007} is used to solve the Smoluchowski equation, dividing the drop size range $\SI{64}{nm} - \SI{7.7}{mm}$ into $549$ logarithmically distributed bins. Masses are expressed in units of the water molecule mass. The code is mass conservative. The transfer of mass between bins is computed using a quadratic interpolation of the mass density $n(m)$, biased in the direction of small masses. The scheme is upwind-biased and first order in time. The code has been validated on simple kernels leading to exact self-similar solutions, and on an exact solution derived for a particular kernel and initial solution \cite{golovin_solution_1963} (see Supplementary Material). From the moments ${\mathcal M}_p=\int m^p n(m) dm$, the average drop mass $m_m\equiv{\mathcal M}_1/{\mathcal M}_0=\frac 43 \pi \rho_\ell R_m^3$ describes the core of the distribution, and the tail is characterized by the typical drop mass $m_t\equiv{\mathcal M}_3/{\mathcal M}_2=\frac 43 \pi \rho_\ell R_t^3$. The initial number density of drops, $\psi$, controlled by the density of nuclei is used to normalize the drop density $n(m)$ so that the evolution depends on the initial shape of the distribution but not on the initial density $ \psi$, once time is rescaled by $1/\psi$.
\begin{figure}[t!]
\centering
\includegraphics{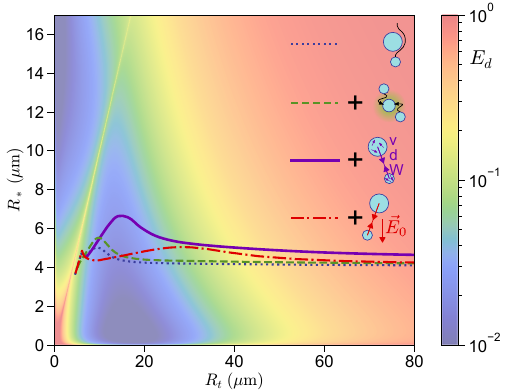}
\caption{Collisional trajectory as defined by the reaction size $R_*$ as a function of the typical tail size $R_t$, for an initial distribution centered at $4~{\rm \mu m}$. $R_*$ is the size that contributes half the growth rate of droplets of size $R_t$ through collisions with smaller droplets (see Appendix~\ref{sec:reactiontrajectory} for a formal definition). Color field: collisional efficiency $E_d$ [Eq.~(\ref{eq:ed})], i.e. the ratio between the kernel $K$ and an idealized, analytical kernel only taking into account the geometric effects of diffusion and gravity. The region of low efficiency (green to blue) corresponds to the regime where the reaction size $R_*$ is not the most populated size, but instead a larger, less populated size that can produce more efficient collisions.}
\label{fig:trajectory}
\end{figure}

\section{Route to raindrops}
\label{sec:pathway}

Figure \ref{fig:distributionmoments} shows the typical time evolution of the drop density $n(m)$, starting from a "cloud" of $4\;{\rm \mu m}$ drops formed by condensation. Consistent with the collision rate shown in Fig.~\ref{fig:collisionrate}, the drop density $n(m)$ evolves slowly until the tail starts presenting drops with a radius above $30~{\rm \mu m}$. Drops larger than this cross-over size efficiently absorb smaller drops as they fall in the air due to their surface and inertia, accelerating their growth over a much shorter time-scale. Figure~\ref{fig:distributionmoments} shows this extremely rapid growth of the drops in the distribution tail once the low collision rate gap is crossed. To understand the statistical trajectory for the growth of drops across the low collision rate gap and the drop sizes involved in the coagulation process, we define the reaction mass $m_*$ as the mass for which half the collisions leading to a mass $m$ involve a small drop with a mass $x$ smaller than $m_*$ (see Appendix \ref{sec:reactiontrajectory} for a formal definition). The reaction mass $m_*=\frac 43 \pi \rho_\ell R_*^3$ gives the typical size of the small drop involved in collisions producing drops of mass $m$. This reaction size is shown for the typical size of drops in the distribution tails $m_t$ in Fig.~\ref{fig:trajectory}. During the fast cascade producing raindrops, the most numerous collisions take place between the dominant species ($4\;{\rm \mu m}$ in the graph) and the largest drops. However, the situation is different during the crossing of the low collision rate gap, between $1$ to $30\,{\rm \mu m}$: the larger drops are formed by collisions inside the tails themselves, and not between drops in the tails and drops in the core of the distribution. This indicates that, despite drops larger than average being less numerous, the collision frequency is higher due to the sharp increase in collisional efficiency. The route to rain is therefore not controlled by a single dynamical mechanism that can easily be abstracted in an analytical model, but by subtle processes associated with the non-homogeneous structure of the collision kernel in the range of size surrounding the minimum of the collision efficiency.
\begin{figure}[t!]
\centering
\includegraphics{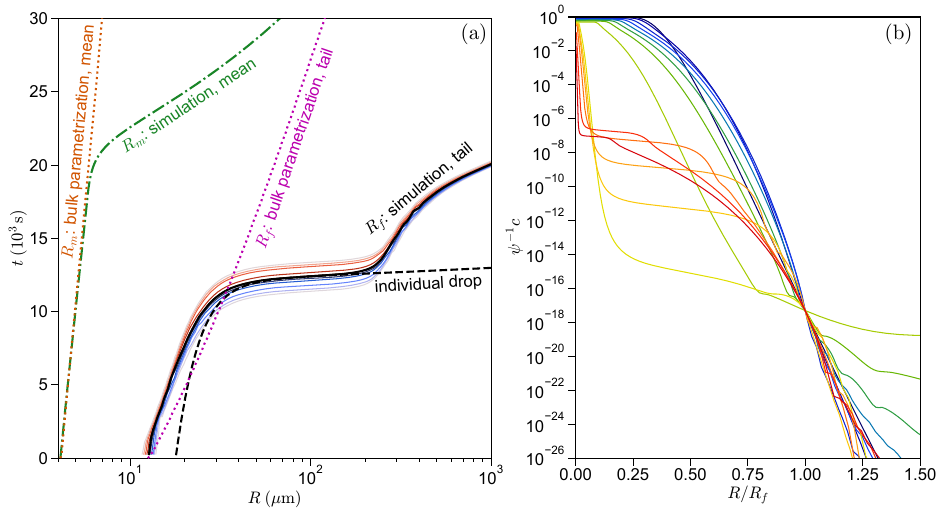}
\caption{(a) Characteristic sizes of the drop distribution as a function of time. For the full computation, dashed dotted green line: size in the distribution core $R_m$. Solid colored lines: position of the front $R_f$ for values of $c$ ranging from $10^{-15}$ (red) to $10^{-20}$ (blue). Black line: $R_f$ for $c=5\,10^{-18}$. Dashed black line: continuous growth for one droplet given by Eq.~(\ref{eq:growthrate2}). Results for the bulk parameterization are in dotted lines; orange: core size $R_m$, pink: position of the front $R_f$. (b) Cumulative distribution $c=\int_m^\infty n(x) \mathrm{d}x$ as a function of $R/R_f$, where the front size $R_f$ is given by $c = 5\,10^{-18}$. Colors and distributions are the same as those of Fig.~\ref{fig:distributionmoments}. The curves collapse at all times for $R\gtrsim R_f$, except right at the point of gap crossing.}
\label{fig:asymptoticmoments}
\end{figure}

This result suggests a transition along the trajectory from nucleation to raindrops, shifting from a collective dynamics involving all possible collisions in the core of the distribution to a dynamics controlled by individual drops collecting mass while falling. Can these dynamics be described by low-dimensional models, such as the bulk parameterization schemes commonly used in cloud microphysics? To test this, we solve for the distribution $n_l(m)$ using a reduced dimensionality model obtained by employing log-normal distributions as test functions. The equation is projected onto equations for the moments of the logarithm $\int \ln(m)n(m)\mathrm{d}m$ and $\int \ln^2(m)n(m)\mathrm{d}m$. Figure~\ref{fig:asymptoticmoments} shows that this approximation quantitatively predicts the slow increase of the mean size $R_m$ at small sizes but fails to capture the cascade towards raindrops. Assuming that the few drops in the tail do not interact with each other, but only with the numerous small drops, we label each drop by a particular value of the cumulative distribution $c=\int_m^\infty n(x) \mathrm{d}x$. Figure~\ref{fig:asymptoticmoments}(b) shows that $c$ falls on a master curve when plotted as a function of the drop radius $R$ rescaled by a radius $R_f(t)$ obtained at a reference value of $c$ (here chosen at $5\,10^{-18}$). This indicates that drops belonging to the distribution tail follow a similar path $R(t)$. Considering a single drop absorbing small drops distributed as predicted by the low dimensional model (bulk parametrization), its mass $m=\frac 43 \pi \rho_\ell R^3$ grows according to:
\begin{equation}
\frac{\mathrm{d} m}{\mathrm{d}t}=4\pi \rho_\ell R^2 \frac{dR}{dt}= \int_0^\infty \mathrm d x K(x, m) x n_l(x).
\label{eq:growthrate2}
\end{equation}
Figure~\ref{fig:asymptoticmoments}(a) shows a good agreement at intermediate times when the mass contained in raindrops is negligible compared to the total mass. However, at longer times, the amount of small drops decreases due to their absorption by raindrops, leading to a new collective regime in which raindrops interact with each other indirectly through the depletion of the number of small drops.

\section{Rain initiation time}
\label{sec:raininitiation}

\subsection{Coalescence pathway}

\begin{figure}[t!]
\centering
\includegraphics{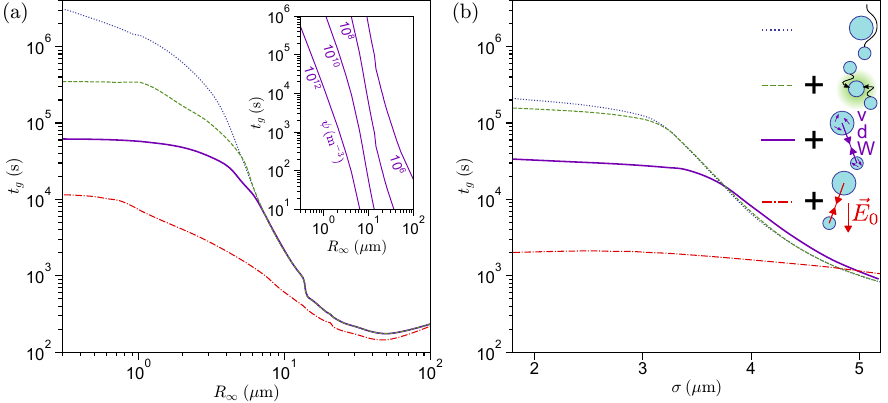}
\caption{(a) Rain initiation time $t_g$ as a function of the initial size $R_\infty$ for a constant liquid water content $\rho_d = 1~{\rm g.m^{-3}}$. The density $\psi$ is determined by the liquid water content $\rho_d = 4\pi \rho_\ell \psi R_\infty^3/3$: larger droplets are less numerous. Inset: $t_g$ as a function of $R_\infty$ for constant density $\psi$. (b) Rain initiation time $t_g$ as a function of the initial distribution width $\sigma$ for $R_\infty = 4\,{\rm \mu m}$. Dotted blue lines: hydrodynamics alone; dashed green lines: adding thermal diffusion; solid purple lines: further including van der Waals interactions; dashed-dotted red lines: additionally incorporating a static electric field large enough to close the low collision rate gap (field strength $50~{\rm kV.m^{-1}}$).}
\label{fig:initialsize}
\end{figure}

The first transition pathway to rain, starting from droplets of a few microns, involves the growth of droplets through coalescence. Figure~\ref{fig:distributionmoments} illustrates that the rain initiation time $t_g$ of this \emph{coalescence pathway} to rain is directly determined by the time required for drops to grow across the low collision efficiency gap. This process is governed by the delicate balance between dynamical mechanisms persisting in the range of scales where both Brownian motion and inertia are small compared to other effects. Indeed, although collisions are driven by gravity and are limited by hydrodynamic effects, particularly by the lubrication layer, small effects, especially electrostatic ones, produce significant variations in efficiency. Figure \ref{fig:initialsize}(a) summarizes the dependence of the rain initiation time $t_g$ on the drop size $R_\infty$. For comparison, in the context of a warm cumulus cloud, $t_g$ is observed to be on the order of $10^3~{\rm s}$ \cite{beard_warm-rain_1993,falkovich_rain_2006}. For this slow pathway to occur within a reasonably short time, enough droplets must be available for this nonlinear process to unfold rapidly enough. The rain initiation time $t_g$ is thus controlled by the liquid water content $\rho_d$, as $t_g \propto 1/\rho_d$. Throughout this paper, the size of the initial distribution is varied while keeping a constant liquid water content, $\rho_d = 1~{\rm g.m^{-3}}$. This implies that larger initial drops are also less numerous. However, it is also feasible to vary the initial size for a given density, $\psi$, which means that larger droplets also correspond to a larger liquid water content. For a given initial droplet size, $R_\infty$, $t_g$ is given by $t_g \propto \rho_d^{-1}$. When $\psi$ is kept constant, $\rho_d$ is proportional to $R_\infty^{-3}$, which explains the rapid decrease of $t_g$ as size increases in this scenario. The curves of $t_g$ vs $R_\infty$ for any liquid water content can be obtained from Fig.~\ref{fig:initialsize}(a) by shifting the curves vertically, towards larger times as $\rho_d$ decreases. Starting from a cloud composed of $4~{\rm \mu m}$ drops, one sees that $t_g \sim 10^3~{\rm s}$ would correspond to $\rho_d \sim 300~{\rm g.m^{-3}}$, which is far above any measured liquid water content. The insert in Fig.~\ref{fig:initialsize}(a) presents the dependence of $t_g$ on the droplet density $\psi$. The rain initiation $t_g$ is not only determined by $R_\infty$ but is also strongly influenced by the width of the initial distribution, as depicted in Fig.~\ref{fig:initialsize}(b) using the standard deviation, $\sigma$. Notably, the presence of a few large droplets in the tail can initiate rain just as rapidly as increasing the average drop size.

\subsection{Mixing pathway}

\begin{figure}[t!]
\centering
\includegraphics{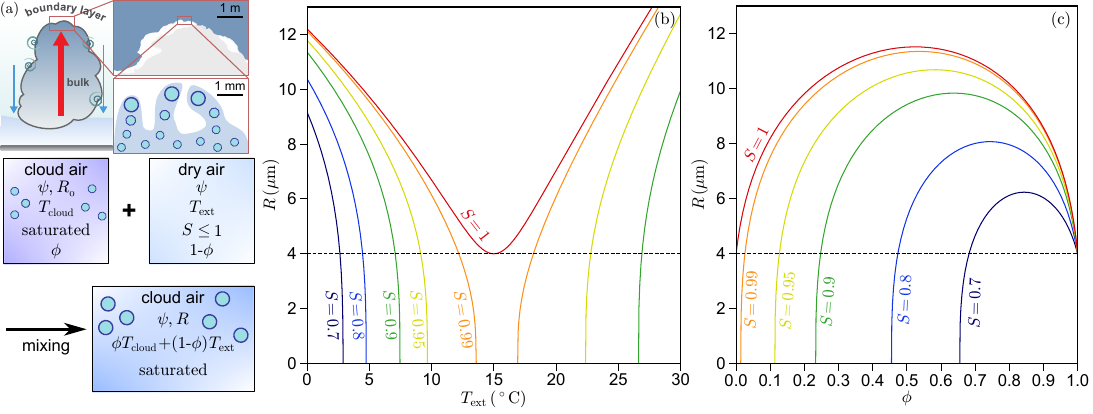}
\caption{(a) Schematic of the mixing process. Wet, cloud air is mixed with unsaturated, drop-free air at a different temperature. The process is represented by the isobaric mixing of two different air volumes on a timescale much shorter than the droplet condensation growth time. Cloudy air is saturated at $T = T_\mathrm{cloud}$, droplet concentration $\psi$, droplet size $R_0$ and volume fraction $\phi$ in the mixture. The air above the cloud is at $T = T_\mathrm{ext}$, at saturation $S \leq 1$ with a volume fraction $1-\phi$. The mixed air relaxes to saturation by either growth or evaporation of the drops, which stay at a constant number concentration $\psi$. (b) Droplet size $R$ after mixing as a function of the external air temperature $T_\mathrm{ext}$, for varying external air saturations $S$. The cloud is initially at $T_\mathrm{cloud} = 15~{\rm ^\circ C}$, with a number concentration $\psi = 10^8~{\rm m^{-3}}$ of droplets of $R_0 = 4~{\rm \mu m}$. Its volume fraction in the mixed air is $\phi = 0.7$. Both mixing with colder or warmer air can lead to droplet growth due to the change in saturation, although mixing with warmer air requires it to be closer to saturation to avoid evaporation. (c) Droplet size as a function of the mixing volume fraction $\phi$ for $T_\mathrm{cloud} = 15~{\rm ^\circ C}$, $T_\mathrm{ext} = 2.5~{\rm ^\circ C}$, $\psi = 10^8~{\rm m^{-3}}$.}
\label{fig:mixingpathway}
\end{figure}

If the mean drop size $R_\infty$ exceeds $10~{\rm \mu m}$, then Fig.~\ref{fig:initialsize}(a) shows that the coalescence dynamics changes completely: the distribution contains drops large enough to enter the zone of efficient collisions, and the time $t_g$ becomes much less sensitive to the nature of the mechanisms determining the collision efficiency near its minimum. However, drops can hardly grow by condensation up to $10~{\rm \mu m}$ in the bulk of the cloud. We therefore hypothesize that the growth of drops across the collisional barrier takes place in the upper part of the cloud, where mixing with humid air at a different temperature can occur. This second transition pathway to rain thus involves creating particular conditions of saturation and CCN concentration \emph{above} the cloud. To illustrate this fast \emph{mixing pathway} to rain, we consider the isobaric mixing of a fraction $\phi$ of saturated air from the cloud with a fraction $(1-\phi)$ of dry, drop-free air [Fig.~\ref{fig:mixingpathway}(a)]. Initially, the cloud air is at temperature $T_\mathrm{cloud}$ and is saturated with a water vapor density $\rho_\mathrm{sat}(T_\mathrm{cloud})$. It contains a monodisperse population of droplets with number concentration $\psi$ and radius $R_0$. The dry air above the cloud is at temperature $T_\mathrm{ext}$ and is subsaturated with a saturation $S \leq 1$, resulting in a water vapor density of $S \rho_\mathrm{sat}(T_\mathrm{ext})$. We assume that the density of cloud condensation nuclei (CCN) is the same inside and outside the cloud, leaving the cloud droplet concentration $\psi$ unchanged. The temperature and water vapor density after complete mixing are given by:
\begin{align}
T^f &= \phi T_\mathrm{cloud} + (1-\phi) T_\mathrm{ext} \\
\rho_v^f &= \phi \rho_\mathrm{sat}(T_\mathrm{cloud}) + (1-\phi) S \rho_\mathrm{sat}(T_\mathrm{ext})
\end{align}
We assume that the mixing process occurs on a timescale much shorter than the droplet condensation growth time. The mixing process can either create supersaturated conditions if $\rho_v^f > \rho_\mathrm{sat}(T^f)$, or evaporate droplets otherwise. Once condensation/evaporation has taken place, the final droplet size $R$ is given by Eq.~(\ref{eq:revolutionpsi}):
\begin{equation}
\frac{4\pi}{3} \psi (R^3-R_0^3) = \frac{\rho_v^f-\rho_\mathrm{sat}(T^f)}{\rho_\ell}.
\end{equation}
Figure~\ref{fig:mixingpathway}(b) shows the final size of droplets initially at $R_0 = 4~{\rm \mu m}$, for a fixed mixing ratio $\phi = 0.7$. Above a threshold temperature of the air above the cloud, which increases with its saturation, droplets fully evaporate. If this air is wet and warm or cold enough, mixing leads to droplet growth. For given initial saturation $S$ and temperatures $T_\mathrm{cloud}$, $T_\mathrm{ext}$, there is an optimal mixing ratio $\phi$ for droplet growth, as shown in Fig.~\ref{fig:mixingpathway}(c). Mixing should incorporate enough dry air to create supersaturated conditions, but not so much that it dries out the upper part of the cloud. Mixing with colder air at the cloud top is only possible during the initial ascent of an air parcel, as it needs to be positively buoyant to find colder air above it. Once the cloud top reaches the inversion layer, it can no longer rise as it becomes capped by warmer air. However, the isobaric mixing process presented here is almost symmetrical with respect to which air volume is the warmest, as temperatures above $T_\mathrm{cloud} = 15~{\rm ^\circ C}$ in Fig.~\ref{fig:mixingpathway}(b) show. Mixing with slightly subsaturated ($S = 0.9$, green curve) hot air can still lead to droplet growth, as long as it is warm enough. The conditions needed for growth above $10~{\rm \mu m}$ from this idealized mixing process alone are seldom fulfilled in clouds, requiring large temperature differences and very wet air above the cloud. Such conditions could be achieved by preconditioning of the atmosphere, through successive convective events that gradually moisten the air \cite{roesner_effect_1990,hohenegger_preconditioning_2013,rapp_interactions_2011,touze-peiffer_cold_2022}.

\subsection{Electrostatic pathway}

Going back to Fig.~\ref{fig:distributionmoments}(b), one observes that van der Waals interactions between droplets results in a fivefold reduction in $t_g$. This suggests a third possible pathway to rain. The gap in collision efficiency can be bypassed by the introduction of a static electric field large enough to close the low collision rate gap (dashed dotted red curve on Fig.~\ref{fig:collisionrate}). The typical field strength needed is $50~{\rm kV.m^{-1}}$ (see \cite{poydenot_gap_2024}), which is significantly ($60$ times) lower than the electric breakdown field of $3000~{\rm kV.m^{-1}}$, leads to a further reduction in $t_g$ by a factor of $10$. Figure~\ref{fig:trajectory} demonstrates that the collisional trajectory remains qualitatively consistent for all dynamical mechanisms considered. This consistency is reflected in the robust decomposition of the dynamics into a collective enlargement of the core of the distribution and individual drops growing in the tails by absorption of smaller drops. For drops on the order of $4~{\rm \mu m}$, a rain initiation time $t_g \sim 10^3~{\rm s}$ requires an external electric field much larger than the values reported in the bulk of non-precipitating clouds \cite{marshall_electric_1995,harrison_microphysical_2015}. However, even if the bulk of the cloud is electrically neutral, the upper layer of the cloud may be subjected to significant electric fields. Recent observations and field campaigns have investigated the charge and electric fields of fair weather clouds, and has shown that in the midlatitudes, stratiform clouds are charged at cloud base \cite{nicoll_stratiform_2016,harrison_evaluating_2017,harrison_measuring_2022}. The magnitude of these effects in the tropics is unclear \cite{kartalev_possible_2006,williams_lightning_2004,williams_global_2009,harrison_carnegie_2013,afreen_fair-weather_2021}. Droplet charging, rather than the static vertical field presented here, could have effects on the droplet coalescence rate \cite{harrison_precipitation_2020,harrison_demonstration_2021}; however, droplets tend to have charges of the same sign, which creates repulsive interactions. Further work is needed to measure the electrical properties of clouds over their entire depth, along with their droplet populations, to determine which, if any, electrostatic mechanism could lead to increased collision rates.
\begin{figure}[b!]
\centering
\includegraphics{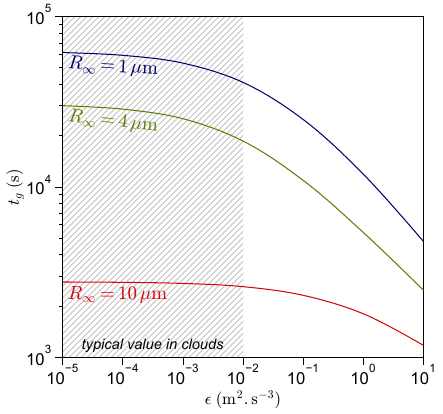}
\caption{Rain initiation time $t_g$ as a function of the turbulent kinetic energy dissipation rate $\epsilon$, for different initial sizes $R_\infty$. The typical range for clouds is shown by hatched gray lines, below $\epsilon \leq 10^{-2}~{\rm m^2.s^{-3}}$ \cite{shaw_particle-turbulence_2003,mellado_cloud-top_2017}. The liquid water content $\rho_d = 1~{\rm g.m^{-3}}$ is identical for all three curves.}
\label{fig:initialsizeturbulence}
\end{figure}

\subsection{Turbulence pathway}

A fourth possible pathway for rain formation would be dominated by turbulence-induced collisions (Fig.~\ref{fig:initialsizeturbulence}). For particles smaller than the Kolmogorov length scale $\ell_K$, which is the scale of the smallest turbulent structures, turbulence is known to increase collision rates through two different mechanisms. These effects are controlled by the turbulent Stokes number $\mathrm{St}_K = \tau_p/\tau_K$, where $\tau_p = 2\rho_\ell R^2/(9\eta_g)$ is the Stokes time of the particle with radius $R$, and $\tau_K$ is the Kolmogorov time of the turbulent flow. At low $\mathrm{St}_K$, droplets follow the underlying flow and behave as passive tracers. When $\mathrm{St}_K$ is large, the velocity fluctuates on timescales much shorter than the particle can respond to, allowing droplets to cross streamlines. $\mathrm{St}_K = 1$ is typically reached around $40 \,{\rm \mu m}$. First, for moderate $\mathrm{St}_K$, particles cluster in regions of low vorticity, leading to enhanced collision rates. At larger $\mathrm{St}_K$, particles can be slung around by turbulent eddies and focused into regions where they meet with large relative velocities. A turbulent collision kernel $K_\mathrm{turb}$ has been proposed by \citet{pumir_collisional_2016}, based on the results of multiple direct numerical simulations of inertial particles in turbulence (see references therein). The total collision kernel is taken additively as a lowest-order approximation: $K_\mathrm{total} = K + K_\mathrm{turb}$. Here, $K$ refers to the kernel introduced previously, which accounts for Brownian motion, hydrodynamics, and van der Waals forces. For water droplets in air, turbulent coalescence dynamics are controlled only by the turbulent energy dissipation rate $\epsilon$, typically around $10^{-3} - 10^{-2} \,{\rm m^2.s^{-3}}$ in clouds \cite{shaw_particle-turbulence_2003,mellado_cloud-top_2017}. Figure~\ref{fig:initialsizeturbulence} shows that turbulence has little effect on the rain initiation time, due to the relatively weak turbulence in clouds. Its effect is larger for smaller drops, as the efficiency gap strongly decreases the Brownian-gravitational-electrostatic collision rate in this size range. Note that the turbulent collision efficiency is assumed to be unity, i.e., all hydrodynamic effects are neglected when the turbulent flow brings droplets together, and they are assumed to always coalesce. There is currently very little knowledge about the fine details of turbulence-induced collisions, which could, as is the case for gravitational collisions, strongly reduce the collision rate \cite{pinsky_turbulence_1997,pinsky_collisions_1999,pinsky_turbulence_1997-1,ayala_effects_2008-1,ayala_effects_2008}. In the present stage of knowledge, this \emph{turbulence pathway} to rain is unlikely in warm clouds.

\section{Conclusion}
In conclusion, the initiation of rain in warm clouds requires two distinct conditions: the formation of rain precursors larger than $30~{\rm \mu m}$ in the upper part of the cloud, and the passage of these drops through a dense spray as they fall, allowing them to grow by collecting enough small drops (on the order of $10^6$). It is not the core of the drop size distribution that triggers the onset of this cascade, but rather its tails, which display much faster and very different dynamics. We have investigated four pathways in this article.

First, the \emph{coalescence pathway} requires initially small droplets to cross the low-efficiency gap only by collision and coalescence, above which they rapidly grow. As the collision process is highly nonlinear, the time to go through this pathway is inversely proportional to the liquid water content. To account for the shortest observed rain initiation time (about $10^3~{\rm s}$), this pathway requires large liquid water contents that may not be achieved in every precipitating cloud (roughly above $1~{\rm g.m^{-3}}$). However, many clouds can sustain themselves for much longer, typically a few hours, in which case this slow pathway becomes more likely.

Second, the \emph{mixing pathway} bypasses the slow collision efficiency gap by growing droplets above $10~{\rm \mu m}$ through mixing of unsaturated, droplet-free air with wet, droplet-laden air at different temperatures at the cloud top. Mixing locally creates supersaturated conditions that grow existing droplets rather than nucleate new ones. This fast pathway depends on the local structure of the water vapor field and the drop size distribution \cite{baker_evolution_1979,baker_influence_1980,baker_effects_1984,roesner_effect_1990,villermaux_fine_2017,allwayin_locally_2024}, and requires sufficiently wet air above the cloud to grow the droplets, rather than evaporate them and dry out the cloud. This suggests that the relative history of successive convective plumes would control rain initiation in this pathway. However, the temperature difference and humidity needed for droplet growth are large, and are unlikely to be realized in usual atmospheric conditions, at least for the mixing model presented here.

Third, the low-efficiency gap can be closed in the presence of a static vertical electric field inducing attractive electrostatic interactions between colliding droplets. This \emph{electrostatic pathway} involves uncharged drops, as charged drops typically have the same charge \cite{harrison_microphysical_2015}, so their electrostatic interactions are repulsive, further increasing the rain formation time. The field strength needed for this pathway to become relevant is rather large (a fraction of the breakdown field), and has not been observed so far in warm precipitating clouds. In the absence of ice through which charges can be separated and large fields can emerge \cite{latham_electrification_1981,williams_mixed-phase_1991,saunders_review_1993,mason_charge_2000,dash_theory_2001,jungwirth_possible_2005}, it is unclear how the required fields could be achieved.

Lastly, droplets can collide due to their relative velocities given by the underlying turbulence in clouds. If the droplets have sufficient inertia with regards to the fluctuations of this velocity field, their collision rate could be greatly enhanced. However, due to the relatively weak turbulence in clouds, this \emph{turbulence pathway} is at this point very unlikely.

These four pathways are idealized scenarios that must now be tested using measurements performed in different types of clouds. We have identified here that droplet size distributions, liquid water content, electric field strengths, and turbulence intensities within various cloud environments should yield the most information on rain initiation. Additionally, large eddy simulations and cloud-resolving models \cite{guichard_short_2017} can help integrate these mechanisms and explore their interplay under different atmospheric conditions. By combining observations with the theoretical insights present here, we can achieve a better understanding of rain initiation processes and improve predictions of precipitation patterns. This multifaceted approach will not only enhance our fundamental knowledge of cloud physics but also contribute to more accurate weather forecasting and climate modeling.

\begin{appendix}
\section{Self-similar kernels}
\label{sec:selfsimilarkernels}
Different kernels of physical interest present self-similar properties. The relative Brownian motion between two diffusing particles in a medium with viscosity $\eta_g$ results in a collision kernel, which has originally been studied by Smoluchowski \cite{smoluchowski_versuch_1918}
\begin{figure}[h]
\centering
\includegraphics{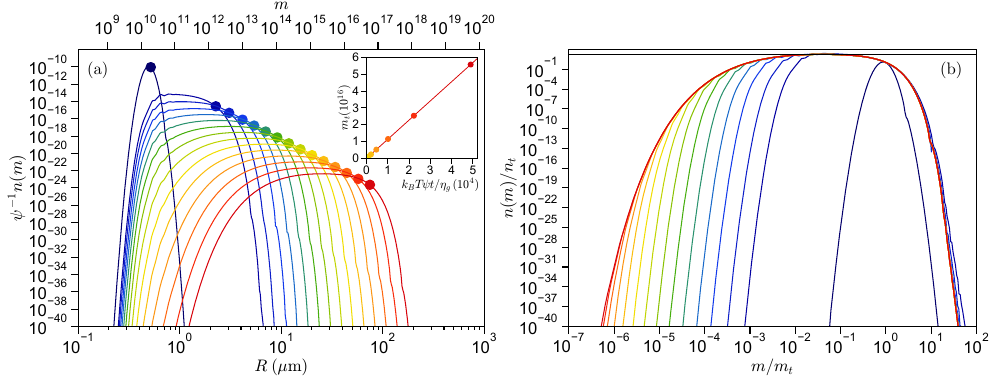}
\caption{(a) Mass distribution $n(m)$ at different times (colors from blue to red), obtained by integration of the Smoluchowski equation with the homogeneous Brownian kernel. Masses are expressed in units of the mass of a water molecule. Insert: typical mass $m_t$ as a function of time. Times are sampled logarithmically, so that the distribution, of which its typical mass grows linearly in time, appears to be moving at a constant rate in log scale. (b) Rescaled distribution $n/n_t$ as a function of the rescaled mass $m/m_t$ at different times [the same as panel (a)]. The mass used for rescaling is the typical mass in the tails $m_t = \int \mathrm d m n(m) m^3/\int \mathrm d m n(m) m^2$, and $n_t = n(m_t)$. Solid black line: apparent scaling law $n/n_t \sim \mathrm{constant}$.}
\label{fig:distribs_mecanismes_diffusion}
\end{figure}
\begin{equation}
K = \frac{2}{3} \frac{k_B T}{\eta_g}\left(x^{-1/3}+y^{-1/3}\right)\left(x^{1/3}+y^{1/3}\right),
\label{eq:browniankernel}
\end{equation}
Here, the exponents are $\mu = -\nu = -1/3$ and $\lambda = 0$. Similarly, differential gravitational settling between two particles of radii $R_i$, falling under their own weight at terminal velocity $U^t_i$, results in the kernel:
\begin{equation}
K = \pi (R_1+R_2)^2 \left|U^t_1-U^t_2\right|.
\label{eq:kernel_eff1}
\end{equation}
There are two regimes for the fallspeed of droplets. At small Reynolds number (below $50\,{\mathrm \mu m}$ in air), the drag on the droplets is governed by Stokes' law, resulting in $U^t_i \propto R_i^2$, $\mu = 0$ and $\lambda = \nu = 4/3$. At larger Reynolds number, the force on the droplets is quadratic with velocity so that $U^t_i \propto \sqrt{R_i}$, $\mu = 0$ and $\lambda = \nu = 5/6$. Lastly, in uniform shear flow at rate $\dot\gamma$, the kernel is given by:
\begin{equation}
K = \frac{4}{3} \dot\gamma (R_1+R_2)^3
\label{eq:shearkernel}
\end{equation}
and $\mu = 0$, $\lambda = \nu = 1$. This mechanism is relevant for small cloud droplets with low inertia, which collide due to local turbulent shear at sub-Kolmogorov scales \cite{saffman_collision_1956}. For certain initial conditions and collision kernels, there are analytical solutions to the Smoluchowski equation. These include the constant kernel $K = K_s$, the sum kernel $K(x, y) = K_s(x+y)$ and the product kernel $K(x, y) = K_s xy$ \cite{smoluchowski_versuch_1918,davies_self-similar_1999,fernandez-diaz_exact_2007,leyvraz_scaling_2003,krapivsky_kinetic_2010}. However, only the constant kernel is of physical relevance, as it corresponds to the limit of the Brownian kernel where particles have the same size.

The values of the exponents $\mu$, $\nu$, $\lambda$ control the qualitative behavior of the growth.
\begin{figure}[h]
\centering
\includegraphics{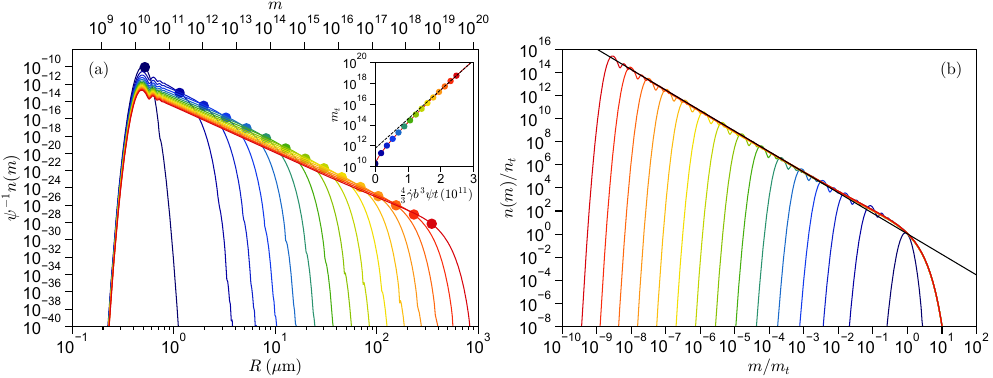}
\caption{(a) Mass distribution $n(m)$ at different times (colors from blue to red), obtained by integration of the Smoluchowski equation with the homogeneous shear kernel. Insert: typical mass $m_t$ as a function of time. Times are sampled linearly, so that the distribution, of which its typical mass grows exponentially, appears to be moving at at constant rate in log scale. Masses are expressed in units of the mass of a water molecule. (b) Rescaled distribution $n/n_t$ as a function of the rescaled mass $m/m_t$ at different times [the same as panel (a)]. The mass used for rescaling is the typical mass in the tails $m_t = \int \mathrm d m n(m) m^3/\int \mathrm d m n(m) m^2$, and $n_t = n(m_t)$. Solid black line: apparent scaling law $n/n_t \propto (m/m_t)^{-1.8}$.}
\label{fig:distribs_mecanismes_shear}
\end{figure}

Figures~\ref{fig:distribs_mecanismes_diffusion}(a) and~\ref{fig:distribs_mecanismes_shear}(a) show the evolution of the distribution over time for two homogeneous kernels, respectively the Brownian kernel (\ref{eq:browniankernel}) and the shear kernel (\ref{eq:shearkernel}). With the Brownian kernel [Fig.~\ref{fig:distribs_mecanismes_diffusion}(a)], the initially peaked distribution broadens and moves up in size, the typical size $m_t$ growing linearly in time. With the shear kernel [Fig.~\ref{fig:distribs_mecanismes_shear}(a)], a power-law regime develops from the initial peak and grows towards larger sizes, and is cut off exponentially above $m_t$. In both cases, after a short transient, the evolution enters a self-similar regime [Figs.~\ref{fig:distribs_mecanismes_diffusion}(b) and~\ref{fig:distribs_mecanismes_shear}(b)]: the rescaled distribution $n/n(m_t)$ as a function of the rescaled size $m/m_t$ collapses onto the same curve at large $m/m_t$, and extends over time as a power law towards smaller $m/m_t$: for the Brownian kernel, $n(m)/n(m_t) \sim 1$; for the shear kernel, $n(m)/n(m_t) \propto (m/m_t)^{-1.8}$. To understand these kinetics, we consider a single drop of mass $m$, growing by collisions with $\psi$ drops per unit volume that are all of size $m$. We assume that for all sizes, the scaling behavior $K = K_s x^\mu y^\nu$ holds exactly. The growth rate of the drop reads
\begin{equation}
\frac{\mathrm{d} m}{\mathrm d t} = K_s \psi m^\lambda.
\label{eq:scalingsolutionkernel}
\end{equation}
Depending on the value of $\lambda$, three behaviors are possible. For $\lambda < 1$, $m \propto t^{1/(1-\lambda)}$: the drop grows as a power law of time, linearly for Brownian motion ($\lambda = 0$), as Fig.~\ref{fig:distribs_mecanismes_diffusion}(a) shows. For $\lambda = 1$, which is the case for the shear kernel, $m \propto \exp(K_s \psi t)$, as on Fig.~\ref{fig:distribs_mecanismes_shear}(a). There is a remaining edge case: for $\lambda > 1$, $m = [m_0^{-(\lambda-1)} -(1-\lambda)K_s \psi t)]^{1/(1-\lambda)}$. $m$ diverges at finite time $t_c = 1/[K_s\psi(1-\lambda)m_0^{\lambda-1}]$, which is unphysical. Actually, a solution exists at all times \cite{leyvraz_critical_1982,ziff_kinetics_1983,ernst_coagulation_1984-1}. At the critical time $t = t_c$, the dilute phase and its size distribution separate from a cluster of infinite size, referred to as a "gel" by analogy with the gelation transition \cite{white_global_1980,white_form_1982,lushnikov_sol_2004}. The finite gelation time obtained with these heuristics comes from finite size effects of the system; for $\lambda > 1$ and $\mu \leq 0$, the gelation time is rigorously zero \cite{van_dongen_possible_1987}. It has been established that at long times, the size distribution no longer depends on the initial condition, but follows the scaling laws \cite{lushnikov_evolution_1973,van_dongen_dynamic_1985,van_dongen_solutions_1987,van_dongen_scaling_1988}
\begin{equation}
\label{eq:scalingsolution}
n(m) \sim n_0 t^{-2} \phi[m/s(t)].
\end{equation}
$n_0$ is a constant and $s(t)$ is a characteristic size of the system, e.g. the mean size $\mathcal M_1/\mathcal M_0$, or a size constructed with higher order moments such as $m_t = \mathcal M_3/\mathcal M_2$, used here. $s(t)$ is a solution to the one-drop growth Eq.~(\ref{eq:scalingsolutionkernel}). \citet{van_dongen_scaling_1988} have shown that the scaling function $\phi(x)$ has two asymptotic regimes for $\lambda \leq 1$. At large $x$, $\phi(x) \sim x^{-\lambda} e^{-\delta x}$, with $\delta$ a constant. At small $x$, for $\mu = 0$ -- which is the case for the shear kernel -- the scaling function follows a power law $\phi(x) \propto x^{-\tau}$, as shown on Fig.~\ref{fig:distribs_mecanismes_shear}(b). The exponent $\tau$ is smaller than $2$ and for $\lambda = 1$, its value is set by the shape of $\phi$; here we find it to be around $1.8$. For $\mu < 0$, $\phi$ decays exponentially at small $x$, as Fig.~\ref{fig:distribs_mecanismes_diffusion}(b) shows.

\section{Reaction trajectory}
\label{sec:reactiontrajectory}
The growth term in the Smoluchowski equation (\ref{eq:smolu}) can be written as
\begin{equation}
\frac{1}{2} \int_0^m \mathrm d x\, n(m-x) n(x) K(m-x, x) = \int_0^{m/2} \mathrm d x\, n(m-x) n(x) K(m-x, x) = \int_{m/2}^m \mathrm d x\, n(m-x) n(x) K(m-x, x),
\end{equation}
by symmetry of the integrand $n(m-x) n(x) K(m-x, x)$ around the axis $x = m/2$. We are interested in the partial growth term
\begin{equation}
F(y | m)= \int_0^{y} \mathrm d x\, n(m-x) n(x) K(m-x, x),
\end{equation}
which describes the contribution of drops of size smaller than $y$ to the growth of drops of size $m$. We define the reaction size $m_*$ as
\begin{equation}
F(m_* | m)= \frac{1}{2} F\left(\frac{m}{2} \big| m\right).
\end{equation}
In other words, the reaction size $m_*$ is the size for which the drops contribute half the growth rate of all the drops of sizes smaller than $m$. The reaction size is defined with respect to another size; we take this size to be the typical size in the tails $m_t$. The results shown here are not very sensitive to this choice, and are almost identical using instead the mean size $m_m$.

$m_*$ is computed numerically in this paper, using the distributions $n(m, t)$ obtained over time and the full expression of the kernel $K$. To illustrate its physical meaning, we can compute it analytically in the ideal case of the constant kernel. For the kernel $K = 1$ and an initial condition localized in mass space, the long-term behavior of the solution is [Eq.~(\ref{eq:scalingsolution})]
\begin{equation}
n(x, t) = \frac{1}{t^2} \exp\left(-\frac{x}{2t}\right).
\end{equation}
The typical mass is $m_m = 2t$. The partial growth rate is
\begin{equation}
F(y | m)= \frac{1}{t^4} y \exp\left(-\frac{m}{2t}\right),
\end{equation}
such that the reaction size is
\begin{equation}
m_* = \frac{m}{4},
\end{equation}
meaning that the droplets of typical size $m$ mostly react with each other. Numerically, we recover the $1/4$ prefactor at long times.
%
%

\end{appendix}


\begin{thebibliography}{119}%
\makeatletter
\providecommand \@ifxundefined [1]{%
 \@ifx{#1\undefined}
}%
\providecommand \@ifnum [1]{%
 \ifnum #1\expandafter \@firstoftwo
 \else \expandafter \@secondoftwo
 \fi
}%
\providecommand \@ifx [1]{%
 \ifx #1\expandafter \@firstoftwo
 \else \expandafter \@secondoftwo
 \fi
}%
\providecommand \natexlab [1]{#1}%
\providecommand \enquote  [1]{``#1''}%
\providecommand \bibnamefont  [1]{#1}%
\providecommand \bibfnamefont [1]{#1}%
\providecommand \citenamefont [1]{#1}%
\providecommand \href@noop [0]{\@secondoftwo}%
\providecommand \href [0]{\begingroup \@sanitize@url \@href}%
\providecommand \@href[1]{\@@startlink{#1}\@@href}%
\providecommand \@@href[1]{\endgroup#1\@@endlink}%
\providecommand \@sanitize@url [0]{\catcode `\\12\catcode `\$12\catcode
  `\&12\catcode `\#12\catcode `\^12\catcode `\_12\catcode `\%12\relax}%
\providecommand \@@startlink[1]{}%
\providecommand \@@endlink[0]{}%
\providecommand \url  [0]{\begingroup\@sanitize@url \@url }%
\providecommand \@url [1]{\endgroup\@href {#1}{\urlprefix }}%
\providecommand \urlprefix  [0]{URL }%
\providecommand \Eprint [0]{\href }%
\providecommand \doibase [0]{https://doi.org/}%
\providecommand \selectlanguage [0]{\@gobble}%
\providecommand \bibinfo  [0]{\@secondoftwo}%
\providecommand \bibfield  [0]{\@secondoftwo}%
\providecommand \translation [1]{[#1]}%
\providecommand \BibitemOpen [0]{}%
\providecommand \bibitemStop [0]{}%
\providecommand \bibitemNoStop [0]{.\EOS\space}%
\providecommand \EOS [0]{\spacefactor3000\relax}%
\providecommand \BibitemShut  [1]{\csname bibitem#1\endcsname}%
\let\auto@bib@innerbib\@empty
\bibitem [{\citenamefont {Calvin}\ \emph {et~al.}(2023)\citenamefont {Calvin},
  \citenamefont {Dasgupta}, \citenamefont {Krinner}, \citenamefont {Mukherji},
  \citenamefont {Thorne}, \citenamefont {Trisos}, \citenamefont {Romero},
  \citenamefont {Aldunce}, \citenamefont {Barrett}, \citenamefont {Blanco},
  \citenamefont {Cheung}, \citenamefont {Connors}, \citenamefont {Denton},
  \citenamefont {{Diongue-Niang}}, \citenamefont {Dodman}, \citenamefont
  {Garschagen}, \citenamefont {Geden}, \citenamefont {Hayward}, \citenamefont
  {Jones}, \citenamefont {Jotzo}, \citenamefont {Krug}, \citenamefont {Lasco},
  \citenamefont {Lee}, \citenamefont {{Masson-Delmotte}}, \citenamefont
  {Meinshausen}, \citenamefont {Mintenbeck}, \citenamefont {Mokssit},
  \citenamefont {Otto}, \citenamefont {Pathak}, \citenamefont {Pirani},
  \citenamefont {Poloczanska}, \citenamefont {P{\"o}rtner}, \citenamefont
  {Revi}, \citenamefont {Roberts}, \citenamefont {Roy}, \citenamefont {Ruane},
  \citenamefont {Skea}, \citenamefont {Shukla}, \citenamefont {Slade},
  \citenamefont {Slangen}, \citenamefont {Sokona}, \citenamefont
  {S{\"o}rensson}, \citenamefont {Tignor}, \citenamefont {Van~Vuuren},
  \citenamefont {Wei}, \citenamefont {Winkler}, \citenamefont {Zhai},
  \citenamefont {Zommers}, \citenamefont {Hourcade}, \citenamefont {Johnson},
  \citenamefont {Pachauri}, \citenamefont {Simpson}, \citenamefont {Singh},
  \citenamefont {Thomas}, \citenamefont {Totin}, \citenamefont {Arias},
  \citenamefont {Bustamante}, \citenamefont {Elgizouli}, \citenamefont {Flato},
  \citenamefont {Howden}, \citenamefont {{M{\'e}ndez-Vallejo}}, \citenamefont
  {Pereira}, \citenamefont {{Pichs-Madruga}}, \citenamefont {Rose},
  \citenamefont {Saheb}, \citenamefont {S{\'a}nchez~Rodr{\'i}guez},
  \citenamefont {{\"U}rge-Vorsatz}, \citenamefont {Xiao}, \citenamefont
  {Yassaa}, \citenamefont {Alegr{\'i}a}, \citenamefont {Armour}, \citenamefont
  {{Bednar-Friedl}}, \citenamefont {Blok}, \citenamefont {Ciss{\'e}},
  \citenamefont {Dentener}, \citenamefont {Eriksen}, \citenamefont {Fischer},
  \citenamefont {Garner}, \citenamefont {Guivarch}, \citenamefont {Haasnoot},
  \citenamefont {Hansen}, \citenamefont {Hauser}, \citenamefont {Hawkins},
  \citenamefont {Hermans}, \citenamefont {Kopp}, \citenamefont
  {{Leprince-Ringuet}}, \citenamefont {Lewis}, \citenamefont {Ley},
  \citenamefont {Ludden}, \citenamefont {Niamir}, \citenamefont {Nicholls},
  \citenamefont {Some}, \citenamefont {Szopa}, \citenamefont {Trewin},
  \citenamefont {Van Der~Wijst}, \citenamefont {Winter}, \citenamefont
  {Witting}, \citenamefont {Birt}, \citenamefont {Ha}, \citenamefont {Romero},
  \citenamefont {Kim}, \citenamefont {Haites}, \citenamefont {Jung},
  \citenamefont {Stavins}, \citenamefont {Birt}, \citenamefont {Ha},
  \citenamefont {Orendain}, \citenamefont {Ignon}, \citenamefont {Park},
  \citenamefont {Park}, \citenamefont {Reisinger}, \citenamefont {Cammaramo},
  \citenamefont {Fischlin}, \citenamefont {Fuglestvedt}, \citenamefont
  {Hansen}, \citenamefont {Ludden}, \citenamefont {{Masson-Delmotte}},
  \citenamefont {Matthews}, \citenamefont {Mintenbeck}, \citenamefont {Pirani},
  \citenamefont {Poloczanska}, \citenamefont {{Leprince-Ringuet}},\ and\
  \citenamefont {P{\'e}an}}]{lee_ipcc_2023}%
  \BibitemOpen
  \bibfield  {author} {\bibinfo {author} {\bibfnamefont {K.}~\bibnamefont
  {Calvin}}, \bibinfo {author} {\bibfnamefont {D.}~\bibnamefont {Dasgupta}},
  \bibinfo {author} {\bibfnamefont {G.}~\bibnamefont {Krinner}}, \bibinfo
  {author} {\bibfnamefont {A.}~\bibnamefont {Mukherji}}, \bibinfo {author}
  {\bibfnamefont {P.~W.}\ \bibnamefont {Thorne}}, \bibinfo {author}
  {\bibfnamefont {C.}~\bibnamefont {Trisos}}, \bibinfo {author} {\bibfnamefont
  {J.}~\bibnamefont {Romero}}, \bibinfo {author} {\bibfnamefont
  {P.}~\bibnamefont {Aldunce}}, \bibinfo {author} {\bibfnamefont
  {K.}~\bibnamefont {Barrett}}, \bibinfo {author} {\bibfnamefont
  {G.}~\bibnamefont {Blanco}}, \bibinfo {author} {\bibfnamefont {W.~W.}\
  \bibnamefont {Cheung}}, \bibinfo {author} {\bibfnamefont {S.}~\bibnamefont
  {Connors}}, \bibinfo {author} {\bibfnamefont {F.}~\bibnamefont {Denton}},
  \bibinfo {author} {\bibfnamefont {A.}~\bibnamefont {{Diongue-Niang}}},
  \bibinfo {author} {\bibfnamefont {D.}~\bibnamefont {Dodman}}, \bibinfo
  {author} {\bibfnamefont {M.}~\bibnamefont {Garschagen}}, \bibinfo {author}
  {\bibfnamefont {O.}~\bibnamefont {Geden}}, \bibinfo {author} {\bibfnamefont
  {B.}~\bibnamefont {Hayward}}, \bibinfo {author} {\bibfnamefont
  {C.}~\bibnamefont {Jones}}, \bibinfo {author} {\bibfnamefont
  {F.}~\bibnamefont {Jotzo}}, \bibinfo {author} {\bibfnamefont
  {T.}~\bibnamefont {Krug}}, \bibinfo {author} {\bibfnamefont {R.}~\bibnamefont
  {Lasco}}, \bibinfo {author} {\bibfnamefont {Y.-Y.}\ \bibnamefont {Lee}},
  \bibinfo {author} {\bibfnamefont {V.}~\bibnamefont {{Masson-Delmotte}}},
  \bibinfo {author} {\bibfnamefont {M.}~\bibnamefont {Meinshausen}}, \bibinfo
  {author} {\bibfnamefont {K.}~\bibnamefont {Mintenbeck}}, \bibinfo {author}
  {\bibfnamefont {A.}~\bibnamefont {Mokssit}}, \bibinfo {author} {\bibfnamefont
  {F.~E.}\ \bibnamefont {Otto}}, \bibinfo {author} {\bibfnamefont
  {M.}~\bibnamefont {Pathak}}, \bibinfo {author} {\bibfnamefont
  {A.}~\bibnamefont {Pirani}}, \bibinfo {author} {\bibfnamefont
  {E.}~\bibnamefont {Poloczanska}}, \bibinfo {author} {\bibfnamefont {H.-O.}\
  \bibnamefont {P{\"o}rtner}}, \bibinfo {author} {\bibfnamefont
  {A.}~\bibnamefont {Revi}}, \bibinfo {author} {\bibfnamefont {D.~C.}\
  \bibnamefont {Roberts}}, \bibinfo {author} {\bibfnamefont {J.}~\bibnamefont
  {Roy}}, \bibinfo {author} {\bibfnamefont {A.~C.}\ \bibnamefont {Ruane}},
  \bibinfo {author} {\bibfnamefont {J.}~\bibnamefont {Skea}}, \bibinfo {author}
  {\bibfnamefont {P.~R.}\ \bibnamefont {Shukla}}, \bibinfo {author}
  {\bibfnamefont {R.}~\bibnamefont {Slade}}, \bibinfo {author} {\bibfnamefont
  {A.}~\bibnamefont {Slangen}}, \bibinfo {author} {\bibfnamefont
  {Y.}~\bibnamefont {Sokona}}, \bibinfo {author} {\bibfnamefont {A.~A.}\
  \bibnamefont {S{\"o}rensson}}, \bibinfo {author} {\bibfnamefont
  {M.}~\bibnamefont {Tignor}}, \bibinfo {author} {\bibfnamefont
  {D.}~\bibnamefont {Van~Vuuren}}, \bibinfo {author} {\bibfnamefont {Y.-M.}\
  \bibnamefont {Wei}}, \bibinfo {author} {\bibfnamefont {H.}~\bibnamefont
  {Winkler}}, \bibinfo {author} {\bibfnamefont {P.}~\bibnamefont {Zhai}},
  \bibinfo {author} {\bibfnamefont {Z.}~\bibnamefont {Zommers}}, \bibinfo
  {author} {\bibfnamefont {J.-C.}\ \bibnamefont {Hourcade}}, \bibinfo {author}
  {\bibfnamefont {F.~X.}\ \bibnamefont {Johnson}}, \bibinfo {author}
  {\bibfnamefont {S.}~\bibnamefont {Pachauri}}, \bibinfo {author}
  {\bibfnamefont {N.~P.}\ \bibnamefont {Simpson}}, \bibinfo {author}
  {\bibfnamefont {C.}~\bibnamefont {Singh}}, \bibinfo {author} {\bibfnamefont
  {A.}~\bibnamefont {Thomas}}, \bibinfo {author} {\bibfnamefont
  {E.}~\bibnamefont {Totin}}, \bibinfo {author} {\bibfnamefont
  {P.}~\bibnamefont {Arias}}, \bibinfo {author} {\bibfnamefont
  {M.}~\bibnamefont {Bustamante}}, \bibinfo {author} {\bibfnamefont
  {I.}~\bibnamefont {Elgizouli}}, \bibinfo {author} {\bibfnamefont
  {G.}~\bibnamefont {Flato}}, \bibinfo {author} {\bibfnamefont
  {M.}~\bibnamefont {Howden}}, \bibinfo {author} {\bibfnamefont
  {C.}~\bibnamefont {{M{\'e}ndez-Vallejo}}}, \bibinfo {author} {\bibfnamefont
  {J.~J.}\ \bibnamefont {Pereira}}, \bibinfo {author} {\bibfnamefont
  {R.}~\bibnamefont {{Pichs-Madruga}}}, \bibinfo {author} {\bibfnamefont
  {S.~K.}\ \bibnamefont {Rose}}, \bibinfo {author} {\bibfnamefont
  {Y.}~\bibnamefont {Saheb}}, \bibinfo {author} {\bibfnamefont
  {R.}~\bibnamefont {S{\'a}nchez~Rodr{\'i}guez}}, \bibinfo {author}
  {\bibfnamefont {D.}~\bibnamefont {{\"U}rge-Vorsatz}}, \bibinfo {author}
  {\bibfnamefont {C.}~\bibnamefont {Xiao}}, \bibinfo {author} {\bibfnamefont
  {N.}~\bibnamefont {Yassaa}}, \bibinfo {author} {\bibfnamefont
  {A.}~\bibnamefont {Alegr{\'i}a}}, \bibinfo {author} {\bibfnamefont
  {K.}~\bibnamefont {Armour}}, \bibinfo {author} {\bibfnamefont
  {B.}~\bibnamefont {{Bednar-Friedl}}}, \bibinfo {author} {\bibfnamefont
  {K.}~\bibnamefont {Blok}}, \bibinfo {author} {\bibfnamefont {G.}~\bibnamefont
  {Ciss{\'e}}}, \bibinfo {author} {\bibfnamefont {F.}~\bibnamefont {Dentener}},
  \bibinfo {author} {\bibfnamefont {S.}~\bibnamefont {Eriksen}}, \bibinfo
  {author} {\bibfnamefont {E.}~\bibnamefont {Fischer}}, \bibinfo {author}
  {\bibfnamefont {G.}~\bibnamefont {Garner}}, \bibinfo {author} {\bibfnamefont
  {C.}~\bibnamefont {Guivarch}}, \bibinfo {author} {\bibfnamefont
  {M.}~\bibnamefont {Haasnoot}}, \bibinfo {author} {\bibfnamefont
  {G.}~\bibnamefont {Hansen}}, \bibinfo {author} {\bibfnamefont
  {M.}~\bibnamefont {Hauser}}, \bibinfo {author} {\bibfnamefont
  {E.}~\bibnamefont {Hawkins}}, \bibinfo {author} {\bibfnamefont
  {T.}~\bibnamefont {Hermans}}, \bibinfo {author} {\bibfnamefont
  {R.}~\bibnamefont {Kopp}}, \bibinfo {author} {\bibfnamefont {N.}~\bibnamefont
  {{Leprince-Ringuet}}}, \bibinfo {author} {\bibfnamefont {J.}~\bibnamefont
  {Lewis}}, \bibinfo {author} {\bibfnamefont {D.}~\bibnamefont {Ley}}, \bibinfo
  {author} {\bibfnamefont {C.}~\bibnamefont {Ludden}}, \bibinfo {author}
  {\bibfnamefont {L.}~\bibnamefont {Niamir}}, \bibinfo {author} {\bibfnamefont
  {Z.}~\bibnamefont {Nicholls}}, \bibinfo {author} {\bibfnamefont
  {S.}~\bibnamefont {Some}}, \bibinfo {author} {\bibfnamefont {S.}~\bibnamefont
  {Szopa}}, \bibinfo {author} {\bibfnamefont {B.}~\bibnamefont {Trewin}},
  \bibinfo {author} {\bibfnamefont {K.-I.}\ \bibnamefont {Van Der~Wijst}},
  \bibinfo {author} {\bibfnamefont {G.}~\bibnamefont {Winter}}, \bibinfo
  {author} {\bibfnamefont {M.}~\bibnamefont {Witting}}, \bibinfo {author}
  {\bibfnamefont {A.}~\bibnamefont {Birt}}, \bibinfo {author} {\bibfnamefont
  {M.}~\bibnamefont {Ha}}, \bibinfo {author} {\bibfnamefont {J.}~\bibnamefont
  {Romero}}, \bibinfo {author} {\bibfnamefont {J.}~\bibnamefont {Kim}},
  \bibinfo {author} {\bibfnamefont {E.~F.}\ \bibnamefont {Haites}}, \bibinfo
  {author} {\bibfnamefont {Y.}~\bibnamefont {Jung}}, \bibinfo {author}
  {\bibfnamefont {R.}~\bibnamefont {Stavins}}, \bibinfo {author} {\bibfnamefont
  {A.}~\bibnamefont {Birt}}, \bibinfo {author} {\bibfnamefont {M.}~\bibnamefont
  {Ha}}, \bibinfo {author} {\bibfnamefont {D.~J.~A.}\ \bibnamefont {Orendain}},
  \bibinfo {author} {\bibfnamefont {L.}~\bibnamefont {Ignon}}, \bibinfo
  {author} {\bibfnamefont {S.}~\bibnamefont {Park}}, \bibinfo {author}
  {\bibfnamefont {Y.}~\bibnamefont {Park}}, \bibinfo {author} {\bibfnamefont
  {A.}~\bibnamefont {Reisinger}}, \bibinfo {author} {\bibfnamefont
  {D.}~\bibnamefont {Cammaramo}}, \bibinfo {author} {\bibfnamefont
  {A.}~\bibnamefont {Fischlin}}, \bibinfo {author} {\bibfnamefont {J.~S.}\
  \bibnamefont {Fuglestvedt}}, \bibinfo {author} {\bibfnamefont
  {G.}~\bibnamefont {Hansen}}, \bibinfo {author} {\bibfnamefont
  {C.}~\bibnamefont {Ludden}}, \bibinfo {author} {\bibfnamefont
  {V.}~\bibnamefont {{Masson-Delmotte}}}, \bibinfo {author} {\bibfnamefont
  {J.~R.}\ \bibnamefont {Matthews}}, \bibinfo {author} {\bibfnamefont
  {K.}~\bibnamefont {Mintenbeck}}, \bibinfo {author} {\bibfnamefont
  {A.}~\bibnamefont {Pirani}}, \bibinfo {author} {\bibfnamefont
  {E.}~\bibnamefont {Poloczanska}}, \bibinfo {author} {\bibfnamefont
  {N.}~\bibnamefont {{Leprince-Ringuet}}},\ and\ \bibinfo {author}
  {\bibfnamefont {C.}~\bibnamefont {P{\'e}an}},\ }\href
  {https://doi.org/10.59327/IPCC/AR6-9789291691647} {\emph {\bibinfo {title}
  {{{IPCC}}, 2023: {{Climate Change}} 2023: {{Synthesis Report}}.
  {{Contribution}} of {{Working Groups I}}, {{II}} and {{III}} to the {{Sixth
  Assessment Report}} of the {{Intergovernmental Panel}} on {{Climate Change}}
  [{{Core Writing Team}}, {{H}}. {{Lee}} and {{J}}. {{Romero}} (Eds.)].
  {{IPCC}}, {{Geneva}}, {{Switzerland}}.}}},\ \bibinfo {type} {Tech. Rep.}\
  (\bibinfo  {institution} {Intergovernmental Panel on Climate Change (IPCC)},\
  \bibinfo {year} {2023})\BibitemShut {NoStop}%
\bibitem [{\citenamefont {Tobin}\ \emph {et~al.}(2012)\citenamefont {Tobin},
  \citenamefont {Bony},\ and\ \citenamefont {Roca}}]{tobin_observational_2012}%
  \BibitemOpen
  \bibfield  {author} {\bibinfo {author} {\bibfnamefont {I.}~\bibnamefont
  {Tobin}}, \bibinfo {author} {\bibfnamefont {S.}~\bibnamefont {Bony}},\ and\
  \bibinfo {author} {\bibfnamefont {R.}~\bibnamefont {Roca}},\ }\bibfield
  {title} {\bibinfo {title} {Observational {{Evidence}} for {{Relationships}}
  between the {{Degree}} of {{Aggregation}} of {{Deep Convection}}, {{Water
  Vapor}}, {{Surface Fluxes}}, and {{Radiation}}},\ }\href
  {https://doi.org/10.1175/JCLI-D-11-00258.1} {\bibfield  {journal} {\bibinfo
  {journal} {Journal of Climate}\ }\textbf {\bibinfo {volume} {25}},\ \bibinfo
  {pages} {6885} (\bibinfo {year} {2012})}\BibitemShut {NoStop}%
\bibitem [{\citenamefont {Bony}\ \emph {et~al.}(2015)\citenamefont {Bony},
  \citenamefont {Stevens}, \citenamefont {Frierson}, \citenamefont {Jakob},
  \citenamefont {Kageyama}, \citenamefont {Pincus}, \citenamefont {Shepherd},
  \citenamefont {Sherwood}, \citenamefont {Siebesma}, \citenamefont {Sobel},
  \citenamefont {Watanabe},\ and\ \citenamefont {Webb}}]{bony_clouds_2015}%
  \BibitemOpen
  \bibfield  {author} {\bibinfo {author} {\bibfnamefont {S.}~\bibnamefont
  {Bony}}, \bibinfo {author} {\bibfnamefont {B.}~\bibnamefont {Stevens}},
  \bibinfo {author} {\bibfnamefont {D.~M.~W.}\ \bibnamefont {Frierson}},
  \bibinfo {author} {\bibfnamefont {C.}~\bibnamefont {Jakob}}, \bibinfo
  {author} {\bibfnamefont {M.}~\bibnamefont {Kageyama}}, \bibinfo {author}
  {\bibfnamefont {R.}~\bibnamefont {Pincus}}, \bibinfo {author} {\bibfnamefont
  {T.~G.}\ \bibnamefont {Shepherd}}, \bibinfo {author} {\bibfnamefont {S.~C.}\
  \bibnamefont {Sherwood}}, \bibinfo {author} {\bibfnamefont {A.~P.}\
  \bibnamefont {Siebesma}}, \bibinfo {author} {\bibfnamefont {A.~H.}\
  \bibnamefont {Sobel}}, \bibinfo {author} {\bibfnamefont {M.}~\bibnamefont
  {Watanabe}},\ and\ \bibinfo {author} {\bibfnamefont {M.~J.}\ \bibnamefont
  {Webb}},\ }\bibfield  {title} {\bibinfo {title} {Clouds, circulation and
  climate sensitivity},\ }\href {https://doi.org/10.1038/ngeo2398} {\bibfield
  {journal} {\bibinfo  {journal} {Nature Geoscience}\ }\textbf {\bibinfo
  {volume} {8}},\ \bibinfo {pages} {261} (\bibinfo {year} {2015})}\BibitemShut
  {NoStop}%
\bibitem [{\citenamefont {Boucher}\ \emph {et~al.}(2020)\citenamefont
  {Boucher}, \citenamefont {Servonnat}, \citenamefont {Albright}, \citenamefont
  {Aumont}, \citenamefont {Balkanski}, \citenamefont {Bastrikov}, \citenamefont
  {Bekki}, \citenamefont {Bonnet}, \citenamefont {Bony}, \citenamefont {Bopp},
  \citenamefont {Braconnot}, \citenamefont {Brockmann}, \citenamefont {Cadule},
  \citenamefont {Caubel}, \citenamefont {Cheruy}, \citenamefont {Codron},
  \citenamefont {Cozic}, \citenamefont {Cugnet}, \citenamefont {D'Andrea},
  \citenamefont {Davini}, \citenamefont {{de Lavergne}}, \citenamefont
  {Denvil}, \citenamefont {Deshayes}, \citenamefont {Devilliers}, \citenamefont
  {Ducharne}, \citenamefont {Dufresne}, \citenamefont {Dupont}, \citenamefont
  {{\'E}th{\'e}}, \citenamefont {Fairhead}, \citenamefont {Falletti},
  \citenamefont {Flavoni}, \citenamefont {Foujols}, \citenamefont {Gardoll},
  \citenamefont {Gastineau}, \citenamefont {Ghattas}, \citenamefont
  {Grandpeix}, \citenamefont {Guenet}, \citenamefont {Guez}, \citenamefont
  {Guilyardi}, \citenamefont {Guimberteau}, \citenamefont {Hauglustaine},
  \citenamefont {Hourdin}, \citenamefont {Idelkadi}, \citenamefont {Joussaume},
  \citenamefont {Kageyama}, \citenamefont {Khodri}, \citenamefont {Krinner},
  \citenamefont {Lebas}, \citenamefont {Levavasseur}, \citenamefont {L{\'e}vy},
  \citenamefont {Li}, \citenamefont {Lott}, \citenamefont {Lurton},
  \citenamefont {Luyssaert}, \citenamefont {Madec}, \citenamefont {Madeleine},
  \citenamefont {Maignan}, \citenamefont {Marchand}, \citenamefont {Marti},
  \citenamefont {Mellul}, \citenamefont {Meurdesoif}, \citenamefont {Mignot},
  \citenamefont {Musat}, \citenamefont {Ottl{\'e}}, \citenamefont {Peylin},
  \citenamefont {Planton}, \citenamefont {Polcher}, \citenamefont {Rio},
  \citenamefont {Rochetin}, \citenamefont {Rousset}, \citenamefont {Sepulchre},
  \citenamefont {Sima}, \citenamefont {Swingedouw}, \citenamefont
  {Thi{\'e}blemont}, \citenamefont {Traore}, \citenamefont {Vancoppenolle},
  \citenamefont {Vial}, \citenamefont {Vialard}, \citenamefont {Viovy},\ and\
  \citenamefont {Vuichard}}]{boucher_presentation_2020}%
  \BibitemOpen
  \bibfield  {author} {\bibinfo {author} {\bibfnamefont {O.}~\bibnamefont
  {Boucher}}, \bibinfo {author} {\bibfnamefont {J.}~\bibnamefont {Servonnat}},
  \bibinfo {author} {\bibfnamefont {A.~L.}\ \bibnamefont {Albright}}, \bibinfo
  {author} {\bibfnamefont {O.}~\bibnamefont {Aumont}}, \bibinfo {author}
  {\bibfnamefont {Y.}~\bibnamefont {Balkanski}}, \bibinfo {author}
  {\bibfnamefont {V.}~\bibnamefont {Bastrikov}}, \bibinfo {author}
  {\bibfnamefont {S.}~\bibnamefont {Bekki}}, \bibinfo {author} {\bibfnamefont
  {R.}~\bibnamefont {Bonnet}}, \bibinfo {author} {\bibfnamefont
  {S.}~\bibnamefont {Bony}}, \bibinfo {author} {\bibfnamefont {L.}~\bibnamefont
  {Bopp}}, \bibinfo {author} {\bibfnamefont {P.}~\bibnamefont {Braconnot}},
  \bibinfo {author} {\bibfnamefont {P.}~\bibnamefont {Brockmann}}, \bibinfo
  {author} {\bibfnamefont {P.}~\bibnamefont {Cadule}}, \bibinfo {author}
  {\bibfnamefont {A.}~\bibnamefont {Caubel}}, \bibinfo {author} {\bibfnamefont
  {F.}~\bibnamefont {Cheruy}}, \bibinfo {author} {\bibfnamefont
  {F.}~\bibnamefont {Codron}}, \bibinfo {author} {\bibfnamefont
  {A.}~\bibnamefont {Cozic}}, \bibinfo {author} {\bibfnamefont
  {D.}~\bibnamefont {Cugnet}}, \bibinfo {author} {\bibfnamefont
  {F.}~\bibnamefont {D'Andrea}}, \bibinfo {author} {\bibfnamefont
  {P.}~\bibnamefont {Davini}}, \bibinfo {author} {\bibfnamefont
  {C.}~\bibnamefont {{de Lavergne}}}, \bibinfo {author} {\bibfnamefont
  {S.}~\bibnamefont {Denvil}}, \bibinfo {author} {\bibfnamefont
  {J.}~\bibnamefont {Deshayes}}, \bibinfo {author} {\bibfnamefont
  {M.}~\bibnamefont {Devilliers}}, \bibinfo {author} {\bibfnamefont
  {A.}~\bibnamefont {Ducharne}}, \bibinfo {author} {\bibfnamefont {J.-L.}\
  \bibnamefont {Dufresne}}, \bibinfo {author} {\bibfnamefont {E.}~\bibnamefont
  {Dupont}}, \bibinfo {author} {\bibfnamefont {C.}~\bibnamefont
  {{\'E}th{\'e}}}, \bibinfo {author} {\bibfnamefont {L.}~\bibnamefont
  {Fairhead}}, \bibinfo {author} {\bibfnamefont {L.}~\bibnamefont {Falletti}},
  \bibinfo {author} {\bibfnamefont {S.}~\bibnamefont {Flavoni}}, \bibinfo
  {author} {\bibfnamefont {M.-A.}\ \bibnamefont {Foujols}}, \bibinfo {author}
  {\bibfnamefont {S.}~\bibnamefont {Gardoll}}, \bibinfo {author} {\bibfnamefont
  {G.}~\bibnamefont {Gastineau}}, \bibinfo {author} {\bibfnamefont
  {J.}~\bibnamefont {Ghattas}}, \bibinfo {author} {\bibfnamefont {J.-Y.}\
  \bibnamefont {Grandpeix}}, \bibinfo {author} {\bibfnamefont {B.}~\bibnamefont
  {Guenet}}, \bibinfo {author} {\bibfnamefont {E.}~\bibnamefont {Guez},
  \bibfnamefont {Lionel}}, \bibinfo {author} {\bibfnamefont {E.}~\bibnamefont
  {Guilyardi}}, \bibinfo {author} {\bibfnamefont {M.}~\bibnamefont
  {Guimberteau}}, \bibinfo {author} {\bibfnamefont {D.}~\bibnamefont
  {Hauglustaine}}, \bibinfo {author} {\bibfnamefont {F.}~\bibnamefont
  {Hourdin}}, \bibinfo {author} {\bibfnamefont {A.}~\bibnamefont {Idelkadi}},
  \bibinfo {author} {\bibfnamefont {S.}~\bibnamefont {Joussaume}}, \bibinfo
  {author} {\bibfnamefont {M.}~\bibnamefont {Kageyama}}, \bibinfo {author}
  {\bibfnamefont {M.}~\bibnamefont {Khodri}}, \bibinfo {author} {\bibfnamefont
  {G.}~\bibnamefont {Krinner}}, \bibinfo {author} {\bibfnamefont
  {N.}~\bibnamefont {Lebas}}, \bibinfo {author} {\bibfnamefont
  {G.}~\bibnamefont {Levavasseur}}, \bibinfo {author} {\bibfnamefont
  {C.}~\bibnamefont {L{\'e}vy}}, \bibinfo {author} {\bibfnamefont
  {L.}~\bibnamefont {Li}}, \bibinfo {author} {\bibfnamefont {F.}~\bibnamefont
  {Lott}}, \bibinfo {author} {\bibfnamefont {T.}~\bibnamefont {Lurton}},
  \bibinfo {author} {\bibfnamefont {S.}~\bibnamefont {Luyssaert}}, \bibinfo
  {author} {\bibfnamefont {G.}~\bibnamefont {Madec}}, \bibinfo {author}
  {\bibfnamefont {J.-B.}\ \bibnamefont {Madeleine}}, \bibinfo {author}
  {\bibfnamefont {F.}~\bibnamefont {Maignan}}, \bibinfo {author} {\bibfnamefont
  {M.}~\bibnamefont {Marchand}}, \bibinfo {author} {\bibfnamefont
  {O.}~\bibnamefont {Marti}}, \bibinfo {author} {\bibfnamefont
  {L.}~\bibnamefont {Mellul}}, \bibinfo {author} {\bibfnamefont
  {Y.}~\bibnamefont {Meurdesoif}}, \bibinfo {author} {\bibfnamefont
  {J.}~\bibnamefont {Mignot}}, \bibinfo {author} {\bibfnamefont
  {I.}~\bibnamefont {Musat}}, \bibinfo {author} {\bibfnamefont
  {C.}~\bibnamefont {Ottl{\'e}}}, \bibinfo {author} {\bibfnamefont
  {P.}~\bibnamefont {Peylin}}, \bibinfo {author} {\bibfnamefont
  {Y.}~\bibnamefont {Planton}}, \bibinfo {author} {\bibfnamefont
  {J.}~\bibnamefont {Polcher}}, \bibinfo {author} {\bibfnamefont
  {C.}~\bibnamefont {Rio}}, \bibinfo {author} {\bibfnamefont {N.}~\bibnamefont
  {Rochetin}}, \bibinfo {author} {\bibfnamefont {C.}~\bibnamefont {Rousset}},
  \bibinfo {author} {\bibfnamefont {P.}~\bibnamefont {Sepulchre}}, \bibinfo
  {author} {\bibfnamefont {A.}~\bibnamefont {Sima}}, \bibinfo {author}
  {\bibfnamefont {D.}~\bibnamefont {Swingedouw}}, \bibinfo {author}
  {\bibfnamefont {R.}~\bibnamefont {Thi{\'e}blemont}}, \bibinfo {author}
  {\bibfnamefont {A.~K.}\ \bibnamefont {Traore}}, \bibinfo {author}
  {\bibfnamefont {M.}~\bibnamefont {Vancoppenolle}}, \bibinfo {author}
  {\bibfnamefont {J.}~\bibnamefont {Vial}}, \bibinfo {author} {\bibfnamefont
  {J.}~\bibnamefont {Vialard}}, \bibinfo {author} {\bibfnamefont
  {N.}~\bibnamefont {Viovy}},\ and\ \bibinfo {author} {\bibfnamefont
  {N.}~\bibnamefont {Vuichard}},\ }\bibfield  {title} {\bibinfo {title}
  {Presentation and {{Evaluation}} of the {{IPSL-CM6A-LR Climate Model}}},\
  }\href {https://doi.org/10.1029/2019MS002010} {\bibfield  {journal} {\bibinfo
   {journal} {Journal of Advances in Modeling Earth Systems}\ }\textbf
  {\bibinfo {volume} {12}},\ \bibinfo {pages} {e2019MS002010} (\bibinfo {year}
  {2020})}\BibitemShut {NoStop}%
\bibitem [{\citenamefont {Zhang}\ and\ \citenamefont
  {McFarlane}(1995)}]{zhang_sensitivity_1995}%
  \BibitemOpen
  \bibfield  {author} {\bibinfo {author} {\bibfnamefont {G.}~\bibnamefont
  {Zhang}}\ and\ \bibinfo {author} {\bibfnamefont {N.~A.}\ \bibnamefont
  {McFarlane}},\ }\bibfield  {title} {\bibinfo {title} {Sensitivity of climate
  simulations to the parameterization of cumulus convection in the {{Canadian}}
  climate centre general circulation model},\ }\href
  {https://doi.org/10.1080/07055900.1995.9649539} {\bibfield  {journal}
  {\bibinfo  {journal} {Atmosphere-Ocean}\ }\textbf {\bibinfo {volume} {33}},\
  \bibinfo {pages} {407} (\bibinfo {year} {1995})}\BibitemShut {NoStop}%
\bibitem [{\citenamefont {Schmidt}\ \emph {et~al.}(2023)\citenamefont
  {Schmidt}, \citenamefont {Andrews}, \citenamefont {Bauer}, \citenamefont
  {Durack}, \citenamefont {Loeb}, \citenamefont {Ramaswamy}, \citenamefont
  {Arnold}, \citenamefont {Bosilovich}, \citenamefont {Cole}, \citenamefont
  {Horowitz}, \citenamefont {Johnson}, \citenamefont {Lyman}, \citenamefont
  {Medeiros}, \citenamefont {Michibata}, \citenamefont {Olonscheck},
  \citenamefont {Paynter}, \citenamefont {Raghuraman}, \citenamefont {Schulz},
  \citenamefont {Takasuka}, \citenamefont {Tallapragada}, \citenamefont
  {Taylor},\ and\ \citenamefont {Ziehn}}]{schmidt_ceresmip_2023}%
  \BibitemOpen
  \bibfield  {author} {\bibinfo {author} {\bibfnamefont {G.~A.}\ \bibnamefont
  {Schmidt}}, \bibinfo {author} {\bibfnamefont {T.}~\bibnamefont {Andrews}},
  \bibinfo {author} {\bibfnamefont {S.~E.}\ \bibnamefont {Bauer}}, \bibinfo
  {author} {\bibfnamefont {P.~J.}\ \bibnamefont {Durack}}, \bibinfo {author}
  {\bibfnamefont {N.~G.}\ \bibnamefont {Loeb}}, \bibinfo {author}
  {\bibfnamefont {V.}~\bibnamefont {Ramaswamy}}, \bibinfo {author}
  {\bibfnamefont {N.~P.}\ \bibnamefont {Arnold}}, \bibinfo {author}
  {\bibfnamefont {M.~G.}\ \bibnamefont {Bosilovich}}, \bibinfo {author}
  {\bibfnamefont {J.}~\bibnamefont {Cole}}, \bibinfo {author} {\bibfnamefont
  {L.~W.}\ \bibnamefont {Horowitz}}, \bibinfo {author} {\bibfnamefont {G.~C.}\
  \bibnamefont {Johnson}}, \bibinfo {author} {\bibfnamefont {J.~M.}\
  \bibnamefont {Lyman}}, \bibinfo {author} {\bibfnamefont {B.}~\bibnamefont
  {Medeiros}}, \bibinfo {author} {\bibfnamefont {T.}~\bibnamefont {Michibata}},
  \bibinfo {author} {\bibfnamefont {D.}~\bibnamefont {Olonscheck}}, \bibinfo
  {author} {\bibfnamefont {D.}~\bibnamefont {Paynter}}, \bibinfo {author}
  {\bibfnamefont {S.~P.}\ \bibnamefont {Raghuraman}}, \bibinfo {author}
  {\bibfnamefont {M.}~\bibnamefont {Schulz}}, \bibinfo {author} {\bibfnamefont
  {D.}~\bibnamefont {Takasuka}}, \bibinfo {author} {\bibfnamefont
  {V.}~\bibnamefont {Tallapragada}}, \bibinfo {author} {\bibfnamefont {P.~C.}\
  \bibnamefont {Taylor}},\ and\ \bibinfo {author} {\bibfnamefont
  {T.}~\bibnamefont {Ziehn}},\ }\bibfield  {title} {\bibinfo {title}
  {{{CERESMIP}}: A climate modeling protocol to investigate recent trends in
  the {{Earth}}'s {{Energy Imbalance}}},\ }\href@noop {} {\bibfield  {journal}
  {\bibinfo  {journal} {Frontiers in Climate}\ }\textbf {\bibinfo {volume} {5}}
  (\bibinfo {year} {2023})}\BibitemShut {NoStop}%
\bibitem [{\citenamefont {Stevens}\ \emph {et~al.}(2020)\citenamefont
  {Stevens}, \citenamefont {Bony}, \citenamefont {Brogniez}, \citenamefont
  {Hentgen}, \citenamefont {Hohenegger}, \citenamefont {Kiemle}, \citenamefont
  {L'Ecuyer}, \citenamefont {Naumann}, \citenamefont {Schulz}, \citenamefont
  {Siebesma}, \citenamefont {Vial}, \citenamefont {Winker},\ and\ \citenamefont
  {Zuidema}}]{stevens_sugar_2020}%
  \BibitemOpen
  \bibfield  {author} {\bibinfo {author} {\bibfnamefont {B.}~\bibnamefont
  {Stevens}}, \bibinfo {author} {\bibfnamefont {S.}~\bibnamefont {Bony}},
  \bibinfo {author} {\bibfnamefont {H.}~\bibnamefont {Brogniez}}, \bibinfo
  {author} {\bibfnamefont {L.}~\bibnamefont {Hentgen}}, \bibinfo {author}
  {\bibfnamefont {C.}~\bibnamefont {Hohenegger}}, \bibinfo {author}
  {\bibfnamefont {C.}~\bibnamefont {Kiemle}}, \bibinfo {author} {\bibfnamefont
  {T.~S.}\ \bibnamefont {L'Ecuyer}}, \bibinfo {author} {\bibfnamefont {A.~K.}\
  \bibnamefont {Naumann}}, \bibinfo {author} {\bibfnamefont {H.}~\bibnamefont
  {Schulz}}, \bibinfo {author} {\bibfnamefont {P.~A.}\ \bibnamefont
  {Siebesma}}, \bibinfo {author} {\bibfnamefont {J.}~\bibnamefont {Vial}},
  \bibinfo {author} {\bibfnamefont {D.~M.}\ \bibnamefont {Winker}},\ and\
  \bibinfo {author} {\bibfnamefont {P.}~\bibnamefont {Zuidema}},\ }\bibfield
  {title} {\bibinfo {title} {Sugar, gravel, fish and flowers: {{Mesoscale}}
  cloud patterns in the trade winds},\ }\href {https://doi.org/10.1002/qj.3662}
  {\bibfield  {journal} {\bibinfo  {journal} {Quarterly Journal of the Royal
  Meteorological Society}\ }\textbf {\bibinfo {volume} {146}},\ \bibinfo
  {pages} {141} (\bibinfo {year} {2020})}\BibitemShut {NoStop}%
\bibitem [{\citenamefont {Schneider}\ \emph {et~al.}(2017)\citenamefont
  {Schneider}, \citenamefont {Teixeira}, \citenamefont {Bretherton},
  \citenamefont {Brient}, \citenamefont {Pressel}, \citenamefont {Sch{\"a}r},\
  and\ \citenamefont {Siebesma}}]{schneider_climate_2017}%
  \BibitemOpen
  \bibfield  {author} {\bibinfo {author} {\bibfnamefont {T.}~\bibnamefont
  {Schneider}}, \bibinfo {author} {\bibfnamefont {J.}~\bibnamefont {Teixeira}},
  \bibinfo {author} {\bibfnamefont {C.~S.}\ \bibnamefont {Bretherton}},
  \bibinfo {author} {\bibfnamefont {F.}~\bibnamefont {Brient}}, \bibinfo
  {author} {\bibfnamefont {K.~G.}\ \bibnamefont {Pressel}}, \bibinfo {author}
  {\bibfnamefont {C.}~\bibnamefont {Sch{\"a}r}},\ and\ \bibinfo {author}
  {\bibfnamefont {A.~P.}\ \bibnamefont {Siebesma}},\ }\bibfield  {title}
  {\bibinfo {title} {Climate goals and computing the future of clouds},\ }\href
  {https://doi.org/10.1038/nclimate3190} {\bibfield  {journal} {\bibinfo
  {journal} {Nature Climate Change}\ }\textbf {\bibinfo {volume} {7}},\
  \bibinfo {pages} {3} (\bibinfo {year} {2017})}\BibitemShut {NoStop}%
\bibitem [{\citenamefont {Stephens}(1978)}]{stephens_radiation_1978}%
  \BibitemOpen
  \bibfield  {author} {\bibinfo {author} {\bibfnamefont {G.~L.}\ \bibnamefont
  {Stephens}},\ }\bibfield  {title} {\bibinfo {title} {Radiation {{Profiles}}
  in {{Extended Water Clouds}}. {{II}}: {{Parameterization Schemes}}},\ }\href
  {https://doi.org/10.1175/1520-0469(1978)035<2123:RPIEWC>2.0.CO;2} {\bibfield
  {journal} {\bibinfo  {journal} {Journal of the Atmospheric Sciences}\
  }\textbf {\bibinfo {volume} {35}},\ \bibinfo {pages} {2123} (\bibinfo {year}
  {1978})}\BibitemShut {NoStop}%
\bibitem [{\citenamefont {Bony}\ \emph {et~al.}(2020)\citenamefont {Bony},
  \citenamefont {Semie}, \citenamefont {Kramer}, \citenamefont {Soden},
  \citenamefont {Tompkins},\ and\ \citenamefont
  {Emanuel}}]{bony_observed_2020}%
  \BibitemOpen
  \bibfield  {author} {\bibinfo {author} {\bibfnamefont {S.}~\bibnamefont
  {Bony}}, \bibinfo {author} {\bibfnamefont {A.}~\bibnamefont {Semie}},
  \bibinfo {author} {\bibfnamefont {R.~J.}\ \bibnamefont {Kramer}}, \bibinfo
  {author} {\bibfnamefont {B.}~\bibnamefont {Soden}}, \bibinfo {author}
  {\bibfnamefont {A.~M.}\ \bibnamefont {Tompkins}},\ and\ \bibinfo {author}
  {\bibfnamefont {K.~A.}\ \bibnamefont {Emanuel}},\ }\bibfield  {title}
  {\bibinfo {title} {Observed {{Modulation}} of the {{Tropical Radiation
  Budget}} by {{Deep Convective Organization}} and {{Lower}}-{{Tropospheric
  Stability}}},\ }\href {https://doi.org/10.1029/2019AV000155} {\bibfield
  {journal} {\bibinfo  {journal} {AGU Advances}\ }\textbf {\bibinfo {volume}
  {1}},\ \bibinfo {pages} {e2019AV000155} (\bibinfo {year} {2020})}\BibitemShut
  {NoStop}%
\bibitem [{\citenamefont {Muller}\ \emph {et~al.}(2022)\citenamefont {Muller},
  \citenamefont {Yang}, \citenamefont {Craig}, \citenamefont {Cronin},
  \citenamefont {Fildier}, \citenamefont {Haerter}, \citenamefont {Hohenegger},
  \citenamefont {Mapes}, \citenamefont {Randall}, \citenamefont {Shamekh},\
  and\ \citenamefont {Sherwood}}]{muller_spontaneous_2022}%
  \BibitemOpen
  \bibfield  {author} {\bibinfo {author} {\bibfnamefont {C.}~\bibnamefont
  {Muller}}, \bibinfo {author} {\bibfnamefont {D.}~\bibnamefont {Yang}},
  \bibinfo {author} {\bibfnamefont {G.}~\bibnamefont {Craig}}, \bibinfo
  {author} {\bibfnamefont {T.}~\bibnamefont {Cronin}}, \bibinfo {author}
  {\bibfnamefont {B.}~\bibnamefont {Fildier}}, \bibinfo {author} {\bibfnamefont
  {J.~O.}\ \bibnamefont {Haerter}}, \bibinfo {author} {\bibfnamefont
  {C.}~\bibnamefont {Hohenegger}}, \bibinfo {author} {\bibfnamefont
  {B.}~\bibnamefont {Mapes}}, \bibinfo {author} {\bibfnamefont
  {D.}~\bibnamefont {Randall}}, \bibinfo {author} {\bibfnamefont
  {S.}~\bibnamefont {Shamekh}},\ and\ \bibinfo {author} {\bibfnamefont {S.~C.}\
  \bibnamefont {Sherwood}},\ }\bibfield  {title} {\bibinfo {title} {Spontaneous
  {{Aggregation}} of {{Convective Storms}}},\ }\href
  {https://doi.org/10.1146/annurev-fluid-022421-011319} {\bibfield  {journal}
  {\bibinfo  {journal} {Annual Review of Fluid Mechanics}\ }\textbf {\bibinfo
  {volume} {54}},\ \bibinfo {pages} {133} (\bibinfo {year} {2022})}\BibitemShut
  {NoStop}%
\bibitem [{\citenamefont {Ahlers}\ \emph {et~al.}(2009)\citenamefont {Ahlers},
  \citenamefont {Grossmann},\ and\ \citenamefont {Lohse}}]{ahlers_heat_2009}%
  \BibitemOpen
  \bibfield  {author} {\bibinfo {author} {\bibfnamefont {G.}~\bibnamefont
  {Ahlers}}, \bibinfo {author} {\bibfnamefont {S.}~\bibnamefont {Grossmann}},\
  and\ \bibinfo {author} {\bibfnamefont {D.}~\bibnamefont {Lohse}},\ }\bibfield
   {title} {\bibinfo {title} {Heat transfer and large scale dynamics in
  turbulent {{Rayleigh-B{\'e}nard}} convection},\ }\href
  {https://doi.org/10.1103/RevModPhys.81.503} {\bibfield  {journal} {\bibinfo
  {journal} {Reviews of Modern Physics}\ }\textbf {\bibinfo {volume} {81}},\
  \bibinfo {pages} {503} (\bibinfo {year} {2009})}\BibitemShut {NoStop}%
\bibitem [{\citenamefont {Bec}(2005)}]{bec_multifractal_2005}%
  \BibitemOpen
  \bibfield  {author} {\bibinfo {author} {\bibfnamefont {J.}~\bibnamefont
  {Bec}},\ }\bibfield  {title} {\bibinfo {title} {Multifractal concentrations
  of inertial particles in smooth random flows},\ }\href
  {https://doi.org/10.1017/S0022112005003368} {\bibfield  {journal} {\bibinfo
  {journal} {Journal of Fluid Mechanics}\ }\textbf {\bibinfo {volume} {528}},\
  \bibinfo {pages} {255} (\bibinfo {year} {2005})}\BibitemShut {NoStop}%
\bibitem [{\citenamefont {Falkovich}\ \emph {et~al.}(2006)\citenamefont
  {Falkovich}, \citenamefont {Stepanov},\ and\ \citenamefont
  {Vucelja}}]{falkovich_rain_2006}%
  \BibitemOpen
  \bibfield  {author} {\bibinfo {author} {\bibfnamefont {G.}~\bibnamefont
  {Falkovich}}, \bibinfo {author} {\bibfnamefont {M.~G.}\ \bibnamefont
  {Stepanov}},\ and\ \bibinfo {author} {\bibfnamefont {M.}~\bibnamefont
  {Vucelja}},\ }\bibfield  {title} {\bibinfo {title} {Rain {{Initiation Time}}
  in {{Turbulent Warm Clouds}}},\ }\href {https://doi.org/10.1175/JAM2364.1}
  {\bibfield  {journal} {\bibinfo  {journal} {Journal of Applied Meteorology
  and Climatology}\ }\textbf {\bibinfo {volume} {45}},\ \bibinfo {pages} {591}
  (\bibinfo {year} {2006})}\BibitemShut {NoStop}%
\bibitem [{\citenamefont {Twomey}(1959)}]{twomey_nuclei_1959}%
  \BibitemOpen
  \bibfield  {author} {\bibinfo {author} {\bibfnamefont {S.}~\bibnamefont
  {Twomey}},\ }\bibfield  {title} {\bibinfo {title} {The nuclei of natural
  cloud formation part {{II}}: {{The}} supersaturation in natural clouds and
  the variation of cloud droplet concentration},\ }\href
  {https://doi.org/10.1007/BF01993560} {\bibfield  {journal} {\bibinfo
  {journal} {Geofisica Pura e Applicata}\ }\textbf {\bibinfo {volume} {43}},\
  \bibinfo {pages} {243} (\bibinfo {year} {1959})}\BibitemShut {NoStop}%
\bibitem [{\citenamefont {Ghan}\ \emph {et~al.}(2011)\citenamefont {Ghan},
  \citenamefont {{Abdul-Razzak}}, \citenamefont {Nenes}, \citenamefont {Ming},
  \citenamefont {Liu}, \citenamefont {Ovchinnikov}, \citenamefont {Shipway},
  \citenamefont {Meskhidze}, \citenamefont {Xu},\ and\ \citenamefont
  {Shi}}]{ghan_droplet_2011}%
  \BibitemOpen
  \bibfield  {author} {\bibinfo {author} {\bibfnamefont {S.~J.}\ \bibnamefont
  {Ghan}}, \bibinfo {author} {\bibfnamefont {H.}~\bibnamefont
  {{Abdul-Razzak}}}, \bibinfo {author} {\bibfnamefont {A.}~\bibnamefont
  {Nenes}}, \bibinfo {author} {\bibfnamefont {Y.}~\bibnamefont {Ming}},
  \bibinfo {author} {\bibfnamefont {X.}~\bibnamefont {Liu}}, \bibinfo {author}
  {\bibfnamefont {M.}~\bibnamefont {Ovchinnikov}}, \bibinfo {author}
  {\bibfnamefont {B.}~\bibnamefont {Shipway}}, \bibinfo {author} {\bibfnamefont
  {N.}~\bibnamefont {Meskhidze}}, \bibinfo {author} {\bibfnamefont
  {J.}~\bibnamefont {Xu}},\ and\ \bibinfo {author} {\bibfnamefont
  {X.}~\bibnamefont {Shi}},\ }\bibfield  {title} {\bibinfo {title} {Droplet
  nucleation: {{Physically-based}} parameterizations and comparative
  evaluation},\ }\bibfield  {journal} {\bibinfo  {journal} {Journal of Advances
  in Modeling Earth Systems}\ }\textbf {\bibinfo {volume} {3}},\ \href
  {https://doi.org/10.1029/2011MS000074} {10.1029/2011MS000074} (\bibinfo
  {year} {2011})\BibitemShut {NoStop}%
\bibitem [{\citenamefont {Pruppacher}\ and\ \citenamefont
  {Klett}(2010)}]{pruppacher_microphysics_2010}%
  \BibitemOpen
  \bibfield  {author} {\bibinfo {author} {\bibfnamefont {H.}~\bibnamefont
  {Pruppacher}}\ and\ \bibinfo {author} {\bibfnamefont {J.}~\bibnamefont
  {Klett}},\ }\href {https://doi.org/10.1007/978-0-306-48100-0} {\emph
  {\bibinfo {title} {Microphysics of {{Clouds}} and {{Precipitation}}}}}\
  (\bibinfo  {publisher} {Springer Netherlands},\ \bibinfo {address}
  {Dordrecht},\ \bibinfo {year} {2010})\BibitemShut {NoStop}%
\bibitem [{\citenamefont {Krueger}(2020)}]{krueger_technical_2020}%
  \BibitemOpen
  \bibfield  {author} {\bibinfo {author} {\bibfnamefont {S.~K.}\ \bibnamefont
  {Krueger}},\ }\bibfield  {title} {\bibinfo {title} {Technical note:
  {{Equilibrium}} droplet size distributions in a turbulent cloud chamber with
  uniform supersaturation},\ }\href {https://doi.org/10.5194/acp-20-7895-2020}
  {\bibfield  {journal} {\bibinfo  {journal} {Atmospheric Chemistry and
  Physics}\ }\textbf {\bibinfo {volume} {20}},\ \bibinfo {pages} {7895}
  (\bibinfo {year} {2020})}\BibitemShut {NoStop}%
\bibitem [{\citenamefont {Hess}\ \emph {et~al.}(1998)\citenamefont {Hess},
  \citenamefont {Koepke},\ and\ \citenamefont {Schult}}]{hess_optical_1998}%
  \BibitemOpen
  \bibfield  {author} {\bibinfo {author} {\bibfnamefont {M.}~\bibnamefont
  {Hess}}, \bibinfo {author} {\bibfnamefont {P.}~\bibnamefont {Koepke}},\ and\
  \bibinfo {author} {\bibfnamefont {I.}~\bibnamefont {Schult}},\ }\bibfield
  {title} {\bibinfo {title} {Optical {{Properties}} of {{Aerosols}} and
  {{Clouds}}: {{The Software Package OPAC}}},\ }\href
  {https://doi.org/10.1175/1520-0477(1998)079<0831:OPOAAC>2.0.CO;2} {\bibfield
  {journal} {\bibinfo  {journal} {Bulletin of the American Meteorological
  Society}\ }\textbf {\bibinfo {volume} {79}},\ \bibinfo {pages} {831}
  (\bibinfo {year} {1998})}\BibitemShut {NoStop}%
\bibitem [{\citenamefont {Khain}\ \emph {et~al.}(2000)\citenamefont {Khain},
  \citenamefont {Ovtchinnikov}, \citenamefont {Pinsky}, \citenamefont
  {Pokrovsky},\ and\ \citenamefont {Krugliak}}]{khain_notes_2000}%
  \BibitemOpen
  \bibfield  {author} {\bibinfo {author} {\bibfnamefont {A.}~\bibnamefont
  {Khain}}, \bibinfo {author} {\bibfnamefont {M.}~\bibnamefont {Ovtchinnikov}},
  \bibinfo {author} {\bibfnamefont {M.}~\bibnamefont {Pinsky}}, \bibinfo
  {author} {\bibfnamefont {A.}~\bibnamefont {Pokrovsky}},\ and\ \bibinfo
  {author} {\bibfnamefont {H.}~\bibnamefont {Krugliak}},\ }\bibfield  {title}
  {\bibinfo {title} {Notes on the state-of-the-art numerical modeling of cloud
  microphysics},\ }\href {https://doi.org/10.1016/S0169-8095(00)00064-8}
  {\bibfield  {journal} {\bibinfo  {journal} {Atmospheric Research}\ }\textbf
  {\bibinfo {volume} {55}},\ \bibinfo {pages} {159} (\bibinfo {year}
  {2000})}\BibitemShut {NoStop}%
\bibitem [{\citenamefont {Seifert}\ \emph {et~al.}(2006)\citenamefont
  {Seifert}, \citenamefont {Khain}, \citenamefont {Pokrovsky},\ and\
  \citenamefont {Beheng}}]{seifert_comparison_2006}%
  \BibitemOpen
  \bibfield  {author} {\bibinfo {author} {\bibfnamefont {A.}~\bibnamefont
  {Seifert}}, \bibinfo {author} {\bibfnamefont {A.}~\bibnamefont {Khain}},
  \bibinfo {author} {\bibfnamefont {A.}~\bibnamefont {Pokrovsky}},\ and\
  \bibinfo {author} {\bibfnamefont {K.~D.}\ \bibnamefont {Beheng}},\ }\bibfield
   {title} {\bibinfo {title} {A comparison of spectral bin and two-moment bulk
  mixed-phase cloud microphysics},\ }\href
  {https://doi.org/10.1016/j.atmosres.2005.06.009} {\bibfield  {journal}
  {\bibinfo  {journal} {Atmospheric Research}\ }\textbf {\bibinfo {volume}
  {80}},\ \bibinfo {pages} {46} (\bibinfo {year} {2006})}\BibitemShut {NoStop}%
\bibitem [{\citenamefont {Khain}\ \emph {et~al.}(2015)\citenamefont {Khain},
  \citenamefont {Beheng}, \citenamefont {Heymsfield}, \citenamefont {Korolev},
  \citenamefont {Krichak}, \citenamefont {Levin}, \citenamefont {Pinsky},
  \citenamefont {Phillips}, \citenamefont {Prabhakaran}, \citenamefont
  {Teller}, \citenamefont {Van Den~Heever},\ and\ \citenamefont
  {Yano}}]{khain_representation_2015}%
  \BibitemOpen
  \bibfield  {author} {\bibinfo {author} {\bibfnamefont {A.~P.}\ \bibnamefont
  {Khain}}, \bibinfo {author} {\bibfnamefont {K.~D.}\ \bibnamefont {Beheng}},
  \bibinfo {author} {\bibfnamefont {A.}~\bibnamefont {Heymsfield}}, \bibinfo
  {author} {\bibfnamefont {A.}~\bibnamefont {Korolev}}, \bibinfo {author}
  {\bibfnamefont {S.~O.}\ \bibnamefont {Krichak}}, \bibinfo {author}
  {\bibfnamefont {Z.}~\bibnamefont {Levin}}, \bibinfo {author} {\bibfnamefont
  {M.}~\bibnamefont {Pinsky}}, \bibinfo {author} {\bibfnamefont
  {V.}~\bibnamefont {Phillips}}, \bibinfo {author} {\bibfnamefont
  {T.}~\bibnamefont {Prabhakaran}}, \bibinfo {author} {\bibfnamefont
  {A.}~\bibnamefont {Teller}}, \bibinfo {author} {\bibfnamefont {S.~C.}\
  \bibnamefont {Van Den~Heever}},\ and\ \bibinfo {author} {\bibfnamefont
  {J.-I.}\ \bibnamefont {Yano}},\ }\bibfield  {title} {\bibinfo {title}
  {Representation of microphysical processes in cloud-resolving models:
  {{Spectral}} (bin) microphysics versus bulk parameterization},\ }\href
  {https://doi.org/10.1002/2014RG000468} {\bibfield  {journal} {\bibinfo
  {journal} {Reviews of Geophysics}\ }\textbf {\bibinfo {volume} {53}},\
  \bibinfo {pages} {247} (\bibinfo {year} {2015})}\BibitemShut {NoStop}%
\bibitem [{\citenamefont {Kreidenweis}\ \emph {et~al.}(2019)\citenamefont
  {Kreidenweis}, \citenamefont {Petters},\ and\ \citenamefont
  {Lohmann}}]{kreidenweis_100_2019}%
  \BibitemOpen
  \bibfield  {author} {\bibinfo {author} {\bibfnamefont {S.~M.}\ \bibnamefont
  {Kreidenweis}}, \bibinfo {author} {\bibfnamefont {M.}~\bibnamefont
  {Petters}},\ and\ \bibinfo {author} {\bibfnamefont {U.}~\bibnamefont
  {Lohmann}},\ }\bibfield  {title} {\bibinfo {title} {100 {{Years}} of
  {{Progress}} in {{Cloud Physics}}, {{Aerosols}}, and {{Aerosol Chemistry
  Research}}},\ }\href {https://doi.org/10.1175/AMSMONOGRAPHS-D-18-0024.1}
  {\bibfield  {journal} {\bibinfo  {journal} {Meteorological Monographs}\
  }\textbf {\bibinfo {volume} {59}},\ \bibinfo {pages} {11.1} (\bibinfo {year}
  {2019})}\BibitemShut {NoStop}%
\bibitem [{\citenamefont {Flossmann}\ and\ \citenamefont
  {Wobrock}(2010)}]{flossmann_review_2010}%
  \BibitemOpen
  \bibfield  {author} {\bibinfo {author} {\bibfnamefont {A.~I.}\ \bibnamefont
  {Flossmann}}\ and\ \bibinfo {author} {\bibfnamefont {W.}~\bibnamefont
  {Wobrock}},\ }\bibfield  {title} {\bibinfo {title} {A review of our
  understanding of the aerosol--cloud interaction from the perspective of a bin
  resolved cloud scale modelling},\ }\href
  {https://doi.org/10.1016/j.atmosres.2010.05.008} {\bibfield  {journal}
  {\bibinfo  {journal} {Atmospheric Research}\ }\bibinfo {series} {From the
  {{Lab}} to {{Models}} and {{Global Observations}}: {{Hans R}}. {{Pruppacher}}
  and {{Cloud Physics}}},\ \textbf {\bibinfo {volume} {97}},\ \bibinfo {pages}
  {478} (\bibinfo {year} {2010})}\BibitemShut {NoStop}%
\bibitem [{\citenamefont {Cotton}\ \emph {et~al.}(2011)\citenamefont {Cotton},
  \citenamefont {Bryan},\ and\ \citenamefont {{Van den
  Heever}}}]{cotton_storm_2011}%
  \BibitemOpen
  \bibfield  {author} {\bibinfo {author} {\bibfnamefont {W.~R.}\ \bibnamefont
  {Cotton}}, \bibinfo {author} {\bibfnamefont {G.~H.}\ \bibnamefont {Bryan}},\
  and\ \bibinfo {author} {\bibfnamefont {S.~C.}\ \bibnamefont {{Van den
  Heever}}},\ }\href@noop {} {\emph {\bibinfo {title} {Storm and Cloud
  Dynamics: The Dynamics of Clouds and Precipitating Mesoscale Systems}}},\
  \bibinfo {edition} {2nd}\ ed.,\ \bibinfo {series} {International Geophysics
  Series}\ No.~\bibinfo {number} {99}\ (\bibinfo  {publisher} {Acad. Press},\
  \bibinfo {address} {Amsterdam},\ \bibinfo {year} {2011})\BibitemShut
  {NoStop}%
\bibitem [{\citenamefont {Hansen}\ \emph {et~al.}(2023)\citenamefont {Hansen},
  \citenamefont {Sato}, \citenamefont {Simons}, \citenamefont {Nazarenko},
  \citenamefont {Sangha}, \citenamefont {Kharecha}, \citenamefont {Zachos},
  \citenamefont {{von Schuckmann}}, \citenamefont {Loeb}, \citenamefont
  {Osman}, \citenamefont {Jin}, \citenamefont {Tselioudis}, \citenamefont
  {Jeong}, \citenamefont {Lacis}, \citenamefont {Ruedy}, \citenamefont
  {Russell}, \citenamefont {Cao},\ and\ \citenamefont
  {Li}}]{hansen_global_2023}%
  \BibitemOpen
  \bibfield  {author} {\bibinfo {author} {\bibfnamefont {J.~E.}\ \bibnamefont
  {Hansen}}, \bibinfo {author} {\bibfnamefont {M.}~\bibnamefont {Sato}},
  \bibinfo {author} {\bibfnamefont {L.}~\bibnamefont {Simons}}, \bibinfo
  {author} {\bibfnamefont {L.~S.}\ \bibnamefont {Nazarenko}}, \bibinfo {author}
  {\bibfnamefont {I.}~\bibnamefont {Sangha}}, \bibinfo {author} {\bibfnamefont
  {P.}~\bibnamefont {Kharecha}}, \bibinfo {author} {\bibfnamefont {J.~C.}\
  \bibnamefont {Zachos}}, \bibinfo {author} {\bibfnamefont {K.}~\bibnamefont
  {{von Schuckmann}}}, \bibinfo {author} {\bibfnamefont {N.~G.}\ \bibnamefont
  {Loeb}}, \bibinfo {author} {\bibfnamefont {M.~B.}\ \bibnamefont {Osman}},
  \bibinfo {author} {\bibfnamefont {Q.}~\bibnamefont {Jin}}, \bibinfo {author}
  {\bibfnamefont {G.}~\bibnamefont {Tselioudis}}, \bibinfo {author}
  {\bibfnamefont {E.}~\bibnamefont {Jeong}}, \bibinfo {author} {\bibfnamefont
  {A.}~\bibnamefont {Lacis}}, \bibinfo {author} {\bibfnamefont
  {R.}~\bibnamefont {Ruedy}}, \bibinfo {author} {\bibfnamefont
  {G.}~\bibnamefont {Russell}}, \bibinfo {author} {\bibfnamefont
  {J.}~\bibnamefont {Cao}},\ and\ \bibinfo {author} {\bibfnamefont
  {J.}~\bibnamefont {Li}},\ }\bibfield  {title} {\bibinfo {title} {Global
  warming in the pipeline},\ }\href {https://doi.org/10.1093/oxfclm/kgad008}
  {\bibfield  {journal} {\bibinfo  {journal} {Oxford Open Climate Change}\
  }\textbf {\bibinfo {volume} {3}},\ \bibinfo {pages} {kgad008} (\bibinfo
  {year} {2023})}\BibitemShut {NoStop}%
\bibitem [{\citenamefont {Kessler}(1969)}]{kessler_distribution_1969}%
  \BibitemOpen
  \bibfield  {author} {\bibinfo {author} {\bibfnamefont {E.}~\bibnamefont
  {Kessler}},\ }\href {https://doi.org/10.1007/978-1-935704-36-2} {\emph
  {\bibinfo {title} {On the {{Distribution}} and {{Continuity}} of {{Water
  Substance}} in {{Atmospheric Circulations}}}}}\ (\bibinfo  {publisher}
  {American Meteorological Society},\ \bibinfo {address} {Boston, MA},\
  \bibinfo {year} {1969})\BibitemShut {NoStop}%
\bibitem [{\citenamefont {Berry}\ and\ \citenamefont
  {Reinhardt}(1974{\natexlab{a}})}]{berry_analysis_1974}%
  \BibitemOpen
  \bibfield  {author} {\bibinfo {author} {\bibfnamefont {E.~X.}\ \bibnamefont
  {Berry}}\ and\ \bibinfo {author} {\bibfnamefont {R.~L.}\ \bibnamefont
  {Reinhardt}},\ }\bibfield  {title} {\bibinfo {title} {An {{Analysis}} of
  {{Cloud Drop Growth}} by {{Collection Part II}}. {{Single Initial
  Distributions}}},\ }\href
  {https://doi.org/10.1175/1520-0469(1974)031<1825:AAOCDG>2.0.CO;2} {\bibfield
  {journal} {\bibinfo  {journal} {Journal of the Atmospheric Sciences}\
  }\textbf {\bibinfo {volume} {31}},\ \bibinfo {pages} {1825} (\bibinfo {year}
  {1974}{\natexlab{a}})}\BibitemShut {NoStop}%
\bibitem [{\citenamefont {Berry}\ and\ \citenamefont
  {Reinhardt}(1974{\natexlab{b}})}]{berry_analysis_1974-3}%
  \BibitemOpen
  \bibfield  {author} {\bibinfo {author} {\bibfnamefont {E.~X.}\ \bibnamefont
  {Berry}}\ and\ \bibinfo {author} {\bibfnamefont {R.~L.}\ \bibnamefont
  {Reinhardt}},\ }\bibfield  {title} {\bibinfo {title} {An {{Analysis}} of
  {{Cloud Drop Growth}} by {{Collection}}: {{Part IV}}. {{A New
  Parameterization}}},\ }\href
  {https://doi.org/10.1175/1520-0469(1974)031<2127:AAOCDG>2.0.CO;2} {\bibfield
  {journal} {\bibinfo  {journal} {Journal of the Atmospheric Sciences}\
  }\textbf {\bibinfo {volume} {31}},\ \bibinfo {pages} {2127} (\bibinfo {year}
  {1974}{\natexlab{b}})}\BibitemShut {NoStop}%
\bibitem [{\citenamefont {Hall}(1980)}]{hall_detailed_1980}%
  \BibitemOpen
  \bibfield  {author} {\bibinfo {author} {\bibfnamefont {W.~D.}\ \bibnamefont
  {Hall}},\ }\bibfield  {title} {\bibinfo {title} {A {{Detailed Microphysical
  Model Within}} a {{Two-Dimensional Dynamic Framework}}: {{Model Description}}
  and {{Preliminary Results}}},\ }\href
  {https://doi.org/10.1175/1520-0469(1980)037<2486:ADMMWA>2.0.CO;2} {\bibfield
  {journal} {\bibinfo  {journal} {Journal of the Atmospheric Sciences}\
  }\textbf {\bibinfo {volume} {37}},\ \bibinfo {pages} {2486} (\bibinfo {year}
  {1980})}\BibitemShut {NoStop}%
\bibitem [{\citenamefont {Seifert}\ and\ \citenamefont
  {Beheng}(2001)}]{seifert_double-moment_2001}%
  \BibitemOpen
  \bibfield  {author} {\bibinfo {author} {\bibfnamefont {A.}~\bibnamefont
  {Seifert}}\ and\ \bibinfo {author} {\bibfnamefont {K.~D.}\ \bibnamefont
  {Beheng}},\ }\bibfield  {title} {\bibinfo {title} {A double-moment
  parameterization for simulating autoconversion, accretion and
  selfcollection},\ }\href {https://doi.org/10.1016/S0169-8095(01)00126-0}
  {\bibfield  {journal} {\bibinfo  {journal} {Atmospheric Research}\ }\textbf
  {\bibinfo {volume} {59--60}},\ \bibinfo {pages} {265} (\bibinfo {year}
  {2001})}\BibitemShut {NoStop}%
\bibitem [{\citenamefont {Morrison}\ \emph {et~al.}(2020)\citenamefont
  {Morrison}, \citenamefont {{van Lier-Walqui}}, \citenamefont {Fridlind},
  \citenamefont {Grabowski}, \citenamefont {Harrington}, \citenamefont {Hoose},
  \citenamefont {Korolev}, \citenamefont {Kumjian}, \citenamefont {Milbrandt},
  \citenamefont {Pawlowska}, \citenamefont {Posselt}, \citenamefont {Prat},
  \citenamefont {Reimel}, \citenamefont {Shima}, \citenamefont {{van
  Diedenhoven}},\ and\ \citenamefont {Xue}}]{morrison_confronting_2020}%
  \BibitemOpen
  \bibfield  {author} {\bibinfo {author} {\bibfnamefont {H.}~\bibnamefont
  {Morrison}}, \bibinfo {author} {\bibfnamefont {M.}~\bibnamefont {{van
  Lier-Walqui}}}, \bibinfo {author} {\bibfnamefont {A.~M.}\ \bibnamefont
  {Fridlind}}, \bibinfo {author} {\bibfnamefont {W.~W.}\ \bibnamefont
  {Grabowski}}, \bibinfo {author} {\bibfnamefont {J.~Y.}\ \bibnamefont
  {Harrington}}, \bibinfo {author} {\bibfnamefont {C.}~\bibnamefont {Hoose}},
  \bibinfo {author} {\bibfnamefont {A.}~\bibnamefont {Korolev}}, \bibinfo
  {author} {\bibfnamefont {M.~R.}\ \bibnamefont {Kumjian}}, \bibinfo {author}
  {\bibfnamefont {J.~A.}\ \bibnamefont {Milbrandt}}, \bibinfo {author}
  {\bibfnamefont {H.}~\bibnamefont {Pawlowska}}, \bibinfo {author}
  {\bibfnamefont {D.~J.}\ \bibnamefont {Posselt}}, \bibinfo {author}
  {\bibfnamefont {O.~P.}\ \bibnamefont {Prat}}, \bibinfo {author}
  {\bibfnamefont {K.~J.}\ \bibnamefont {Reimel}}, \bibinfo {author}
  {\bibfnamefont {S.-I.}\ \bibnamefont {Shima}}, \bibinfo {author}
  {\bibfnamefont {B.}~\bibnamefont {{van Diedenhoven}}},\ and\ \bibinfo
  {author} {\bibfnamefont {L.}~\bibnamefont {Xue}},\ }\bibfield  {title}
  {\bibinfo {title} {Confronting the {{Challenge}} of {{Modeling Cloud}} and
  {{Precipitation Microphysics}}},\ }\href
  {https://doi.org/10.1029/2019MS001689} {\bibfield  {journal} {\bibinfo
  {journal} {Journal of Advances in Modeling Earth Systems}\ }\textbf {\bibinfo
  {volume} {12}},\ \bibinfo {pages} {e2019MS001689} (\bibinfo {year}
  {2020})}\BibitemShut {NoStop}%
\bibitem [{\citenamefont {Shima}\ \emph {et~al.}(2009)\citenamefont {Shima},
  \citenamefont {Kusano}, \citenamefont {Kawano}, \citenamefont {Sugiyama},\
  and\ \citenamefont {Kawahara}}]{shima_super-droplet_2009}%
  \BibitemOpen
  \bibfield  {author} {\bibinfo {author} {\bibfnamefont {S.-i.}\ \bibnamefont
  {Shima}}, \bibinfo {author} {\bibfnamefont {K.}~\bibnamefont {Kusano}},
  \bibinfo {author} {\bibfnamefont {A.}~\bibnamefont {Kawano}}, \bibinfo
  {author} {\bibfnamefont {T.}~\bibnamefont {Sugiyama}},\ and\ \bibinfo
  {author} {\bibfnamefont {S.}~\bibnamefont {Kawahara}},\ }\bibfield  {title}
  {\bibinfo {title} {Super-{{Droplet Method}} for the {{Numerical Simulation}}
  of {{Clouds}} and {{Precipitation}}: A {{Particle-Based Microphysics Model
  Coupled}} with {{Non-hydrostatic Model}}},\ }\href
  {https://doi.org/10.1002/qj.441} {\bibfield  {journal} {\bibinfo  {journal}
  {Quarterly Journal of the Royal Meteorological Society}\ }\textbf {\bibinfo
  {volume} {135}},\ \bibinfo {pages} {1307} (\bibinfo {year} {2009})},\ \Eprint
  {https://arxiv.org/abs/physics/0701103} {arXiv:physics/0701103} \BibitemShut
  {NoStop}%
\bibitem [{\citenamefont {Grabowski}\ \emph {et~al.}(2019)\citenamefont
  {Grabowski}, \citenamefont {Morrison}, \citenamefont {Shima}, \citenamefont
  {Abade}, \citenamefont {Dziekan},\ and\ \citenamefont
  {Pawlowska}}]{grabowski_modeling_2019}%
  \BibitemOpen
  \bibfield  {author} {\bibinfo {author} {\bibfnamefont {W.~W.}\ \bibnamefont
  {Grabowski}}, \bibinfo {author} {\bibfnamefont {H.}~\bibnamefont {Morrison}},
  \bibinfo {author} {\bibfnamefont {S.-I.}\ \bibnamefont {Shima}}, \bibinfo
  {author} {\bibfnamefont {G.~C.}\ \bibnamefont {Abade}}, \bibinfo {author}
  {\bibfnamefont {P.}~\bibnamefont {Dziekan}},\ and\ \bibinfo {author}
  {\bibfnamefont {H.}~\bibnamefont {Pawlowska}},\ }\bibfield  {title} {\bibinfo
  {title} {Modeling of {{Cloud Microphysics}}: {{Can We Do Better}}?},\ }\href
  {https://doi.org/10.1175/BAMS-D-18-0005.1} {\bibfield  {journal} {\bibinfo
  {journal} {Bulletin of the American Meteorological Society}\ }\textbf
  {\bibinfo {volume} {100}},\ \bibinfo {pages} {655} (\bibinfo {year}
  {2019})}\BibitemShut {NoStop}%
\bibitem [{\citenamefont {Shima}\ \emph {et~al.}(2020)\citenamefont {Shima},
  \citenamefont {Sato}, \citenamefont {Hashimoto},\ and\ \citenamefont
  {Misumi}}]{shima_predicting_2020}%
  \BibitemOpen
  \bibfield  {author} {\bibinfo {author} {\bibfnamefont {S.-i.}\ \bibnamefont
  {Shima}}, \bibinfo {author} {\bibfnamefont {Y.}~\bibnamefont {Sato}},
  \bibinfo {author} {\bibfnamefont {A.}~\bibnamefont {Hashimoto}},\ and\
  \bibinfo {author} {\bibfnamefont {R.}~\bibnamefont {Misumi}},\ }\bibfield
  {title} {\bibinfo {title} {Predicting the morphology of ice particles in deep
  convection using the super-droplet method: Development and evaluation of
  {{SCALE-SDM}} 0.2.5-2.2.0, -2.2.1, and -2.2.2},\ }\href
  {https://doi.org/10.5194/gmd-13-4107-2020} {\bibfield  {journal} {\bibinfo
  {journal} {Geoscientific Model Development}\ }\textbf {\bibinfo {volume}
  {13}},\ \bibinfo {pages} {4107} (\bibinfo {year} {2020})}\BibitemShut
  {NoStop}%
\bibitem [{\citenamefont {Bowen}(1950)}]{bowen_fomation_1950}%
  \BibitemOpen
  \bibfield  {author} {\bibinfo {author} {\bibfnamefont {E.~G.}\ \bibnamefont
  {Bowen}},\ }\bibfield  {title} {\bibinfo {title} {The {{Fomation}} of
  {{Rain}} by {{Coalescence}}},\ }\href {https://doi.org/10.1071/ch9500193}
  {\bibfield  {journal} {\bibinfo  {journal} {Australian Journal of Chemistry}\
  }\textbf {\bibinfo {volume} {3}},\ \bibinfo {pages} {193} (\bibinfo {year}
  {1950})}\BibitemShut {NoStop}%
\bibitem [{\citenamefont {Telford}(1955)}]{telford_new_1955}%
  \BibitemOpen
  \bibfield  {author} {\bibinfo {author} {\bibfnamefont {J.~W.}\ \bibnamefont
  {Telford}},\ }\bibfield  {title} {\bibinfo {title} {A {{New Aspect}} of
  {{Coalescence Theory}}},\ }\href
  {https://doi.org/10.1175/1520-0469(1955)012<0436:ANAOCT>2.0.CO;2} {\bibfield
  {journal} {\bibinfo  {journal} {Journal of the Atmospheric Sciences}\
  }\textbf {\bibinfo {volume} {12}},\ \bibinfo {pages} {436} (\bibinfo {year}
  {1955})}\BibitemShut {NoStop}%
\bibitem [{\citenamefont {Kostinski}\ and\ \citenamefont
  {Shaw}(2005)}]{kostinski_fluctuations_2005}%
  \BibitemOpen
  \bibfield  {author} {\bibinfo {author} {\bibfnamefont {A.~B.}\ \bibnamefont
  {Kostinski}}\ and\ \bibinfo {author} {\bibfnamefont {R.~A.}\ \bibnamefont
  {Shaw}},\ }\bibfield  {title} {\bibinfo {title} {Fluctuations and {{Luck}} in
  {{Droplet Growth}} by {{Coalescence}}},\ }\href
  {https://doi.org/10.1175/BAMS-86-2-235} {\bibfield  {journal} {\bibinfo
  {journal} {Bulletin of the American Meteorological Society}\ }\textbf
  {\bibinfo {volume} {86}},\ \bibinfo {pages} {235} (\bibinfo {year}
  {2005})}\BibitemShut {NoStop}%
\bibitem [{\citenamefont {Wilkinson}(2016)}]{wilkinson_large_2016}%
  \BibitemOpen
  \bibfield  {author} {\bibinfo {author} {\bibfnamefont {M.}~\bibnamefont
  {Wilkinson}},\ }\bibfield  {title} {\bibinfo {title} {Large {{Deviation
  Analysis}} of {{Rapid Onset}} of {{Rain Showers}}},\ }\href
  {https://doi.org/10.1103/PhysRevLett.116.018501} {\bibfield  {journal}
  {\bibinfo  {journal} {Physical Review Letters}\ }\textbf {\bibinfo {volume}
  {116}},\ \bibinfo {pages} {018501} (\bibinfo {year} {2016})}\BibitemShut
  {NoStop}%
\bibitem [{\citenamefont {Ghosh}\ \emph {et~al.}(2005)\citenamefont {Ghosh},
  \citenamefont {D{\'a}vila}, \citenamefont {Hunt}, \citenamefont {Srdic},
  \citenamefont {Fernando},\ and\ \citenamefont {Jonas}}]{ghosh_how_2005}%
  \BibitemOpen
  \bibfield  {author} {\bibinfo {author} {\bibfnamefont {S.}~\bibnamefont
  {Ghosh}}, \bibinfo {author} {\bibfnamefont {J.}~\bibnamefont {D{\'a}vila}},
  \bibinfo {author} {\bibfnamefont {J.}~\bibnamefont {Hunt}}, \bibinfo {author}
  {\bibfnamefont {A.}~\bibnamefont {Srdic}}, \bibinfo {author} {\bibfnamefont
  {H.}~\bibnamefont {Fernando}},\ and\ \bibinfo {author} {\bibfnamefont
  {P.}~\bibnamefont {Jonas}},\ }\bibfield  {title} {\bibinfo {title} {How
  turbulence enhances coalescence of settling particles with applications to
  rain in clouds},\ }\href {https://doi.org/10.1098/rspa.2005.1490} {\bibfield
  {journal} {\bibinfo  {journal} {Proceedings of the Royal Society A:
  Mathematical, Physical and Engineering Sciences}\ }\textbf {\bibinfo {volume}
  {461}},\ \bibinfo {pages} {3059} (\bibinfo {year} {2005})}\BibitemShut
  {NoStop}%
\bibitem [{\citenamefont {Wilkinson}\ \emph {et~al.}(2006)\citenamefont
  {Wilkinson}, \citenamefont {Mehlig},\ and\ \citenamefont
  {Bezuglyy}}]{wilkinson_caustic_2006}%
  \BibitemOpen
  \bibfield  {author} {\bibinfo {author} {\bibfnamefont {M.}~\bibnamefont
  {Wilkinson}}, \bibinfo {author} {\bibfnamefont {B.}~\bibnamefont {Mehlig}},\
  and\ \bibinfo {author} {\bibfnamefont {V.}~\bibnamefont {Bezuglyy}},\
  }\bibfield  {title} {\bibinfo {title} {Caustic {{Activation}} of {{Rain
  Showers}}},\ }\href {https://doi.org/10.1103/PhysRevLett.97.048501}
  {\bibfield  {journal} {\bibinfo  {journal} {Physical Review Letters}\
  }\textbf {\bibinfo {volume} {97}},\ \bibinfo {pages} {048501} (\bibinfo
  {year} {2006})}\BibitemShut {NoStop}%
\bibitem [{\citenamefont {Wang}\ and\ \citenamefont
  {Grabowski}(2009)}]{wang_role_2009}%
  \BibitemOpen
  \bibfield  {author} {\bibinfo {author} {\bibfnamefont {L.-P.}\ \bibnamefont
  {Wang}}\ and\ \bibinfo {author} {\bibfnamefont {W.~W.}\ \bibnamefont
  {Grabowski}},\ }\bibfield  {title} {\bibinfo {title} {The role of air
  turbulence in warm rain initiation},\ }\href
  {https://doi.org/10.1002/asl.210} {\bibfield  {journal} {\bibinfo  {journal}
  {Atmospheric Science Letters}\ }\textbf {\bibinfo {volume} {10}},\ \bibinfo
  {pages} {1} (\bibinfo {year} {2009})}\BibitemShut {NoStop}%
\bibitem [{\citenamefont {Devenish}\ \emph {et~al.}(2012)\citenamefont
  {Devenish}, \citenamefont {Bartello}, \citenamefont {Brenguier},
  \citenamefont {Collins}, \citenamefont {Grabowski}, \citenamefont
  {IJzermans}, \citenamefont {Malinowski}, \citenamefont {Reeks}, \citenamefont
  {Vassilicos}, \citenamefont {Wang},\ and\ \citenamefont
  {Warhaft}}]{devenish_droplet_2012}%
  \BibitemOpen
  \bibfield  {author} {\bibinfo {author} {\bibfnamefont {B.~J.}\ \bibnamefont
  {Devenish}}, \bibinfo {author} {\bibfnamefont {P.}~\bibnamefont {Bartello}},
  \bibinfo {author} {\bibfnamefont {J.-L.}\ \bibnamefont {Brenguier}}, \bibinfo
  {author} {\bibfnamefont {L.~R.}\ \bibnamefont {Collins}}, \bibinfo {author}
  {\bibfnamefont {W.~W.}\ \bibnamefont {Grabowski}}, \bibinfo {author}
  {\bibfnamefont {R.~H.~A.}\ \bibnamefont {IJzermans}}, \bibinfo {author}
  {\bibfnamefont {S.~P.}\ \bibnamefont {Malinowski}}, \bibinfo {author}
  {\bibfnamefont {M.~W.}\ \bibnamefont {Reeks}}, \bibinfo {author}
  {\bibfnamefont {J.~C.}\ \bibnamefont {Vassilicos}}, \bibinfo {author}
  {\bibfnamefont {L.-P.}\ \bibnamefont {Wang}},\ and\ \bibinfo {author}
  {\bibfnamefont {Z.}~\bibnamefont {Warhaft}},\ }\bibfield  {title} {\bibinfo
  {title} {Droplet growth in warm turbulent clouds},\ }\href
  {https://doi.org/10.1002/qj.1897} {\bibfield  {journal} {\bibinfo  {journal}
  {Quarterly Journal of the Royal Meteorological Society}\ }\textbf {\bibinfo
  {volume} {138}},\ \bibinfo {pages} {1401} (\bibinfo {year}
  {2012})}\BibitemShut {NoStop}%
\bibitem [{\citenamefont {Grabowski}\ and\ \citenamefont
  {Wang}(2013)}]{grabowski_growth_2013}%
  \BibitemOpen
  \bibfield  {author} {\bibinfo {author} {\bibfnamefont {W.~W.}\ \bibnamefont
  {Grabowski}}\ and\ \bibinfo {author} {\bibfnamefont {L.-P.}\ \bibnamefont
  {Wang}},\ }\bibfield  {title} {\bibinfo {title} {Growth of {{Cloud Droplets}}
  in a {{Turbulent Environment}}},\ }\href
  {https://doi.org/10.1146/annurev-fluid-011212-140750} {\bibfield  {journal}
  {\bibinfo  {journal} {Annual Review of Fluid Mechanics}\ }\textbf {\bibinfo
  {volume} {45}},\ \bibinfo {pages} {293} (\bibinfo {year} {2013})}\BibitemShut
  {NoStop}%
\bibitem [{\citenamefont {Pumir}\ and\ \citenamefont
  {Wilkinson}(2016)}]{pumir_collisional_2016}%
  \BibitemOpen
  \bibfield  {author} {\bibinfo {author} {\bibfnamefont {A.}~\bibnamefont
  {Pumir}}\ and\ \bibinfo {author} {\bibfnamefont {M.}~\bibnamefont
  {Wilkinson}},\ }\bibfield  {title} {\bibinfo {title} {Collisional
  {{Aggregation Due}} to {{Turbulence}}},\ }\href
  {https://doi.org/10.1146/annurev-conmatphys-031115-011538} {\bibfield
  {journal} {\bibinfo  {journal} {Annual Review of Condensed Matter Physics}\
  }\textbf {\bibinfo {volume} {7}},\ \bibinfo {pages} {141} (\bibinfo {year}
  {2016})}\BibitemShut {NoStop}%
\bibitem [{\citenamefont {Szumowski}\ \emph {et~al.}(1999)\citenamefont
  {Szumowski}, \citenamefont {Rauber},\ and\ \citenamefont
  {Ochs}}]{szumowski_microphysical_1999}%
  \BibitemOpen
  \bibfield  {author} {\bibinfo {author} {\bibfnamefont {M.~J.}\ \bibnamefont
  {Szumowski}}, \bibinfo {author} {\bibfnamefont {R.~M.}\ \bibnamefont
  {Rauber}},\ and\ \bibinfo {author} {\bibfnamefont {H.~T.}\ \bibnamefont
  {Ochs}},\ }\bibfield  {title} {\bibinfo {title} {The {{Microphysical
  Structure}} and {{Evolution}} of {{Hawaiian Rainband Clouds}}. {{Part III}}:
  {{A Test}} of the {{Ultragiant Nuclei Hypothesis}}},\ }\href
  {https://doi.org/10.1175/1520-0469(1999)056<1980:TMSAEO>2.0.CO;2} {\bibfield
  {journal} {\bibinfo  {journal} {Journal of the Atmospheric Sciences}\
  }\textbf {\bibinfo {volume} {56}},\ \bibinfo {pages} {1980} (\bibinfo {year}
  {1999})}\BibitemShut {NoStop}%
\bibitem [{\citenamefont {Mechem}\ and\ \citenamefont
  {Kogan}(2008)}]{mechem_bulk_2008}%
  \BibitemOpen
  \bibfield  {author} {\bibinfo {author} {\bibfnamefont {D.~B.}\ \bibnamefont
  {Mechem}}\ and\ \bibinfo {author} {\bibfnamefont {Y.~L.}\ \bibnamefont
  {Kogan}},\ }\bibfield  {title} {\bibinfo {title} {A {{Bulk Parameterization}}
  of {{Giant CCN}}},\ }\href {https://doi.org/10.1175/2007JAS2502.1} {\bibfield
   {journal} {\bibinfo  {journal} {Journal of the Atmospheric Sciences}\
  }\textbf {\bibinfo {volume} {65}},\ \bibinfo {pages} {2458} (\bibinfo {year}
  {2008})}\BibitemShut {NoStop}%
\bibitem [{\citenamefont {Posselt}\ and\ \citenamefont
  {Lohmann}(2008)}]{posselt_influence_2008}%
  \BibitemOpen
  \bibfield  {author} {\bibinfo {author} {\bibfnamefont {R.}~\bibnamefont
  {Posselt}}\ and\ \bibinfo {author} {\bibfnamefont {U.}~\bibnamefont
  {Lohmann}},\ }\bibfield  {title} {\bibinfo {title} {Influence of {{Giant
  CCN}} on warm rain processes in the {{ECHAM5 GCM}}},\ }\href
  {https://doi.org/10.5194/acp-8-3769-2008} {\bibfield  {journal} {\bibinfo
  {journal} {Atmospheric Chemistry and Physics}\ }\textbf {\bibinfo {volume}
  {8}},\ \bibinfo {pages} {3769} (\bibinfo {year} {2008})}\BibitemShut
  {NoStop}%
\bibitem [{\citenamefont {Barahona}\ \emph {et~al.}(2010)\citenamefont
  {Barahona}, \citenamefont {West}, \citenamefont {Stier}, \citenamefont
  {Romakkaniemi}, \citenamefont {Kokkola},\ and\ \citenamefont
  {Nenes}}]{barahona_comprehensively_2010}%
  \BibitemOpen
  \bibfield  {author} {\bibinfo {author} {\bibfnamefont {D.}~\bibnamefont
  {Barahona}}, \bibinfo {author} {\bibfnamefont {R.~E.~L.}\ \bibnamefont
  {West}}, \bibinfo {author} {\bibfnamefont {P.}~\bibnamefont {Stier}},
  \bibinfo {author} {\bibfnamefont {S.}~\bibnamefont {Romakkaniemi}}, \bibinfo
  {author} {\bibfnamefont {H.}~\bibnamefont {Kokkola}},\ and\ \bibinfo {author}
  {\bibfnamefont {A.}~\bibnamefont {Nenes}},\ }\bibfield  {title} {\bibinfo
  {title} {Comprehensively accounting for the effect of giant {{CCN}} in cloud
  activation parameterizations},\ }\href
  {https://doi.org/10.5194/acp-10-2467-2010} {\bibfield  {journal} {\bibinfo
  {journal} {Atmospheric Chemistry and Physics}\ }\textbf {\bibinfo {volume}
  {10}},\ \bibinfo {pages} {2467} (\bibinfo {year} {2010})}\BibitemShut
  {NoStop}%
\bibitem [{\citenamefont {Blanchard}(1950)}]{blanchard_behavior_1950}%
  \BibitemOpen
  \bibfield  {author} {\bibinfo {author} {\bibfnamefont {D.~C.}\ \bibnamefont
  {Blanchard}},\ }\bibfield  {title} {\bibinfo {title} {The behavior of water
  drops at terminal velocity in air},\ }\href
  {https://doi.org/10.1029/TR031i006p00836} {\bibfield  {journal} {\bibinfo
  {journal} {Transactions, American Geophysical Union}\ }\textbf {\bibinfo
  {volume} {31}},\ \bibinfo {pages} {836} (\bibinfo {year} {1950})}\BibitemShut
  {NoStop}%
\bibitem [{\citenamefont {Villermaux}(2020)}]{villermaux_fragmentation_2020}%
  \BibitemOpen
  \bibfield  {author} {\bibinfo {author} {\bibfnamefont {E.}~\bibnamefont
  {Villermaux}},\ }\bibfield  {title} {\bibinfo {title} {Fragmentation versus
  {{Cohesion}}},\ }\href {https://doi.org/10.1017/jfm.2020.366} {\bibfield
  {journal} {\bibinfo  {journal} {Journal of Fluid Mechanics}\ }\textbf
  {\bibinfo {volume} {898}},\ \bibinfo {pages} {P1} (\bibinfo {year}
  {2020})}\BibitemShut {NoStop}%
\bibitem [{\citenamefont {Zeng}(2018)}]{zeng_modeling_2018}%
  \BibitemOpen
  \bibfield  {author} {\bibinfo {author} {\bibfnamefont {X.}~\bibnamefont
  {Zeng}},\ }\bibfield  {title} {\bibinfo {title} {Modeling the {{Effect}} of
  {{Radiation}} on {{Warm Rain Initiation}}},\ }\href
  {https://doi.org/10.1029/2018JD028354} {\bibfield  {journal} {\bibinfo
  {journal} {Journal of Geophysical Research: Atmospheres}\ }\textbf {\bibinfo
  {volume} {123}},\ \bibinfo {pages} {6896} (\bibinfo {year}
  {2018})}\BibitemShut {NoStop}%
\bibitem [{\citenamefont {Beard}\ and\ \citenamefont
  {Ochs}(1993)}]{beard_warm-rain_1993}%
  \BibitemOpen
  \bibfield  {author} {\bibinfo {author} {\bibfnamefont {K.~V.}\ \bibnamefont
  {Beard}}\ and\ \bibinfo {author} {\bibfnamefont {H.~T.}\ \bibnamefont
  {Ochs}},\ }\bibfield  {title} {\bibinfo {title} {Warm-{{Rain Initiation}}:
  {{An Overview}} of {{Microphysical Mechanisms}}},\ }\href
  {https://doi.org/10.1175/1520-0450(1993)032<0608:WRIAOO>2.0.CO;2} {\bibfield
  {journal} {\bibinfo  {journal} {Journal of Applied Meteorology and
  Climatology}\ }\textbf {\bibinfo {volume} {32}},\ \bibinfo {pages} {608}
  (\bibinfo {year} {1993})}\BibitemShut {NoStop}%
\bibitem [{\citenamefont {McFarquhar}(2022)}]{mcfarquhar_rainfall_2022}%
  \BibitemOpen
  \bibfield  {author} {\bibinfo {author} {\bibfnamefont {G.~M.}\ \bibnamefont
  {McFarquhar}},\ }\bibfield  {title} {\bibinfo {title} {Rainfall
  microphysics},\ }in\ \href
  {https://doi.org/10.1016/B978-0-12-822544-8.00009-3} {\emph {\bibinfo
  {booktitle} {Rainfall: Modeling, Measurement and Applications}}},\ \bibinfo
  {editor} {edited by\ \bibinfo {editor} {\bibfnamefont {R.}~\bibnamefont
  {Morbidelli}}}\ (\bibinfo  {publisher} {Elsevier},\ \bibinfo {year} {2022})\
  pp.\ \bibinfo {pages} {1--26}\BibitemShut {NoStop}%
\bibitem [{\citenamefont {{von
  Smoluchowski}}(1916)}]{von_smoluchowski_three_1916}%
  \BibitemOpen
  \bibfield  {author} {\bibinfo {author} {\bibfnamefont {M.}~\bibnamefont {{von
  Smoluchowski}}},\ }\bibfield  {title} {\bibinfo {title} {{Three Presentations
  on Diffusion, Molecular Movement According to Brown and Coagulation of
  Colloid Particles.}},\ }\href@noop {} {\bibfield  {journal} {\bibinfo
  {journal} {Physikalische Zeitschrift}\ }\textbf {\bibinfo {volume} {17}},\
  \bibinfo {pages} {557} (\bibinfo {year} {1916})}\BibitemShut {NoStop}%
\bibitem [{\citenamefont {Gillespie}(1972)}]{gillespie_stochastic_1972}%
  \BibitemOpen
  \bibfield  {author} {\bibinfo {author} {\bibfnamefont {D.~T.}\ \bibnamefont
  {Gillespie}},\ }\bibfield  {title} {\bibinfo {title} {The {{Stochastic
  Coalescence Model}} for {{Cloud Droplet Growth}}},\ }\href
  {https://doi.org/10.1175/1520-0469(1972)029<1496:TSCMFC>2.0.CO;2} {\bibfield
  {journal} {\bibinfo  {journal} {Journal of the Atmospheric Sciences}\
  }\textbf {\bibinfo {volume} {29}},\ \bibinfo {pages} {1496} (\bibinfo {year}
  {1972})}\BibitemShut {NoStop}%
\bibitem [{\citenamefont {Gillespie}(1975)}]{gillespie_three_1975}%
  \BibitemOpen
  \bibfield  {author} {\bibinfo {author} {\bibfnamefont {D.~T.}\ \bibnamefont
  {Gillespie}},\ }\bibfield  {title} {\bibinfo {title} {Three {{Models}} for
  the {{Coalescence Growth}} of {{Cloud Drops}}},\ }\href
  {https://doi.org/10.1175/1520-0469(1975)032<0600:TMFTCG>2.0.CO;2} {\bibfield
  {journal} {\bibinfo  {journal} {Journal of the Atmospheric Sciences}\
  }\textbf {\bibinfo {volume} {32}},\ \bibinfo {pages} {600} (\bibinfo {year}
  {1975})}\BibitemShut {NoStop}%
\bibitem [{\citenamefont {Khain}\ and\ \citenamefont
  {Pinsky}(2018)}]{khain_physical_2018}%
  \BibitemOpen
  \bibfield  {author} {\bibinfo {author} {\bibfnamefont {A.~P.}\ \bibnamefont
  {Khain}}\ and\ \bibinfo {author} {\bibfnamefont {M.}~\bibnamefont {Pinsky}},\
  }\href {https://doi.org/10.1017/9781139049481} {\emph {\bibinfo {title}
  {Physical {{Processes}} in {{Clouds}} and {{Cloud Modeling}}}}},\ \bibinfo
  {edition} {1st}\ ed.\ (\bibinfo  {publisher} {Cambridge University Press},\
  \bibinfo {year} {2018})\BibitemShut {NoStop}%
\bibitem [{\citenamefont {Brenguier}\ and\ \citenamefont
  {Chaumat}(2001)}]{brenguier_droplet_2001}%
  \BibitemOpen
  \bibfield  {author} {\bibinfo {author} {\bibfnamefont {J.-L.}\ \bibnamefont
  {Brenguier}}\ and\ \bibinfo {author} {\bibfnamefont {L.}~\bibnamefont
  {Chaumat}},\ }\bibfield  {title} {\bibinfo {title} {Droplet {{Spectra
  Broadening}} in {{Cumulus Clouds}}. {{Part I}}: {{Broadening}} in {{Adiabatic
  Cores}}},\ }\href
  {https://doi.org/10.1175/1520-0469(2001)058<0628:DSBICC>2.0.CO;2} {\bibfield
  {journal} {\bibinfo  {journal} {Journal of the Atmospheric Sciences}\
  }\textbf {\bibinfo {volume} {58}},\ \bibinfo {pages} {628} (\bibinfo {year}
  {2001})}\BibitemShut {NoStop}%
\bibitem [{\citenamefont {Mazoyer}\ \emph {et~al.}(2022)\citenamefont
  {Mazoyer}, \citenamefont {Burnet},\ and\ \citenamefont
  {Denjean}}]{mazoyer_experimental_2022}%
  \BibitemOpen
  \bibfield  {author} {\bibinfo {author} {\bibfnamefont {M.}~\bibnamefont
  {Mazoyer}}, \bibinfo {author} {\bibfnamefont {F.}~\bibnamefont {Burnet}},\
  and\ \bibinfo {author} {\bibfnamefont {C.}~\bibnamefont {Denjean}},\
  }\bibfield  {title} {\bibinfo {title} {Experimental study on the evolution of
  droplet size distribution during the fog life cycle},\ }\href
  {https://doi.org/10.5194/acp-22-11305-2022} {\bibfield  {journal} {\bibinfo
  {journal} {Atmospheric Chemistry and Physics}\ }\textbf {\bibinfo {volume}
  {22}},\ \bibinfo {pages} {11305} (\bibinfo {year} {2022})}\BibitemShut
  {NoStop}%
\bibitem [{\citenamefont
  {Van~Dongen}(1987{\natexlab{a}})}]{van_dongen_solutions_1987}%
  \BibitemOpen
  \bibfield  {author} {\bibinfo {author} {\bibfnamefont {P.}~\bibnamefont
  {Van~Dongen}},\ }\bibfield  {title} {\bibinfo {title} {Solutions of
  {{Smoluchowski}}'s coagulation equation at large cluster sizes},\ }\href
  {https://doi.org/10.1016/0378-4371(87)90240-8} {\bibfield  {journal}
  {\bibinfo  {journal} {Physica A: Statistical Mechanics and its Applications}\
  }\textbf {\bibinfo {volume} {145}},\ \bibinfo {pages} {15} (\bibinfo {year}
  {1987}{\natexlab{a}})}\BibitemShut {NoStop}%
\bibitem [{\citenamefont {Aldous}(1999)}]{aldous_deterministic_1999}%
  \BibitemOpen
  \bibfield  {author} {\bibinfo {author} {\bibfnamefont {D.~J.}\ \bibnamefont
  {Aldous}},\ }\bibfield  {title} {\bibinfo {title} {Deterministic and
  stochastic models for coalescence (aggregation and coagulation): A review of
  the mean-field theory for probabilists},\ }\href@noop {} {\bibfield
  {journal} {\bibinfo  {journal} {Bernoulli}\ }\textbf {\bibinfo {volume}
  {5}},\ \bibinfo {pages} {3} (\bibinfo {year} {1999})}\BibitemShut {NoStop}%
\bibitem [{\citenamefont {Fournier}\ and\ \citenamefont {Lauren{\c
  c}ot}(2005)}]{fournier_existence_2005}%
  \BibitemOpen
  \bibfield  {author} {\bibinfo {author} {\bibfnamefont {N.}~\bibnamefont
  {Fournier}}\ and\ \bibinfo {author} {\bibfnamefont {P.}~\bibnamefont
  {Lauren{\c c}ot}},\ }\bibfield  {title} {\bibinfo {title} {Existence of
  {{Self-Similar Solutions}} to {{Smoluchowski}}'s {{Coagulation Equation}}},\
  }\href {https://doi.org/10.1007/s00220-004-1258-5} {\bibfield  {journal}
  {\bibinfo  {journal} {Communications in Mathematical Physics}\ }\textbf
  {\bibinfo {volume} {256}},\ \bibinfo {pages} {589} (\bibinfo {year}
  {2005})}\BibitemShut {NoStop}%
\bibitem [{\citenamefont {Herrmann}\ \emph {et~al.}(2016)\citenamefont
  {Herrmann}, \citenamefont {Niethammer},\ and\ \citenamefont
  {Vel{\'a}zquez}}]{herrmann_instabilities_2016}%
  \BibitemOpen
  \bibfield  {author} {\bibinfo {author} {\bibfnamefont {M.}~\bibnamefont
  {Herrmann}}, \bibinfo {author} {\bibfnamefont {B.}~\bibnamefont
  {Niethammer}},\ and\ \bibinfo {author} {\bibfnamefont {J.~J.~L.}\
  \bibnamefont {Vel{\'a}zquez}},\ }\bibfield  {title} {\bibinfo {title}
  {Instabilities and oscillations in coagulation equations with kernels of
  homogeneity one},\ }\href {https://doi.org/10.1090/qam/1454} {\bibfield
  {journal} {\bibinfo  {journal} {Quarterly of Applied Mathematics}\ }\textbf
  {\bibinfo {volume} {75}},\ \bibinfo {pages} {105} (\bibinfo {year}
  {2016})}\BibitemShut {NoStop}%
\bibitem [{\citenamefont {Langmuir}(1948)}]{langmuir_production_1948}%
  \BibitemOpen
  \bibfield  {author} {\bibinfo {author} {\bibfnamefont {I.}~\bibnamefont
  {Langmuir}},\ }\bibfield  {title} {\bibinfo {title} {The {{Production}} of
  {{Rain}} by a {{Chain Reaction}} in {{Cumulus Clouds}} at {{Temperatures
  Above Freezing}}},\ }\href
  {https://doi.org/10.1175/1520-0469(1948)005<0175:TPORBA>2.0.CO;2} {\bibfield
  {journal} {\bibinfo  {journal} {Journal of the Atmospheric Sciences}\
  }\textbf {\bibinfo {volume} {5}},\ \bibinfo {pages} {175} (\bibinfo {year}
  {1948})}\BibitemShut {NoStop}%
\bibitem [{\citenamefont {Poydenot}\ and\ \citenamefont
  {Andreotti}(2024)}]{poydenot_gap_2024}%
  \BibitemOpen
  \bibfield  {author} {\bibinfo {author} {\bibfnamefont {F.}~\bibnamefont
  {Poydenot}}\ and\ \bibinfo {author} {\bibfnamefont {B.}~\bibnamefont
  {Andreotti}},\ }\bibfield  {title} {\bibinfo {title} {Gap in drop collision
  rate between diffusive and inertial regimes explains the stability of fogs
  and non-precipitating clouds},\ }\href {https://doi.org/10.1017/jfm.2024.413}
  {\bibfield  {journal} {\bibinfo  {journal} {Journal of Fluid Mechanics}\
  }\textbf {\bibinfo {volume} {987}},\ \bibinfo {pages} {A37} (\bibinfo {year}
  {2024})}\BibitemShut {NoStop}%
\bibitem [{\citenamefont {Debry}\ and\ \citenamefont
  {Sportisse}(2007)}]{debry_solving_2007}%
  \BibitemOpen
  \bibfield  {author} {\bibinfo {author} {\bibfnamefont {E.}~\bibnamefont
  {Debry}}\ and\ \bibinfo {author} {\bibfnamefont {B.}~\bibnamefont
  {Sportisse}},\ }\bibfield  {title} {\bibinfo {title} {Solving aerosol
  coagulation with size-binning methods},\ }\href
  {https://doi.org/10.1016/j.apnum.2006.09.007} {\bibfield  {journal} {\bibinfo
   {journal} {Applied Numerical Mathematics}\ }\textbf {\bibinfo {volume}
  {57}},\ \bibinfo {pages} {1008} (\bibinfo {year} {2007})}\BibitemShut
  {NoStop}%
\bibitem [{\citenamefont {Golovin}(1963)}]{golovin_solution_1963}%
  \BibitemOpen
  \bibfield  {author} {\bibinfo {author} {\bibfnamefont {A.~M.}\ \bibnamefont
  {Golovin}},\ }\bibfield  {title} {\bibinfo {title} {The solution of the
  coagulation equation for cloud droplets in a rising air current},\
  }\href@noop {} {\bibfield  {journal} {\bibinfo  {journal} {Izv. Geophys.
  Ser}\ }\textbf {\bibinfo {volume} {5}},\ \bibinfo {pages} {82} (\bibinfo
  {year} {1963})}\BibitemShut {NoStop}%
\bibitem [{\citenamefont {Roesner}\ \emph {et~al.}(1990)\citenamefont
  {Roesner}, \citenamefont {Flossmann},\ and\ \citenamefont
  {Pruppacher}}]{roesner_effect_1990}%
  \BibitemOpen
  \bibfield  {author} {\bibinfo {author} {\bibfnamefont {S.}~\bibnamefont
  {Roesner}}, \bibinfo {author} {\bibfnamefont {A.~I.}\ \bibnamefont
  {Flossmann}},\ and\ \bibinfo {author} {\bibfnamefont {H.~R.}\ \bibnamefont
  {Pruppacher}},\ }\bibfield  {title} {\bibinfo {title} {The effect on the
  evolution of the drop spectrum in clouds of the preconditioning of air by
  successive convective elements},\ }\href
  {https://doi.org/10.1002/qj.49711649607} {\bibfield  {journal} {\bibinfo
  {journal} {Quarterly Journal of the Royal Meteorological Society}\ }\textbf
  {\bibinfo {volume} {116}},\ \bibinfo {pages} {1389} (\bibinfo {year}
  {1990})}\BibitemShut {NoStop}%
\bibitem [{\citenamefont {Hohenegger}\ and\ \citenamefont
  {Stevens}(2013)}]{hohenegger_preconditioning_2013}%
  \BibitemOpen
  \bibfield  {author} {\bibinfo {author} {\bibfnamefont {C.}~\bibnamefont
  {Hohenegger}}\ and\ \bibinfo {author} {\bibfnamefont {B.}~\bibnamefont
  {Stevens}},\ }\bibfield  {title} {\bibinfo {title} {Preconditioning {{Deep
  Convection}} with {{Cumulus Congestus}}},\ }\href
  {https://doi.org/10.1175/JAS-D-12-089.1} {\bibfield  {journal} {\bibinfo
  {journal} {Journal of the Atmospheric Sciences}\ }\textbf {\bibinfo {volume}
  {70}},\ \bibinfo {pages} {448} (\bibinfo {year} {2013})}\BibitemShut
  {NoStop}%
\bibitem [{\citenamefont {Rapp}\ \emph {et~al.}(2011)\citenamefont {Rapp},
  \citenamefont {Kummerow},\ and\ \citenamefont
  {Fowler}}]{rapp_interactions_2011}%
  \BibitemOpen
  \bibfield  {author} {\bibinfo {author} {\bibfnamefont {A.~D.}\ \bibnamefont
  {Rapp}}, \bibinfo {author} {\bibfnamefont {C.~D.}\ \bibnamefont {Kummerow}},\
  and\ \bibinfo {author} {\bibfnamefont {L.}~\bibnamefont {Fowler}},\
  }\bibfield  {title} {\bibinfo {title} {Interactions between warm rain clouds
  and atmospheric preconditioning for deep convection in the tropics},\
  }\bibfield  {journal} {\bibinfo  {journal} {Journal of Geophysical Research:
  Atmospheres}\ }\textbf {\bibinfo {volume} {116}},\ \href
  {https://doi.org/10.1029/2011JD016143} {10.1029/2011JD016143} (\bibinfo
  {year} {2011})\BibitemShut {NoStop}%
\bibitem [{\citenamefont {{Touz{\'e}-Peiffer}}\ \emph
  {et~al.}(2022)\citenamefont {{Touz{\'e}-Peiffer}}, \citenamefont {Vogel},\
  and\ \citenamefont {Rochetin}}]{touze-peiffer_cold_2022}%
  \BibitemOpen
  \bibfield  {author} {\bibinfo {author} {\bibfnamefont {L.}~\bibnamefont
  {{Touz{\'e}-Peiffer}}}, \bibinfo {author} {\bibfnamefont {R.}~\bibnamefont
  {Vogel}},\ and\ \bibinfo {author} {\bibfnamefont {N.}~\bibnamefont
  {Rochetin}},\ }\bibfield  {title} {\bibinfo {title} {Cold {{Pools Observed}}
  during {{EUREC4A}}: {{Detection}} and {{Characterization}} from {{Atmospheric
  Soundings}}},\ }\href {https://doi.org/10.1175/JAMC-D-21-0048.1} {\bibfield
  {journal} {\bibinfo  {journal} {Journal of Applied Meteorology and
  Climatology}\ }\textbf {\bibinfo {volume} {61}},\ \bibinfo {pages} {593}
  (\bibinfo {year} {2022})}\BibitemShut {NoStop}%
\bibitem [{\citenamefont {Marshall}\ \emph {et~al.}(1995)\citenamefont
  {Marshall}, \citenamefont {McCarthy},\ and\ \citenamefont
  {Rust}}]{marshall_electric_1995}%
  \BibitemOpen
  \bibfield  {author} {\bibinfo {author} {\bibfnamefont {T.~C.}\ \bibnamefont
  {Marshall}}, \bibinfo {author} {\bibfnamefont {M.~P.}\ \bibnamefont
  {McCarthy}},\ and\ \bibinfo {author} {\bibfnamefont {W.~D.}\ \bibnamefont
  {Rust}},\ }\bibfield  {title} {\bibinfo {title} {Electric field magnitudes
  and lightning initiation in thunderstorms},\ }\href
  {https://doi.org/10.1029/95JD00020} {\bibfield  {journal} {\bibinfo
  {journal} {Journal of Geophysical Research: Atmospheres}\ }\textbf {\bibinfo
  {volume} {100}},\ \bibinfo {pages} {7097} (\bibinfo {year}
  {1995})}\BibitemShut {NoStop}%
\bibitem [{\citenamefont {Harrison}\ \emph {et~al.}(2015)\citenamefont
  {Harrison}, \citenamefont {Nicoll},\ and\ \citenamefont
  {Ambaum}}]{harrison_microphysical_2015}%
  \BibitemOpen
  \bibfield  {author} {\bibinfo {author} {\bibfnamefont {R.~G.}\ \bibnamefont
  {Harrison}}, \bibinfo {author} {\bibfnamefont {K.~A.}\ \bibnamefont
  {Nicoll}},\ and\ \bibinfo {author} {\bibfnamefont {M.~H.~P.}\ \bibnamefont
  {Ambaum}},\ }\bibfield  {title} {\bibinfo {title} {On the microphysical
  effects of observed cloud edge charging},\ }\href
  {https://doi.org/10.1002/qj.2554} {\bibfield  {journal} {\bibinfo  {journal}
  {Quarterly Journal of the Royal Meteorological Society}\ }\textbf {\bibinfo
  {volume} {141}},\ \bibinfo {pages} {2690} (\bibinfo {year}
  {2015})}\BibitemShut {NoStop}%
\bibitem [{\citenamefont {Nicoll}\ and\ \citenamefont
  {Harrison}(2016)}]{nicoll_stratiform_2016}%
  \BibitemOpen
  \bibfield  {author} {\bibinfo {author} {\bibfnamefont {K.~A.}\ \bibnamefont
  {Nicoll}}\ and\ \bibinfo {author} {\bibfnamefont {R.~G.}\ \bibnamefont
  {Harrison}},\ }\bibfield  {title} {\bibinfo {title} {Stratiform cloud
  electrification: Comparison of theory with multiple in-cloud measurements},\
  }\href {https://doi.org/10.1002/qj.2858} {\bibfield  {journal} {\bibinfo
  {journal} {Quarterly Journal of the Royal Meteorological Society}\ }\textbf
  {\bibinfo {volume} {142}},\ \bibinfo {pages} {2679} (\bibinfo {year}
  {2016})}\BibitemShut {NoStop}%
\bibitem [{\citenamefont {Harrison}\ \emph {et~al.}(2017)\citenamefont
  {Harrison}, \citenamefont {Nicoll},\ and\ \citenamefont
  {Aplin}}]{harrison_evaluating_2017}%
  \BibitemOpen
  \bibfield  {author} {\bibinfo {author} {\bibfnamefont {R.~G.}\ \bibnamefont
  {Harrison}}, \bibinfo {author} {\bibfnamefont {K.~A.}\ \bibnamefont
  {Nicoll}},\ and\ \bibinfo {author} {\bibfnamefont {K.~L.}\ \bibnamefont
  {Aplin}},\ }\bibfield  {title} {\bibinfo {title} {Evaluating stratiform cloud
  base charge remotely},\ }\href {https://doi.org/10.1002/2017GL073128}
  {\bibfield  {journal} {\bibinfo  {journal} {Geophysical Research Letters}\
  }\textbf {\bibinfo {volume} {44}},\ \bibinfo {pages} {6407} (\bibinfo {year}
  {2017})}\BibitemShut {NoStop}%
\bibitem [{\citenamefont {Harrison}(2022)}]{harrison_measuring_2022}%
  \BibitemOpen
  \bibfield  {author} {\bibinfo {author} {\bibfnamefont {R.~G.}\ \bibnamefont
  {Harrison}},\ }\bibfield  {title} {\bibinfo {title} {Measuring electrical
  properties of the lower troposphere using enhanced meteorological
  radiosondes},\ }\href {https://doi.org/10.5194/gi-11-37-2022} {\bibfield
  {journal} {\bibinfo  {journal} {Geoscientific Instrumentation, Methods and
  Data Systems}\ }\textbf {\bibinfo {volume} {11}},\ \bibinfo {pages} {37}
  (\bibinfo {year} {2022})}\BibitemShut {NoStop}%
\bibitem [{\citenamefont {Kartalev}\ \emph {et~al.}(2006)\citenamefont
  {Kartalev}, \citenamefont {Rycroft}, \citenamefont {Fuellekrug},
  \citenamefont {Papitashvili},\ and\ \citenamefont
  {Keremidarska}}]{kartalev_possible_2006}%
  \BibitemOpen
  \bibfield  {author} {\bibinfo {author} {\bibfnamefont {M.~D.}\ \bibnamefont
  {Kartalev}}, \bibinfo {author} {\bibfnamefont {M.~J.}\ \bibnamefont
  {Rycroft}}, \bibinfo {author} {\bibfnamefont {M.}~\bibnamefont {Fuellekrug}},
  \bibinfo {author} {\bibfnamefont {V.~O.}\ \bibnamefont {Papitashvili}},\ and\
  \bibinfo {author} {\bibfnamefont {V.~I.}\ \bibnamefont {Keremidarska}},\
  }\bibfield  {title} {\bibinfo {title} {A possible explanation for the
  dominant effect of {{South American}} thunderstorms on the {{Carnegie}}
  curve},\ }\href {https://doi.org/10.1016/j.jastp.2005.05.012} {\bibfield
  {journal} {\bibinfo  {journal} {Journal of Atmospheric and Solar-Terrestrial
  Physics}\ }\bibinfo {series} {Vertical {{Coupling}} in the
  {{Atmosphere}}/{{Ionosphere System}}},\ \textbf {\bibinfo {volume} {68}},\
  \bibinfo {pages} {457} (\bibinfo {year} {2006})}\BibitemShut {NoStop}%
\bibitem [{\citenamefont {Williams}\ and\ \citenamefont
  {S{\'a}tori}(2004)}]{williams_lightning_2004}%
  \BibitemOpen
  \bibfield  {author} {\bibinfo {author} {\bibfnamefont {E.~R.}\ \bibnamefont
  {Williams}}\ and\ \bibinfo {author} {\bibfnamefont {G.}~\bibnamefont
  {S{\'a}tori}},\ }\bibfield  {title} {\bibinfo {title} {Lightning,
  thermodynamic and hydrological comparison of the two tropical continental
  chimneys},\ }\href {https://doi.org/10.1016/j.jastp.2004.05.015} {\bibfield
  {journal} {\bibinfo  {journal} {Journal of Atmospheric and Solar-Terrestrial
  Physics}\ }\bibinfo {series} {{{SPECIAL}} - {{Space Processes}} and
  {{Electrical Changes}} in {{Atmospheric L}} Ayers},\ \textbf {\bibinfo
  {volume} {66}},\ \bibinfo {pages} {1213} (\bibinfo {year}
  {2004})}\BibitemShut {NoStop}%
\bibitem [{\citenamefont {Williams}(2009)}]{williams_global_2009}%
  \BibitemOpen
  \bibfield  {author} {\bibinfo {author} {\bibfnamefont {E.~R.}\ \bibnamefont
  {Williams}},\ }\bibfield  {title} {\bibinfo {title} {The global electrical
  circuit: {{A}} review},\ }\href
  {https://doi.org/10.1016/j.atmosres.2008.05.018} {\bibfield  {journal}
  {\bibinfo  {journal} {Atmospheric Research}\ }\bibinfo {series} {13th
  {{International Conference}} on {{Atmospheric Electricity}}},\ \textbf
  {\bibinfo {volume} {91}},\ \bibinfo {pages} {140} (\bibinfo {year}
  {2009})}\BibitemShut {NoStop}%
\bibitem [{\citenamefont {Harrison}(2013)}]{harrison_carnegie_2013}%
  \BibitemOpen
  \bibfield  {author} {\bibinfo {author} {\bibfnamefont {R.~G.}\ \bibnamefont
  {Harrison}},\ }\bibfield  {title} {\bibinfo {title} {The {{Carnegie
  Curve}}},\ }\href {https://doi.org/10.1007/s10712-012-9210-2} {\bibfield
  {journal} {\bibinfo  {journal} {Surveys in Geophysics}\ }\textbf {\bibinfo
  {volume} {34}},\ \bibinfo {pages} {209} (\bibinfo {year} {2013})}\BibitemShut
  {NoStop}%
\bibitem [{\citenamefont {Afreen}\ \emph {et~al.}(2021)\citenamefont {Afreen},
  \citenamefont {Victor}, \citenamefont {Nazir}, \citenamefont {Siingh},
  \citenamefont {Bashir}, \citenamefont {Ahmad}, \citenamefont {Javid~Ahmad},\
  and\ \citenamefont {Singh}}]{afreen_fair-weather_2021}%
  \BibitemOpen
  \bibfield  {author} {\bibinfo {author} {\bibfnamefont {S.}~\bibnamefont
  {Afreen}}, \bibinfo {author} {\bibfnamefont {N.~J.}\ \bibnamefont {Victor}},
  \bibinfo {author} {\bibfnamefont {S.}~\bibnamefont {Nazir}}, \bibinfo
  {author} {\bibfnamefont {D.}~\bibnamefont {Siingh}}, \bibinfo {author}
  {\bibfnamefont {G.}~\bibnamefont {Bashir}}, \bibinfo {author} {\bibfnamefont
  {N.}~\bibnamefont {Ahmad}}, \bibinfo {author} {\bibfnamefont
  {S.}~\bibnamefont {Javid~Ahmad}},\ and\ \bibinfo {author} {\bibfnamefont
  {R.~P.}\ \bibnamefont {Singh}},\ }\bibfield  {title} {\bibinfo {title}
  {Fair-weather atmospheric electric field measurements at {{Gulmarg}},
  {{India}}},\ }\href {https://doi.org/10.1007/s12040-021-01745-5} {\bibfield
  {journal} {\bibinfo  {journal} {Journal of Earth System Science}\ }\textbf
  {\bibinfo {volume} {131}},\ \bibinfo {pages} {7} (\bibinfo {year}
  {2021})}\BibitemShut {NoStop}%
\bibitem [{\citenamefont {Harrison}\ \emph {et~al.}(2020)\citenamefont
  {Harrison}, \citenamefont {Nicoll}, \citenamefont {Ambaum}, \citenamefont
  {Marlton}, \citenamefont {Aplin},\ and\ \citenamefont
  {Lockwood}}]{harrison_precipitation_2020}%
  \BibitemOpen
  \bibfield  {author} {\bibinfo {author} {\bibfnamefont {R.~G.}\ \bibnamefont
  {Harrison}}, \bibinfo {author} {\bibfnamefont {K.~A.}\ \bibnamefont
  {Nicoll}}, \bibinfo {author} {\bibfnamefont {M.~H.~P.}\ \bibnamefont
  {Ambaum}}, \bibinfo {author} {\bibfnamefont {G.~J.}\ \bibnamefont {Marlton}},
  \bibinfo {author} {\bibfnamefont {K.~L.}\ \bibnamefont {Aplin}},\ and\
  \bibinfo {author} {\bibfnamefont {M.}~\bibnamefont {Lockwood}},\ }\bibfield
  {title} {\bibinfo {title} {Precipitation {{Modification}} by
  {{Ionization}}},\ }\href {https://doi.org/10.1103/PhysRevLett.124.198701}
  {\bibfield  {journal} {\bibinfo  {journal} {Physical Review Letters}\
  }\textbf {\bibinfo {volume} {124}},\ \bibinfo {pages} {198701} (\bibinfo
  {year} {2020})}\BibitemShut {NoStop}%
\bibitem [{\citenamefont {Harrison}\ \emph {et~al.}(2021)\citenamefont
  {Harrison}, \citenamefont {Nicoll}, \citenamefont {Tilley}, \citenamefont
  {Marlton}, \citenamefont {Chindea}, \citenamefont {Dingley}, \citenamefont
  {Iravani}, \citenamefont {Cleaver}, \citenamefont {du~Bois},\ and\
  \citenamefont {Brus}}]{harrison_demonstration_2021}%
  \BibitemOpen
  \bibfield  {author} {\bibinfo {author} {\bibfnamefont {R.~G.}\ \bibnamefont
  {Harrison}}, \bibinfo {author} {\bibfnamefont {K.~A.}\ \bibnamefont
  {Nicoll}}, \bibinfo {author} {\bibfnamefont {D.~J.}\ \bibnamefont {Tilley}},
  \bibinfo {author} {\bibfnamefont {G.~J.}\ \bibnamefont {Marlton}}, \bibinfo
  {author} {\bibfnamefont {S.}~\bibnamefont {Chindea}}, \bibinfo {author}
  {\bibfnamefont {G.~P.}\ \bibnamefont {Dingley}}, \bibinfo {author}
  {\bibfnamefont {P.}~\bibnamefont {Iravani}}, \bibinfo {author} {\bibfnamefont
  {D.~J.}\ \bibnamefont {Cleaver}}, \bibinfo {author} {\bibfnamefont {J.~L.}\
  \bibnamefont {du~Bois}},\ and\ \bibinfo {author} {\bibfnamefont
  {D.}~\bibnamefont {Brus}},\ }\bibfield  {title} {\bibinfo {title}
  {Demonstration of a {{Remotely Piloted Atmospheric Measurement}} and {{Charge
  Release Platform}} for {{Geoengineering}}},\ }\href
  {https://doi.org/10.1175/JTECH-D-20-0092.1} {\bibfield  {journal} {\bibinfo
  {journal} {Journal of Atmospheric and Oceanic Technology}\ }\textbf {\bibinfo
  {volume} {38}},\ \bibinfo {pages} {63} (\bibinfo {year} {2021})}\BibitemShut
  {NoStop}%
\bibitem [{\citenamefont {Shaw}(2003)}]{shaw_particle-turbulence_2003}%
  \BibitemOpen
  \bibfield  {author} {\bibinfo {author} {\bibfnamefont {R.~A.}\ \bibnamefont
  {Shaw}},\ }\bibfield  {title} {\bibinfo {title} {Particle-turbulence
  interactions in atmospheric clouds},\ }\href@noop {} {\bibfield  {journal}
  {\bibinfo  {journal} {Annual Review of Fluid Mechanics}\ }\textbf {\bibinfo
  {volume} {35}},\ \bibinfo {pages} {183} (\bibinfo {year} {2003})}\BibitemShut
  {NoStop}%
\bibitem [{\citenamefont {Mellado}(2017)}]{mellado_cloud-top_2017}%
  \BibitemOpen
  \bibfield  {author} {\bibinfo {author} {\bibfnamefont {J.~P.}\ \bibnamefont
  {Mellado}},\ }\bibfield  {title} {\bibinfo {title} {Cloud-{{Top Entrainment}}
  in {{Stratocumulus Clouds}}},\ }\href
  {https://doi.org/10.1146/annurev-fluid-010816-060231} {\bibfield  {journal}
  {\bibinfo  {journal} {Annual Review of Fluid Mechanics}\ }\textbf {\bibinfo
  {volume} {49}},\ \bibinfo {pages} {145} (\bibinfo {year} {2017})}\BibitemShut
  {NoStop}%
\bibitem [{\citenamefont {Pinsky}\ and\ \citenamefont
  {Khain}(1997{\natexlab{a}})}]{pinsky_turbulence_1997}%
  \BibitemOpen
  \bibfield  {author} {\bibinfo {author} {\bibfnamefont {M.~B.}\ \bibnamefont
  {Pinsky}}\ and\ \bibinfo {author} {\bibfnamefont {A.~P.}\ \bibnamefont
  {Khain}},\ }\bibfield  {title} {\bibinfo {title} {Turbulence effects on the
  collision kernel. {{I}}: {{Formation}} of velocity deviations of drops
  falling within a turbulent three-dimensional flow},\ }\href
  {https://doi.org/10.1002/qj.49712354204} {\bibfield  {journal} {\bibinfo
  {journal} {Quarterly Journal of the Royal Meteorological Society}\ }\textbf
  {\bibinfo {volume} {123}},\ \bibinfo {pages} {1517} (\bibinfo {year}
  {1997}{\natexlab{a}})}\BibitemShut {NoStop}%
\bibitem [{\citenamefont {Pinsky}\ \emph {et~al.}(1999)\citenamefont {Pinsky},
  \citenamefont {Khain},\ and\ \citenamefont
  {Shapiro}}]{pinsky_collisions_1999}%
  \BibitemOpen
  \bibfield  {author} {\bibinfo {author} {\bibfnamefont {M.}~\bibnamefont
  {Pinsky}}, \bibinfo {author} {\bibfnamefont {A.}~\bibnamefont {Khain}},\ and\
  \bibinfo {author} {\bibfnamefont {M.}~\bibnamefont {Shapiro}},\ }\bibfield
  {title} {\bibinfo {title} {Collisions of {{Small Drops}} in a {{Turbulent
  Flow}}. {{Part I}}: {{Collision Efficiency}}. {{Problem Formulation}} and
  {{Preliminary Results}}},\ }\href
  {https://doi.org/10.1175/1520-0469(1999)056<2585:COSDIA>2.0.CO;2} {\bibfield
  {journal} {\bibinfo  {journal} {Journal of the Atmospheric Sciences}\
  }\textbf {\bibinfo {volume} {56}},\ \bibinfo {pages} {2585} (\bibinfo {year}
  {1999})}\BibitemShut {NoStop}%
\bibitem [{\citenamefont {Pinsky}\ and\ \citenamefont
  {Khain}(1997{\natexlab{b}})}]{pinsky_turbulence_1997-1}%
  \BibitemOpen
  \bibfield  {author} {\bibinfo {author} {\bibfnamefont {M.~B.}\ \bibnamefont
  {Pinsky}}\ and\ \bibinfo {author} {\bibfnamefont {A.~P.}\ \bibnamefont
  {Khain}},\ }\bibfield  {title} {\bibinfo {title} {Turbulence effects on
  droplet growth and size distribution in clouds---{{A}} review},\ }\href
  {https://doi.org/10.1016/S0021-8502(97)00005-0} {\bibfield  {journal}
  {\bibinfo  {journal} {Journal of Aerosol Science}\ }\textbf {\bibinfo
  {volume} {28}},\ \bibinfo {pages} {1177} (\bibinfo {year}
  {1997}{\natexlab{b}})}\BibitemShut {NoStop}%
\bibitem [{\citenamefont {Ayala}\ \emph
  {et~al.}(2008{\natexlab{a}})\citenamefont {Ayala}, \citenamefont {Rosa},\
  and\ \citenamefont {Wang}}]{ayala_effects_2008-1}%
  \BibitemOpen
  \bibfield  {author} {\bibinfo {author} {\bibfnamefont {O.}~\bibnamefont
  {Ayala}}, \bibinfo {author} {\bibfnamefont {B.}~\bibnamefont {Rosa}},\ and\
  \bibinfo {author} {\bibfnamefont {L.-P.}\ \bibnamefont {Wang}},\ }\bibfield
  {title} {\bibinfo {title} {Effects of turbulence on the geometric collision
  rate of sedimenting droplets. {{Part}} 2. {{Theory}} and parameterization},\
  }\href {https://doi.org/10.1088/1367-2630/10/7/075016} {\bibfield  {journal}
  {\bibinfo  {journal} {New Journal of Physics}\ }\textbf {\bibinfo {volume}
  {10}},\ \bibinfo {pages} {075016} (\bibinfo {year}
  {2008}{\natexlab{a}})}\BibitemShut {NoStop}%
\bibitem [{\citenamefont {Ayala}\ \emph
  {et~al.}(2008{\natexlab{b}})\citenamefont {Ayala}, \citenamefont {Rosa},
  \citenamefont {Wang},\ and\ \citenamefont {Grabowski}}]{ayala_effects_2008}%
  \BibitemOpen
  \bibfield  {author} {\bibinfo {author} {\bibfnamefont {O.}~\bibnamefont
  {Ayala}}, \bibinfo {author} {\bibfnamefont {B.}~\bibnamefont {Rosa}},
  \bibinfo {author} {\bibfnamefont {L.-P.}\ \bibnamefont {Wang}},\ and\
  \bibinfo {author} {\bibfnamefont {W.~W.}\ \bibnamefont {Grabowski}},\
  }\bibfield  {title} {\bibinfo {title} {Effects of turbulence on the geometric
  collision rate of sedimenting droplets. {{Part}} 1. {{Results}} from direct
  numerical simulation},\ }\href
  {https://doi.org/10.1088/1367-2630/10/7/075015} {\bibfield  {journal}
  {\bibinfo  {journal} {New Journal of Physics}\ }\textbf {\bibinfo {volume}
  {10}},\ \bibinfo {pages} {075015} (\bibinfo {year}
  {2008}{\natexlab{b}})}\BibitemShut {NoStop}%
\bibitem [{\citenamefont {Baker}\ and\ \citenamefont
  {Latham}(1979)}]{baker_evolution_1979}%
  \BibitemOpen
  \bibfield  {author} {\bibinfo {author} {\bibfnamefont {M.~B.}\ \bibnamefont
  {Baker}}\ and\ \bibinfo {author} {\bibfnamefont {J.}~\bibnamefont {Latham}},\
  }\bibfield  {title} {\bibinfo {title} {The {{Evolution}} of {{Droplet
  Spectra}} and the {{Rate}} of {{Production}} of {{Embryonic Raindrops}} in
  {{Small Cumulus Clouds}}},\ }\href
  {https://doi.org/10.1175/1520-0469(1979)036<1612:TEODSA>2.0.CO;2} {\bibfield
  {journal} {\bibinfo  {journal} {Journal of the Atmospheric Sciences}\
  }\textbf {\bibinfo {volume} {36}},\ \bibinfo {pages} {1612} (\bibinfo {year}
  {1979})}\BibitemShut {NoStop}%
\bibitem [{\citenamefont {Baker}\ \emph {et~al.}(1980)\citenamefont {Baker},
  \citenamefont {Corbin},\ and\ \citenamefont {Latham}}]{baker_influence_1980}%
  \BibitemOpen
  \bibfield  {author} {\bibinfo {author} {\bibfnamefont {M.~B.}\ \bibnamefont
  {Baker}}, \bibinfo {author} {\bibfnamefont {R.~G.}\ \bibnamefont {Corbin}},\
  and\ \bibinfo {author} {\bibfnamefont {J.}~\bibnamefont {Latham}},\
  }\bibfield  {title} {\bibinfo {title} {The influence of entrainment on the
  evolution of cloud droplet spectra: {{I}}. {{A}} model of inhomogeneous
  mixing},\ }\href {https://doi.org/10.1002/qj.49710644914} {\bibfield
  {journal} {\bibinfo  {journal} {Quarterly Journal of the Royal Meteorological
  Society}\ }\textbf {\bibinfo {volume} {106}},\ \bibinfo {pages} {581}
  (\bibinfo {year} {1980})}\BibitemShut {NoStop}%
\bibitem [{\citenamefont {Baker}\ \emph {et~al.}(1984)\citenamefont {Baker},
  \citenamefont {Breidenthal}, \citenamefont {Choularton},\ and\ \citenamefont
  {Latham}}]{baker_effects_1984}%
  \BibitemOpen
  \bibfield  {author} {\bibinfo {author} {\bibfnamefont {M.~B.}\ \bibnamefont
  {Baker}}, \bibinfo {author} {\bibfnamefont {R.~E.}\ \bibnamefont
  {Breidenthal}}, \bibinfo {author} {\bibfnamefont {T.~W.}\ \bibnamefont
  {Choularton}},\ and\ \bibinfo {author} {\bibfnamefont {J.}~\bibnamefont
  {Latham}},\ }\bibfield  {title} {\bibinfo {title} {The {{Effects}} of
  {{Turbulent Mixing}} in {{Clouds}}},\ }\href
  {https://doi.org/10.1175/1520-0469(1984)041<0299:TEOTMI>2.0.CO;2} {\bibfield
  {journal} {\bibinfo  {journal} {Journal of the Atmospheric Sciences}\
  }\textbf {\bibinfo {volume} {41}},\ \bibinfo {pages} {299} (\bibinfo {year}
  {1984})}\BibitemShut {NoStop}%
\bibitem [{\citenamefont {Villermaux}\ \emph {et~al.}(2017)\citenamefont
  {Villermaux}, \citenamefont {Moutte}, \citenamefont {Amielh},\ and\
  \citenamefont {Meunier}}]{villermaux_fine_2017}%
  \BibitemOpen
  \bibfield  {author} {\bibinfo {author} {\bibfnamefont {E.}~\bibnamefont
  {Villermaux}}, \bibinfo {author} {\bibfnamefont {A.}~\bibnamefont {Moutte}},
  \bibinfo {author} {\bibfnamefont {M.}~\bibnamefont {Amielh}},\ and\ \bibinfo
  {author} {\bibfnamefont {P.}~\bibnamefont {Meunier}},\ }\bibfield  {title}
  {\bibinfo {title} {Fine structure of the vapor field in evaporating dense
  sprays},\ }\href {https://doi.org/10.1103/PhysRevFluids.2.074501} {\bibfield
  {journal} {\bibinfo  {journal} {Physical Review Fluids}\ }\textbf {\bibinfo
  {volume} {2}},\ \bibinfo {pages} {074501} (\bibinfo {year}
  {2017})}\BibitemShut {NoStop}%
\bibitem [{\citenamefont {Allwayin}\ \emph {et~al.}(2024)\citenamefont
  {Allwayin}, \citenamefont {Larsen}, \citenamefont {Glienke},\ and\
  \citenamefont {Shaw}}]{allwayin_locally_2024}%
  \BibitemOpen
  \bibfield  {author} {\bibinfo {author} {\bibfnamefont {N.}~\bibnamefont
  {Allwayin}}, \bibinfo {author} {\bibfnamefont {M.~L.}\ \bibnamefont
  {Larsen}}, \bibinfo {author} {\bibfnamefont {S.}~\bibnamefont {Glienke}},\
  and\ \bibinfo {author} {\bibfnamefont {R.~A.}\ \bibnamefont {Shaw}},\
  }\bibfield  {title} {\bibinfo {title} {Locally narrow droplet size
  distributions are ubiquitous in stratocumulus clouds},\ }\href
  {https://doi.org/10.1126/science.adi5550} {\bibfield  {journal} {\bibinfo
  {journal} {Science}\ }\textbf {\bibinfo {volume} {384}},\ \bibinfo {pages}
  {528} (\bibinfo {year} {2024})}\BibitemShut {NoStop}%
\bibitem [{\citenamefont {Latham}(1981)}]{latham_electrification_1981}%
  \BibitemOpen
  \bibfield  {author} {\bibinfo {author} {\bibfnamefont {J.}~\bibnamefont
  {Latham}},\ }\bibfield  {title} {\bibinfo {title} {The electrification of
  thunderstorms},\ }\href {https://doi.org/10.1002/qj.49710745202} {\bibfield
  {journal} {\bibinfo  {journal} {Quarterly Journal of the Royal Meteorological
  Society}\ }\textbf {\bibinfo {volume} {107}},\ \bibinfo {pages} {277}
  (\bibinfo {year} {1981})}\BibitemShut {NoStop}%
\bibitem [{\citenamefont {Williams}\ \emph {et~al.}(1991)\citenamefont
  {Williams}, \citenamefont {Zhang},\ and\ \citenamefont
  {Rydock}}]{williams_mixed-phase_1991}%
  \BibitemOpen
  \bibfield  {author} {\bibinfo {author} {\bibfnamefont {E.~R.}\ \bibnamefont
  {Williams}}, \bibinfo {author} {\bibfnamefont {R.}~\bibnamefont {Zhang}},\
  and\ \bibinfo {author} {\bibfnamefont {J.}~\bibnamefont {Rydock}},\
  }\bibfield  {title} {\bibinfo {title} {Mixed-{{Phase Microphysics}} and
  {{Cloud Electrification}}},\ }\href
  {https://doi.org/10.1175/1520-0469(1991)048<2195:MPMACE>2.0.CO;2} {\bibfield
  {journal} {\bibinfo  {journal} {Journal of the Atmospheric Sciences}\
  }\textbf {\bibinfo {volume} {48}},\ \bibinfo {pages} {2195} (\bibinfo {year}
  {1991})}\BibitemShut {NoStop}%
\bibitem [{\citenamefont {Saunders}(1993)}]{saunders_review_1993}%
  \BibitemOpen
  \bibfield  {author} {\bibinfo {author} {\bibfnamefont {C.~P.~R.}\
  \bibnamefont {Saunders}},\ }\bibfield  {title} {\bibinfo {title} {A
  {{Review}} of {{Thunderstorm Electrification Processes}}},\ }\href
  {https://doi.org/10.1175/1520-0450(1993)032<0642:AROTEP>2.0.CO;2} {\bibfield
  {journal} {\bibinfo  {journal} {Journal of Applied Meteorology and
  Climatology}\ }\textbf {\bibinfo {volume} {32}},\ \bibinfo {pages} {642}
  (\bibinfo {year} {1993})}\BibitemShut {NoStop}%
\bibitem [{\citenamefont {Mason}\ and\ \citenamefont
  {Dash}(2000)}]{mason_charge_2000}%
  \BibitemOpen
  \bibfield  {author} {\bibinfo {author} {\bibfnamefont {B.~L.}\ \bibnamefont
  {Mason}}\ and\ \bibinfo {author} {\bibfnamefont {J.~G.}\ \bibnamefont
  {Dash}},\ }\bibfield  {title} {\bibinfo {title} {Charge and mass transfer in
  ice-ice collisions: {{Experimental}} observations of a mechanism in
  thunderstorm electrification},\ }\href {https://doi.org/10.1029/2000JD900104}
  {\bibfield  {journal} {\bibinfo  {journal} {Journal of Geophysical Research:
  Atmospheres}\ }\textbf {\bibinfo {volume} {105}},\ \bibinfo {pages} {10185}
  (\bibinfo {year} {2000})}\BibitemShut {NoStop}%
\bibitem [{\citenamefont {Dash}\ \emph {et~al.}(2001)\citenamefont {Dash},
  \citenamefont {Mason},\ and\ \citenamefont {Wettlaufer}}]{dash_theory_2001}%
  \BibitemOpen
  \bibfield  {author} {\bibinfo {author} {\bibfnamefont {J.~G.}\ \bibnamefont
  {Dash}}, \bibinfo {author} {\bibfnamefont {B.~L.}\ \bibnamefont {Mason}},\
  and\ \bibinfo {author} {\bibfnamefont {J.~S.}\ \bibnamefont {Wettlaufer}},\
  }\bibfield  {title} {\bibinfo {title} {Theory of charge and mass transfer in
  ice-ice collisions},\ }\href {https://doi.org/10.1029/2001JD900109}
  {\bibfield  {journal} {\bibinfo  {journal} {Journal of Geophysical Research:
  Atmospheres}\ }\textbf {\bibinfo {volume} {106}},\ \bibinfo {pages} {20395}
  (\bibinfo {year} {2001})}\BibitemShut {NoStop}%
\bibitem [{\citenamefont {Jungwirth}\ \emph {et~al.}(2005)\citenamefont
  {Jungwirth}, \citenamefont {Rosenfeld},\ and\ \citenamefont
  {Buch}}]{jungwirth_possible_2005}%
  \BibitemOpen
  \bibfield  {author} {\bibinfo {author} {\bibfnamefont {P.}~\bibnamefont
  {Jungwirth}}, \bibinfo {author} {\bibfnamefont {D.}~\bibnamefont
  {Rosenfeld}},\ and\ \bibinfo {author} {\bibfnamefont {V.}~\bibnamefont
  {Buch}},\ }\bibfield  {title} {\bibinfo {title} {A possible new molecular
  mechanism of thundercloud electrification},\ }\href
  {https://doi.org/10.1016/j.atmosres.2004.11.016} {\bibfield  {journal}
  {\bibinfo  {journal} {Atmospheric Research}\ }\textbf {\bibinfo {volume}
  {76}},\ \bibinfo {pages} {190} (\bibinfo {year} {2005})}\BibitemShut
  {NoStop}%
\bibitem [{\citenamefont {Guichard}\ and\ \citenamefont
  {Couvreux}(2017)}]{guichard_short_2017}%
  \BibitemOpen
  \bibfield  {author} {\bibinfo {author} {\bibfnamefont {F.}~\bibnamefont
  {Guichard}}\ and\ \bibinfo {author} {\bibfnamefont {F.}~\bibnamefont
  {Couvreux}},\ }\bibfield  {title} {\bibinfo {title} {A short review of
  numerical cloud-resolving models},\ }\href
  {https://doi.org/10.1080/16000870.2017.1373578} {\bibfield  {journal}
  {\bibinfo  {journal} {Tellus A: Dynamic Meteorology and Oceanography}\
  }\textbf {\bibinfo {volume} {69}},\ \bibinfo {pages} {1373578} (\bibinfo
  {year} {2017})}\BibitemShut {NoStop}%
\bibitem [{\citenamefont {v~Smoluchowski}(1918)}]{smoluchowski_versuch_1918}%
  \BibitemOpen
  \bibfield  {author} {\bibinfo {author} {\bibfnamefont {M.}~\bibnamefont
  {v~Smoluchowski}},\ }\bibfield  {title} {\bibinfo {title} {Versuch einer
  mathematischen {{Theorie}} der {{Koagulationskinetik}} kolloider
  {{L{\"o}sungen}}},\ }\href@noop {} {\bibfield  {journal} {\bibinfo  {journal}
  {Zeitschrift f{\"u}r physikalische Chemie}\ }\textbf {\bibinfo {volume}
  {92}},\ \bibinfo {pages} {129} (\bibinfo {year} {1918})}\BibitemShut
  {NoStop}%
\bibitem [{\citenamefont {Saffman}\ and\ \citenamefont
  {Turner}(1956)}]{saffman_collision_1956}%
  \BibitemOpen
  \bibfield  {author} {\bibinfo {author} {\bibfnamefont {P.~G.~F.}\
  \bibnamefont {Saffman}}\ and\ \bibinfo {author} {\bibfnamefont {J.~S.}\
  \bibnamefont {Turner}},\ }\bibfield  {title} {\bibinfo {title} {On the
  collision of drops in turbulent clouds},\ }\href@noop {} {\bibfield
  {journal} {\bibinfo  {journal} {Journal of Fluid Mechanics}\ }\textbf
  {\bibinfo {volume} {1}},\ \bibinfo {pages} {16} (\bibinfo {year}
  {1956})}\BibitemShut {NoStop}%
\bibitem [{\citenamefont {Davies}\ \emph {et~al.}(1999)\citenamefont {Davies},
  \citenamefont {King},\ and\ \citenamefont
  {Wattis}}]{davies_self-similar_1999}%
  \BibitemOpen
  \bibfield  {author} {\bibinfo {author} {\bibfnamefont {S.~C.}\ \bibnamefont
  {Davies}}, \bibinfo {author} {\bibfnamefont {J.~R.}\ \bibnamefont {King}},\
  and\ \bibinfo {author} {\bibfnamefont {J.~A.~D.}\ \bibnamefont {Wattis}},\
  }\bibfield  {title} {\bibinfo {title} {Self-similar behaviour in the
  coagulation equations},\ }\href {https://doi.org/10.1023/A:1004589822425}
  {\bibfield  {journal} {\bibinfo  {journal} {Journal of Engineering
  Mathematics}\ }\textbf {\bibinfo {volume} {36}},\ \bibinfo {pages} {57}
  (\bibinfo {year} {1999})}\BibitemShut {NoStop}%
\bibitem [{\citenamefont {{Fern{\'a}ndez-D{\'i}az}}\ and\ \citenamefont
  {{G{\'o}mez-Garc{\'i}a}}(2007)}]{fernandez-diaz_exact_2007}%
  \BibitemOpen
  \bibfield  {author} {\bibinfo {author} {\bibfnamefont {J.~M.}\ \bibnamefont
  {{Fern{\'a}ndez-D{\'i}az}}}\ and\ \bibinfo {author} {\bibfnamefont {G.~J.}\
  \bibnamefont {{G{\'o}mez-Garc{\'i}a}}},\ }\bibfield  {title} {\bibinfo
  {title} {Exact solution of {{Smoluchowski}}'s continuous multi-component
  equation with an additive kernel},\ }\href
  {https://doi.org/10.1209/0295-5075/78/56002} {\bibfield  {journal} {\bibinfo
  {journal} {Europhysics Letters}\ }\textbf {\bibinfo {volume} {78}},\ \bibinfo
  {pages} {56002} (\bibinfo {year} {2007})}\BibitemShut {NoStop}%
\bibitem [{\citenamefont {Leyvraz}(2003)}]{leyvraz_scaling_2003}%
  \BibitemOpen
  \bibfield  {author} {\bibinfo {author} {\bibfnamefont {F.}~\bibnamefont
  {Leyvraz}},\ }\bibfield  {title} {\bibinfo {title} {Scaling {{Theory}} and
  {{Exactly Solved Models In}} the {{Kinetics}} of {{Irreversible
  Aggregation}}},\ }\href {https://doi.org/10.1016/S0370-1573(03)00241-2}
  {\bibfield  {journal} {\bibinfo  {journal} {Physics Reports}\ }\textbf
  {\bibinfo {volume} {383}},\ \bibinfo {pages} {95} (\bibinfo {year}
  {2003})}\BibitemShut {NoStop}%
\bibitem [{\citenamefont {Krapivsky}\ \emph {et~al.}(2010)\citenamefont
  {Krapivsky}, \citenamefont {Redner},\ and\ \citenamefont
  {{Ben-Naim}}}]{krapivsky_kinetic_2010}%
  \BibitemOpen
  \bibfield  {author} {\bibinfo {author} {\bibfnamefont {P.~L.}\ \bibnamefont
  {Krapivsky}}, \bibinfo {author} {\bibfnamefont {S.}~\bibnamefont {Redner}},\
  and\ \bibinfo {author} {\bibfnamefont {E.}~\bibnamefont {{Ben-Naim}}},\
  }\href {https://doi.org/10.1017/CBO9780511780516} {\emph {\bibinfo {title} {A
  {{Kinetic View}} of {{Statistical Physics}}}}}\ (\bibinfo  {publisher}
  {Cambridge University Press},\ \bibinfo {year} {2010})\BibitemShut {NoStop}%
\bibitem [{\citenamefont {Leyvraz}\ and\ \citenamefont
  {Tschudi}(1982)}]{leyvraz_critical_1982}%
  \BibitemOpen
  \bibfield  {author} {\bibinfo {author} {\bibfnamefont {F.}~\bibnamefont
  {Leyvraz}}\ and\ \bibinfo {author} {\bibfnamefont {H.~R.}\ \bibnamefont
  {Tschudi}},\ }\bibfield  {title} {\bibinfo {title} {Critical kinetics near
  gelation},\ }\href {https://doi.org/10.1088/0305-4470/15/6/033} {\bibfield
  {journal} {\bibinfo  {journal} {Journal of Physics A: Mathematical and
  General}\ }\textbf {\bibinfo {volume} {15}},\ \bibinfo {pages} {1951}
  (\bibinfo {year} {1982})}\BibitemShut {NoStop}%
\bibitem [{\citenamefont {Ziff}\ \emph {et~al.}(1983)\citenamefont {Ziff},
  \citenamefont {Ernst},\ and\ \citenamefont {Hendriks}}]{ziff_kinetics_1983}%
  \BibitemOpen
  \bibfield  {author} {\bibinfo {author} {\bibfnamefont {R.~M.}\ \bibnamefont
  {Ziff}}, \bibinfo {author} {\bibfnamefont {M.~H.}\ \bibnamefont {Ernst}},\
  and\ \bibinfo {author} {\bibfnamefont {E.~M.}\ \bibnamefont {Hendriks}},\
  }\bibfield  {title} {\bibinfo {title} {Kinetics of gelation and
  universality},\ }\href {https://doi.org/10.1088/0305-4470/16/10/026}
  {\bibfield  {journal} {\bibinfo  {journal} {Journal of Physics A:
  Mathematical and General}\ }\textbf {\bibinfo {volume} {16}},\ \bibinfo
  {pages} {2293} (\bibinfo {year} {1983})}\BibitemShut {NoStop}%
\bibitem [{\citenamefont {Ernst}\ \emph {et~al.}(1984)\citenamefont {Ernst},
  \citenamefont {Ziff},\ and\ \citenamefont
  {Hendriks}}]{ernst_coagulation_1984-1}%
  \BibitemOpen
  \bibfield  {author} {\bibinfo {author} {\bibfnamefont {M.}~\bibnamefont
  {Ernst}}, \bibinfo {author} {\bibfnamefont {R.}~\bibnamefont {Ziff}},\ and\
  \bibinfo {author} {\bibfnamefont {E.}~\bibnamefont {Hendriks}},\ }\bibfield
  {title} {\bibinfo {title} {Coagulation processes with a phase transition},\
  }\href {https://doi.org/10.1016/0021-9797(84)90292-3} {\bibfield  {journal}
  {\bibinfo  {journal} {Journal of Colloid and Interface Science}\ }\textbf
  {\bibinfo {volume} {97}},\ \bibinfo {pages} {266} (\bibinfo {year}
  {1984})}\BibitemShut {NoStop}%
\bibitem [{\citenamefont {White}(1980)}]{white_global_1980}%
  \BibitemOpen
  \bibfield  {author} {\bibinfo {author} {\bibfnamefont {W.~H.}\ \bibnamefont
  {White}},\ }\bibfield  {title} {\bibinfo {title} {A {{Global Existence
  Theorem}} for {{Smoluchowski}}'s {{Coagulation Equations}}},\ }\href@noop {}
  {\bibfield  {journal} {\bibinfo  {journal} {Proceedings of the American
  Mathematical Society}\ }\textbf {\bibinfo {volume} {80}},\ \bibinfo {pages}
  {273} (\bibinfo {year} {1980})}\BibitemShut {NoStop}%
\bibitem [{\citenamefont {White}(1982)}]{white_form_1982}%
  \BibitemOpen
  \bibfield  {author} {\bibinfo {author} {\bibfnamefont {W.~H.}\ \bibnamefont
  {White}},\ }\bibfield  {title} {\bibinfo {title} {On the form of steady-state
  solutions to the coagulation equations},\ }\href
  {https://doi.org/10.1016/0021-9797(82)90382-4} {\bibfield  {journal}
  {\bibinfo  {journal} {Journal of Colloid and Interface Science}\ }\textbf
  {\bibinfo {volume} {87}},\ \bibinfo {pages} {204} (\bibinfo {year}
  {1982})}\BibitemShut {NoStop}%
\bibitem [{\citenamefont {Lushnikov}(2004)}]{lushnikov_sol_2004}%
  \BibitemOpen
  \bibfield  {author} {\bibinfo {author} {\bibfnamefont {{\relax
  AA}.}~\bibnamefont {Lushnikov}},\ }\bibfield  {title} {\bibinfo {title} {From
  sol to gel exactly},\ }\href@noop {} {\bibfield  {journal} {\bibinfo
  {journal} {Physical Review Letters}\ }\textbf {\bibinfo {volume} {93}},\
  \bibinfo {pages} {198302} (\bibinfo {year} {2004})}\BibitemShut {NoStop}%
\bibitem [{\citenamefont
  {Van~Dongen}(1987{\natexlab{b}})}]{van_dongen_possible_1987}%
  \BibitemOpen
  \bibfield  {author} {\bibinfo {author} {\bibfnamefont {P.~G.~J.}\
  \bibnamefont {Van~Dongen}},\ }\bibfield  {title} {\bibinfo {title} {On the
  possible occurrence of instantaneous gelation in {{Smoluchowski}}'s
  coagulation equation},\ }\href {https://doi.org/10.1088/0305-4470/20/7/033}
  {\bibfield  {journal} {\bibinfo  {journal} {Journal of Physics A:
  Mathematical and General}\ }\textbf {\bibinfo {volume} {20}},\ \bibinfo
  {pages} {1889} (\bibinfo {year} {1987}{\natexlab{b}})}\BibitemShut {NoStop}%
\bibitem [{\citenamefont {Lushnikov}(1973)}]{lushnikov_evolution_1973}%
  \BibitemOpen
  \bibfield  {author} {\bibinfo {author} {\bibfnamefont {{\relax
  AA}.}~\bibnamefont {Lushnikov}},\ }\bibfield  {title} {\bibinfo {title}
  {Evolution of coagulating systems},\ }\href
  {https://doi.org/10.1016/0021-9797(73)90171-9} {\bibfield  {journal}
  {\bibinfo  {journal} {Journal of Colloid and Interface Science}\ }\textbf
  {\bibinfo {volume} {45}},\ \bibinfo {pages} {549} (\bibinfo {year}
  {1973})}\BibitemShut {NoStop}%
\bibitem [{\citenamefont {{van Dongen}}\ and\ \citenamefont
  {Ernst}(1985)}]{van_dongen_dynamic_1985}%
  \BibitemOpen
  \bibfield  {author} {\bibinfo {author} {\bibfnamefont {P.~G.~J.}\
  \bibnamefont {{van Dongen}}}\ and\ \bibinfo {author} {\bibfnamefont {M.~H.}\
  \bibnamefont {Ernst}},\ }\bibfield  {title} {\bibinfo {title} {Dynamic
  {{Scaling}} in the {{Kinetics}} of {{Clustering}}},\ }\href
  {https://doi.org/10.1103/PhysRevLett.54.1396} {\bibfield  {journal} {\bibinfo
   {journal} {Physical Review Letters}\ }\textbf {\bibinfo {volume} {54}},\
  \bibinfo {pages} {1396} (\bibinfo {year} {1985})}\BibitemShut {NoStop}%
\bibitem [{\citenamefont {Van~Dongen}\ and\ \citenamefont
  {Ernst}(1988)}]{van_dongen_scaling_1988}%
  \BibitemOpen
  \bibfield  {author} {\bibinfo {author} {\bibfnamefont {P.~G.~J.}\
  \bibnamefont {Van~Dongen}}\ and\ \bibinfo {author} {\bibfnamefont {M.~H.}\
  \bibnamefont {Ernst}},\ }\bibfield  {title} {\bibinfo {title} {Scaling
  solutions of {{Smoluchowski}}'s coagulation equation},\ }\href@noop {}
  {\bibfield  {journal} {\bibinfo  {journal} {Journal of Statistical Physics}\
  }\textbf {\bibinfo {volume} {50}},\ \bibinfo {pages} {295} (\bibinfo {year}
  {1988})}\BibitemShut {NoStop}%
\end{thebibliography}
\end{document}